\documentclass[aps,twocolumn,floats,pre,superscriptaddress,nofootinbib,longbibliography]{revtex4-2}

\usepackage{graphics,graphicx,epsfig}
\usepackage{amssymb,color}
\usepackage{epsf,epstopdf,wrapfig}
\usepackage{amsmath}
\usepackage{mathtools}
\usepackage{amssymb}
\usepackage{graphicx}
\usepackage{bm}
\usepackage{xr}
\usepackage{scrextend}
\definecolor{purple}{rgb}{0.58, 0.44, 0.86}
\usepackage{lipsum}  

\usepackage[dvipsnames]{xcolor}

\newcommand{\beqa}{\begin{eqnarray}}
\newcommand{\eeqa}{\end{eqnarray}}

\newcommand{\beq}{\begin{equation}}
\newcommand{\eeq}{\end{equation}}

\def\Nina{N_\mathrm{inact}}

\def\Nmiss{N_\mathrm{miss}}

\begin{document}

\title{Inheritance entropy: A model-independent method to probe the hereditary structure \\ of cell lineage trees}

\author{Alessandro Allegrezza}
\affiliation{Dipartimento di Fisica, Sapienza Universit\`{a} di Roma,  Rome, Italy}

\author{Riccardo Beschi}
\affiliation{Dipartimento di Fisica, Sapienza Universit\`{a} di Roma,  Rome, Italy}

\author{Domenico Caudo}
\affiliation{Dipartimento di Fisica, Sapienza Universit\`{a} di Roma,  Rome, Italy}
\affiliation{Center for Life Nano \& Neuro Science, Italian Institute of Technology,  Rome, Italy}

\author{Andrea Cavagna}
\affiliation{Istituto Sistemi Complessi, Consiglio Nazionale delle Ricerche, UOS Sapienza,  Rome, Italy}
\affiliation{Dipartimento di Fisica, Sapienza Universit\`{a} di Roma,  Rome, Italy}
\affiliation{Istituto Nazionale di Fisica Nucleare, Sezione Roma 1, Rome, Italy}

\author{Alessandro Corsi}
\affiliation{Dipartimento di Medicina Molecolare, Sapienza Universit\`{a} di Roma,   Rome, Italy}

\author{Antonio Culla}
\affiliation{Istituto Sistemi Complessi, Consiglio Nazionale delle Ricerche, UOS Sapienza,  Rome, Italy}

\author{Samantha Donsante}
\affiliation{Dipartimento di Medicina Molecolare, Sapienza Universit\`{a} di Roma,   Rome, Italy}
\affiliation{Tettamanti Center, Fondazione IRCCS San Gerardo dei Tintori, Monza, Italy}

\author{Giuseppe Giannicola}
\affiliation{Dipartimento di Scienze Anatomiche, Istologiche, Medico Legali e dell'Apparato Locomotore, Sapienza Universit\`{a} di Roma,  Rome, Italy}

\author{Irene Giardina}
\affiliation{Dipartimento di Fisica, Sapienza Universit\`{a} di Roma,  Rome, Italy}
\affiliation{Istituto Sistemi Complessi, Consiglio Nazionale delle Ricerche, UOS Sapienza,  Rome, Italy}
\affiliation{Istituto Nazionale di Fisica Nucleare, Sezione Roma 1,  Rome, Italy}

\author{Giorgio Gosti}
\affiliation{Center for Life Nano \& Neuro Science, Italian Institute of Technology,  Rome, Italy}
\affiliation{Istituto di Scienze del Patrimonio Culturale, Consiglio Nazionale delle Ricerche,  Montelibretti, Italy}

\author{Tom\'as S. Grigera}
\affiliation{Instituto de F\'\i{}sica de L\'\i{}quidos y Sistemas Biol\'ogicos, CONICET and  Universidad Nacional de La Plata,  La Plata, Argentina}
\affiliation{CCT CONICET La Plata, Consejo Nacional de Investigaciones Cient\'\i{}ficas y T\'ecnicas, Argentina}
\affiliation{Departamento de F\'\i{}sica, Facultad de Ciencias Exactas, Universidad Nacional de La Plata, Argentina}
\affiliation{Istituto Sistemi Complessi, Consiglio Nazionale delle Ricerche, UOS Sapienza,  Rome, Italy}
    
\author{Stefania Melillo}
\affiliation{Istituto Sistemi Complessi, Consiglio Nazionale delle Ricerche, UOS Sapienza,  Rome, Italy}
\affiliation{Dipartimento di Fisica, Sapienza Universit\`{a} di Roma,  Rome, Italy}

\author{Biagio Palmisano}
\affiliation{Dipartimento di Medicina Molecolare, Sapienza Universit\`{a} di Roma,   Rome, Italy}
\affiliation{Istituto Sistemi Complessi, Consiglio Nazionale delle Ricerche, UOS Sapienza,  Rome, Italy}

\author{Leonardo Parisi}
\affiliation{Istituto Sistemi Complessi, Consiglio Nazionale delle Ricerche, UOS Sapienza,  Rome, Italy}
\affiliation{Dipartimento di Fisica, Sapienza Universit\`{a} di Roma,  Rome, Italy}

\author{Lorena Postiglione}
\affiliation{Istituto Sistemi Complessi, Consiglio Nazionale delle Ricerche, UOS Sapienza,  Rome, Italy}
\affiliation{Dipartimento di Ingegneria Chimica, dei Materiali e della Produzione Industriale, Universit\`{a} degli Studi di Napoli Federico II, Napoli, Italy}

\author{Mara Riminucci}
\affiliation{Dipartimento di Medicina Molecolare, Sapienza Universit\`{a} di Roma,   Rome, Italy}

\author{Francesco Saverio Rotondi}
\affiliation{Dipartimento di Fisica, Sapienza Universit\`{a} di Roma,  Rome, Italy}

\begin{abstract}
Quantifying the impact of hereditary transmission within lineage trees remains a fundamental challenge universal to a wide array of biological domains. Here, we introduce the new concept of inheritance entropy, a quantity designed to gauge the hereditary structure of inactive cells across a lineage. We measure this entropy in $32$ human stem cell clonal colonies, obtained from high-definition single-cell lineage tracing experiments, and show that in the greatest majority of clones the entropy is decisively smaller than that of the corresponding non-hereditary ensemble, hence proving that variations in the proliferative power of stem cell lineages are determined by hereditary epigenetic factors that regulate cell-cycle exit.  The method can also be employed to locate the specific node of the tree where a mutation in the probability of inactivity occurs, together with a determination of the lag between the mutation ultimately leading to inactivity and its actual expression. This framework can be used to assess in a robust, simple and model-agnostic way the hereditary origin of differential growth in any type of lineage trees.
\end{abstract}

\maketitle

\section{Introduction}

% studi precedenti di cell inheritance
The role of inheritance in cell replication has been a subject of intense study for a long time \cite{ng2008epigenetic, zion2020asymmetric, sandler2015lineage, mosheiff2018inheritance, hormoz2016inferring, chang2008transcriptome, plambeck2022heritable}. It is believed that cells switch between molecularly and phenotypically distinct states that are passed on to the descendants, thus regulating their fate \cite{hormoz2016inferring,yampolskaya2025}. One way to assess inheritance is therefore to characterize cell states and track how they evolve along a lineage. Single-cell profiling allows to measure the whole transcriptome, as well as proteomes and metabolic signatures, hence characterising the multi-dimensional space of cell states (the `state manifold' or `transcriptional landscape' \cite{wagner2020lineage}); but this type of analysis is hard to combine with classic lineage tracing techniques, which are however important to precisely  track the hereditary ramifications. A powerful alternative is provided by barcoding lineage tracing, which allows to acquire simultaneously cell state information and partial lineage relationships from bulk measurements \cite{baron2019unravelling, weinreb2020lineage, raj2025single, spanjaard2018simultaneous}. However, despite their great potential, barcoding techniques still present limitations in terms of applicability to human cells \cite{ihry2018p53} and accuracy in lineage reconstruction \cite{wagner2020lineage}.

% definizione e rilevanza della non-genetic inheritance
Epigenetic inheritance, namely the transmission of phenotypic traits across generations without changes to DNA sequence, is particularly elusive and difficult to assess: non-genetic transmission can occur through a great variety of molecular mechanisms, among which DNA methylation, histone modifications, and non-coding RNA regulation \cite{Deichmann2016epigenetics, Jablonka2002epigeneticsterm}, which makes it hard to pin down its hereditary nature in a sharp way.  Epigenetic inheritance, though, is also hugely relevant, unfolding its full potential in the realm of stem cells, whose high replicative capacity and promising regenerative potential make them the ideal arena to study how different phenotypic traits emerge and evolve within a clonal population. Besides, differentiation and cell-fate --  which are core issues of stem cell biology -- strongly rely on epigenetic mechanisms \cite{Lunyak2008epigenetic, Meissner2010epigeneticstemcells}. Exploring epigenetic inheritance in cell lineages is therefore a further urgent reason to try and find robust methods of hereditary assessment.

% why inactivity inheritance matters in stem cell research? % and in BMSC in particular?
A particularly general trait, defining the hereditary structure of lineage trees at a very fundamental level, is {\it inactivity}, namely the state the cell enters when it stops dividing. This is primarily for three reasons. First, because in the context of cellular aging, inactivity essentially means senescence, whose epigenetic origin has huge clinical impact \cite{ogrodnik2021cellular, koch2012monitoring, franzen2017senescence, franzen2021dna}. Secondly, because inactivity is in general a more complex state than just senescence: an inactive cell can be reversibly so (quiescent) or irreversibly inactive (differentiated or senescent), a difference that impinges greatly on the fate of cells \cite{nachtwey1969cell, cheung2013molecular, behl2013cell, mens2018cell}. Finally, inactivity is a very general template for any mechanism differentially cutting branch proliferation within a lineage tree, irrespective of the specific nature of the tree. Hence, studying the hereditary ramifications of inactivity has a potentially broad impact on the more general scope of gauging inheritance in {\it any} kind of tree.

% peculiarities of inactivity as an hereditary trait
Inactivity, however, has some peculiarities that make its hereditary character -- if any --  particularly hard to pin down.
Normally, inheritance implies that when a mutation emerges in a cell, thus giving rise to a new phenotype, this mutation is inherited by its progeny, thus amplifying the detectability of that phenotype with increasing generations: when enough doublings have been reached, there is sufficient bulk information in the culture to map the hereditary ramifications of that mutation through bulk methods. The tricky thing about inactivity, though, is that its very impact on the tree development -- namely the obliteration of entire branches -- is also what makes it hard to characterize it at an hereditary level: once inactivity emerges in one cell, this phenotype is {\it not} amplified in its progeny, as the effect of inactivity is rather to delete the descendants of that one cell, thus erasing all the information they potentially carried.  Bluntly put, it is impossible to extract information from cells that do not exist. 

% upstream, not downstream
Since all downstream branches of the inactive cell are missing, it seems instead that one might try to extract information from its {\it upstream} branches --- or progenitors --- which have been alive at some point in the tree development.  This strategy, though, can be pursued only if inactivity manifests itself a few generations {\it after} the mutation causing it, otherwise neither the future nor the past generations would contain any information about the emergence of inactivity. At least in the case of senescence, there is growing evidence that this is indeed the case: senescence is characterized by markers that accumulate in a gradual manner {\it prior} to its expression \cite{koch2012monitoring, franzen2017senescence, franzen2021dna, ogrodnik2021cellular}; hence, it is possible that this happens more generally for the expression of inactivity. And yet, the very absence --- by definition --- of the inactivity trait in the progenitors of the inactive cell, makes it unclear how to phenotypically retrace upstream the origin of the mutation and its hereditary unfolding along the lineage. 
 
% what we do here 
Finding a new method to back-track the ramifications of inactivity, hence establishing beyond doubt its hereditary character, is what we do here. The central idea behind this result is to extract information about the hereditary distribution of inactive cells through the calculation of the Shannon entropy associated to the topology of the entire lineage tree. We test this method through novel experiments of high-resolution single-cell lineage-tracing of bone marrow stromal cells clonal colonies and find that indeed in the greatest majority of lineages the entropy is decisively smaller than that of the corresponding non-hereditary ensemble. Moreover, through our method, the epigenetic change giving rise to inactivity can be traced upstream, thus revealing a map of the mutation events across the colony development. This will allow us to measure the mutation lag, namely the generation gap between the mutation leading to inactivity and the actual expression of its phenotype.

%%%%%%%%%%%%%%%%%%%%%%%%%%%%%%%%%%%%%%
\section{Lineage topology}

\subsection{The system}

Human Bone Marrow Stromal Cells (BMSCs) are a prominent source of skeletal stem cells with ground-breaking the\-ra\-peutic potential \cite{bianco2015skeletal}. However, BMSC colonies have very heterogeneous in vivo behaviour, due to their different potency; this unpredictability is the greatest hurdle to the development of bone regeneration therapies \cite{bianco2013meaning}.

Colony-level heterogeneity urges a fundamental question: how is it possible that one colony as a collective unit behaves differently from another one? If cell-to-cell variability were just an uncorrelated random process, a million cells in a transplant-bound colony would be enough to yield statistical homogeneity, hence washing out any colony-level traits. A possible answer is that the differences between two originating cells are transmitted to their progenies and collectively persist through an epigenetic hereditary mechanism.

If hereditary epigenetic factors play a major role in determining cell-cycle exit, they may be involved also in the regulation of  BMSCs potency, thus shedding some light on that elusive holy grail of stem cell research that is differentiation potential \cite{mareddy2007clonal, mareddy2010stem}. Controlling the hereditary character of inactivity could therefore help developing protocols to control inter-colony heterogeneity, thus advancing the clinical potential of skeletal stem cells. For these reasons, BMSCs are a very important testbed system for the development of new methods to detect inheritance in lineage trees.

%%%%%%%%%%%%%%%%%%%%%%%%%%%%%%%%%%%%%%%%%%
\begin{figure*}
\centering
\includegraphics[width=0.9 \textwidth]{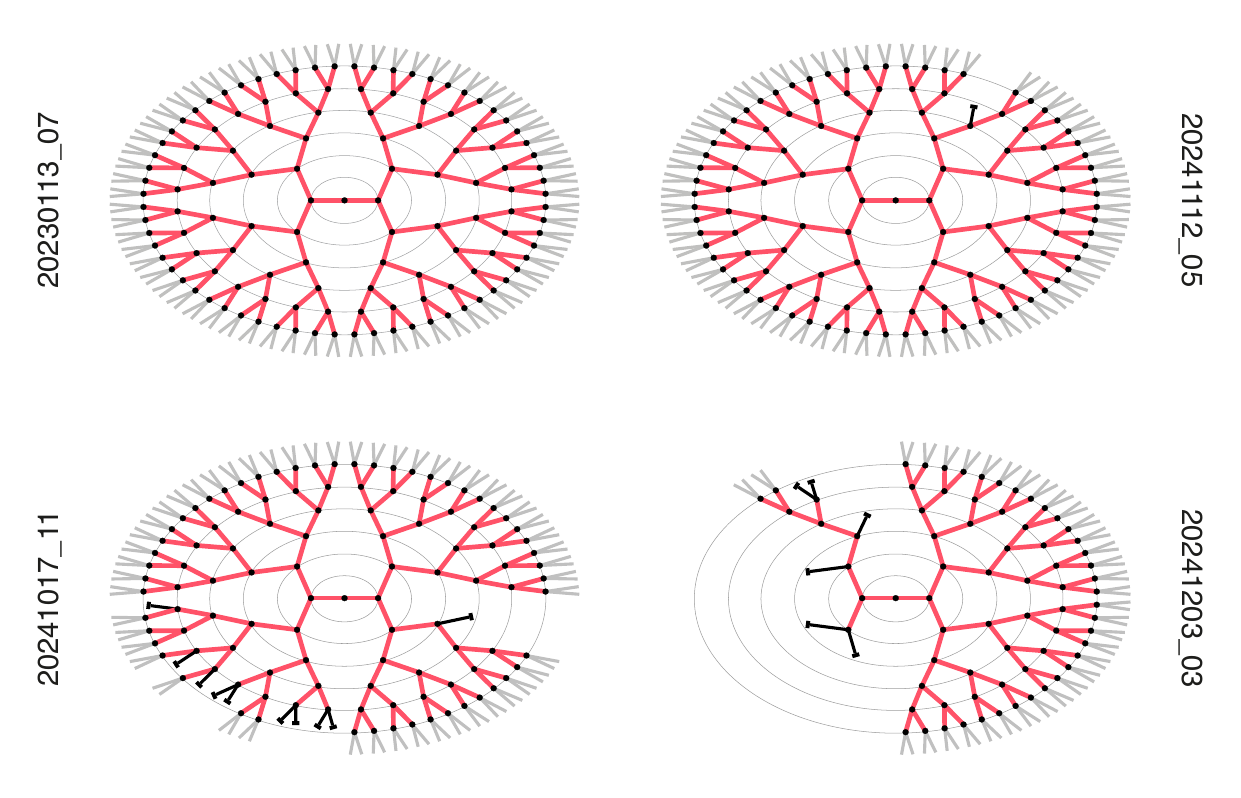}
\caption{{\bf Lineage trees of experimental BMSC colonies.}
Red segments represent active cells and black discs represent mitosis; the length of each segment is fixed, not related to the division time. Inactive cells (which are all G$_0$ in these samples) are represented as black segments ending in a cap. The first cell of the colony is not represented, as we do not record its birth, but only its division (the central disc of the tree). Ovals separate different generations. Cells at the last generation (the $k_7$ leaves), are tracked up to their birth, not up to their division, hence they are neither labeled as active nor inactive and we represent them in grey.
}
\end{figure*}
%%%%%%%%%%%%%%%%%%%%%%%%%%%%%%%%%%%%%%%%%%%

\subsection{Experimental data}

% cells and experiment
Human BMSCs were seeded at very low density, in order to produce clonal colonies derived only from single, spatially isolated cells; our dataset consists of 32 such single-cell-derived clones. The isolated nature of each colony is very important: if different colonies merge, their clonal nature is destroyed, severely confounding the data analysis \cite{kuczek1986importance, rennerfeldt2016concise}. Growth was studied with phase contrast high-resolution time-lapse microscopy. Instead of adopting a fixed duration of the experiments, we fix the maximum generation of the colonies, keeping each experiment going until all replicating cells  have arrived to the seventh generation, $k_\mathrm{max}=7$; in this way we do not need to correct for any type of statistical bias \cite{powell1955some, genthon2023cell}. Details of the experiment and a full characterisation of the biological parameters of all colonies, can be found in \cite{allegrezza2025topology}.

% graphical represenation of the lineage
Samples of four lineage trees are reported in Fig.1, while all trees in the dataset can be found in Section 9 of the Supplemental Material-SM \cite{SM}.
If no branches are interrupted, the number of cells at the seventh generation is $2^7=128$, and a handful of colonies do indeed reach this state (3 out of 32). But often branches {\it are} interrupted, which may happen for two reasons: either the cell commits apoptosis and dies, or the cell stops dividing, yet remaining alive (black capped segments in Fig.1); in this last case the cells enters the G$_0$ phase. We classify a cell as G$_0$ if it does not divide for $84$ hours ($3.5$ days), at which point we stop tracking it; this threshold is to be compared to the mean division time of BMSCs, which is $20\pm 4$ hours (the robustness of this G$_0$ criterion is thoroughly tested in \cite{allegrezza2025topology}). Only $17$ cells die in our whole dataset, against 310 G$_0$ cells; given such paucity of apoptosis and given that the effect of a dead cell on the lineage topology is exactly the same as a G$_0$ cell, we will call {\it inactive} the whole set of G$_0$ and dead cells.\footnote{This is a rather standard (and venerable) convention in the literature on cellular senescence: any difference between dead cells and cells that have ceased to divide is immaterial when dealing with population growth, see for example \cite{kirkwood1975commitment, holliday1977testing, harley1980retesting}.}  Conversely, we will call {\it active} those cells for which we do observe a mitosis. The final `leaves' of the tree, i.e. cells that belong to the last generation ($k_\mathrm{max}=7$), are only tracked up to their birth, not up to their division, hence they are neither labeled as active nor inactive; we call these cells the $k_7$ leaves of the lineage tree.
As we shall see later on, two useful characterisation of a tree are the total number of inactive cells, $N_\mathrm{inact}$, and the total number of cells missing at the last generation, $N_\mathrm{miss}$, that is the number of missing grey leaves in Fig.1.

%%%%%%%%%%%%%%%%%%%%%%%%%%%%%%%%%%%
\subsection{The peculiar distribution of inactive cells}
%%%%%%%%%%%%%%%%%%%%%%%%%%%%%%%%%%%

% different positions of inactive cells - Luria-Delbruck
Inactive cells determine the topology of a lineage  through both their number and their position within the tree. The position of an inactive cell can be described by two `coordinates': its generation $k$ and its radial position.  The first impacts greatly on the number of missing $k_7$ leaves, $N_\mathrm{miss}$: the earlier the generation at which a cell becomes inactive, the larger the number of leaves that are cut; for example, the effect of an inactive cell is very different if it is located at generation $k=2$ ($N_\mathrm{miss}=32$) or at $k=6$ ($N_\mathrm{miss}=2$).
At first sight, this effect seems related to the Luria-Delbr\"uck argument about the uneven impact of hereditary mutations \cite{luria1943mutations}: if the probability of a mutation is homogeneous throughout the lineage, the fluctuations in the occurrence at different generations causes very large heterogeneities in the final number of cells carrying the mutation, which is the conceptual basis for all Luria-Delbr\"uck-inspired fluctuation assays \cite{rosche2000determining}. 
However, because in this case the phenotype is ``inactivity" and because the very definition of inactive cell is that it does not have any progeny, the fact that the inactive phenotype is inherited by the (non-existent) descendants of the inactive cell is completely trivial and uninformative.  Therefore, to check whether or not the mechanism that ultimately leads to inactivity has an hereditary nature, we cannot use a standard fluctuation assay. Instead, we have to employ the lineage information to check whether some kind of mutation occurred {\it before} the emergence of the inactive cell, namely upstream in the tree. 

%%%%%%%%%%%%%%%%%%%%%%%%%%%%%%%%%%%%%%%%%%
\begin{figure}
\centering
\includegraphics[width=0.5 \textwidth]{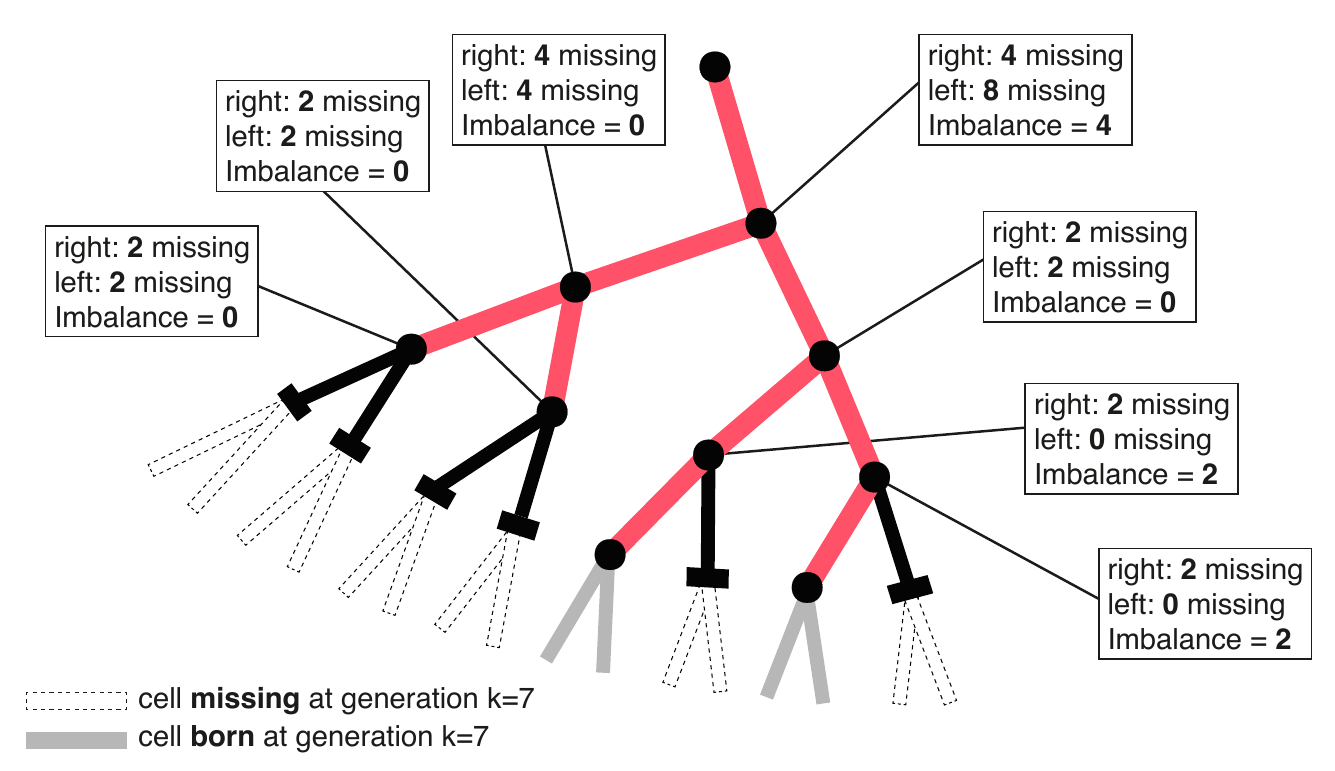}
\caption{{\bf Inactivity imbalance.}
We illustrate here  how the inactivity imbalance is calculated, using a portion of a real lineage (\texttt{20230503\_13}). White dotted segments represent $k_7$ leaves that are missing because of the presence of some inactive cells upstream in that branch.
For each mitosis/node, the inactivity imbalance is defined as the modulus of the difference between the number of missing $k_7$ leaves
in the right branch and the number of missing $k_7$ leaves in the left branch.}
\end{figure}
%%%%%%%%%%%%%%%%%%%%%%%%%%%%%%%%%%%%%%%%%%%

% weird distribution of inactive cells
To make progress, we notice that trees with many inactive cells display something more than the mere amplification of fluctuations due to the different inter-generation placement, namely the fact that the radial distribution of inactive cells across different branches has quite a nontrivial pattern.  Consider for example lineage \texttt{20241017\_11} in Fig.1: there is clearly something peculiar about the locations of the inactive cells in this tree, as they seem more frequent within the same few branches on the south-west side in a definitely non-random way; a similar observation could be made for lineage \texttt{20241203\_03}, where inactive cells are mostly concentrated in the west wing.  This is a very general trait in our dataset: in most lineages we observe significant non-random heterogeneities in the number of inactive cells between branches, indicating that the emergence of these cells is more likely in some branches than in others. If we follow a branch with many inactive cells upstream along the lineage, an hereditary mechanism would suggest that at some point we must find a mitosis at which some factor linked to the probability of inactivity changed, making one of the two daughters more prone to producing inactive cells in its progeny than the other.  We need to put this hypothesis to a quantitative test; to do that we will use the concept of {\it entropy}.

%%%%%%%%%%%%%%%%%%%%%%%%%%%%%
\section{Quantifying inheritance}
%%%%%%%%%%%%%%%%%%%%%%%%%%%%%

\subsection{Inactivity imbalance}

As we have noted, the most conspicuous effect of the presence of an inactive cell is that all cells that would have been born as its progeny are instead missing, so that the number of cells at the final generation, $k_\mathrm{max}=7$, is smaller than what it could have been.  Hence, we will use the number of missing $k_7$ leaves as an indicator of the impact of inactive cells within that branch. We proceed as follows.

Each mitosis in the lineage generates two branches, left and right; in a mitosis connecting generation $(k-1)$ to generation $k$,  each branch can potentially produce up to $2^{(7-k)}$ descendants at generation 7.  However, if inactive cells emerge downstream along these branches, the {\it actual} number of $k_7$ descendants will be smaller (some of these leaves will be missing).  Given a mitosis $m$, we will indicate with $M_m^\mathrm{left}$ the number of missing $k_7$ leaves in its left branch and similarly for $M_m^\mathrm{right}$ (see Fig.2). We stress that the count of missing cells is done only among the cells of the last recorded generation (in our experiments, the $k_7$ leaves).

%%%%%%%%%%%%%%%%%%%%%%%%%%%%%%%%%%%%%%%%%%
\begin{figure*}
\centering
\includegraphics[width=1.0 \textwidth]{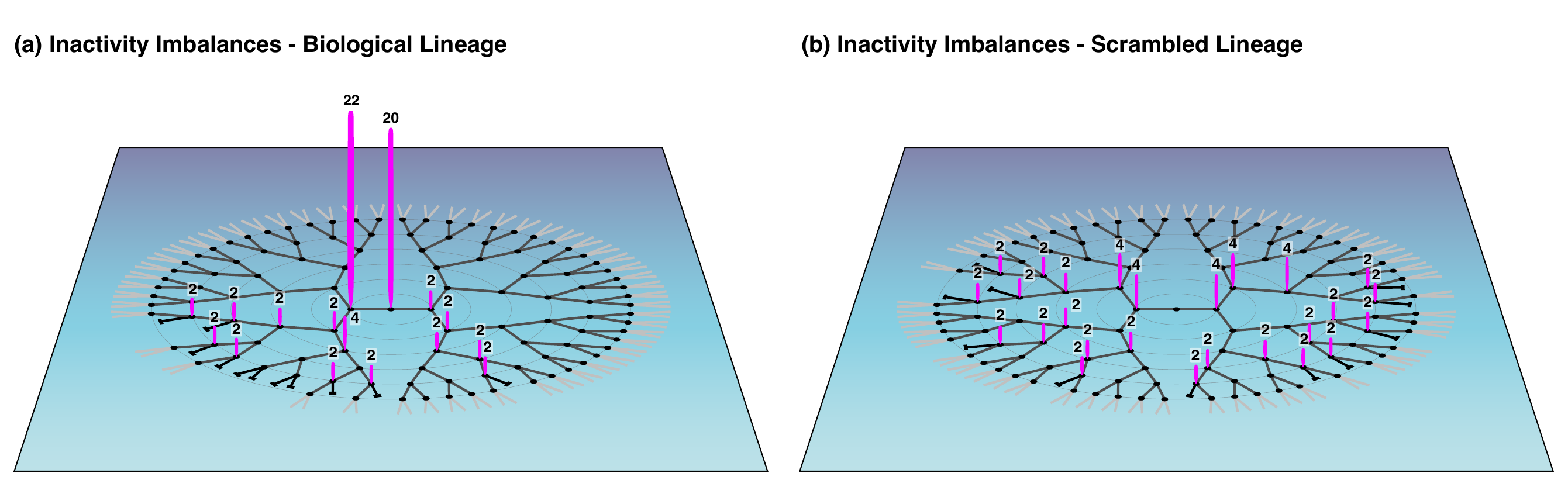}
\caption{{\bf Inactivity imbalance - BMSC vs scrambled.} 
The distribution of the inactivity imbalance across the different mitosis is very different in biological vs.\ randomly scrambled trees.
{\bf(a)} 
For lineage \texttt{20230503\_13}, the inactivity imbalance, $I_m$, is graphically represented as a purple vertical brushstroke on each mitosis/node (zero imbalances are not marked).
In this lineage most of the imbalance is concentrated on just two nodes; of these two mitosis, the one connecting generation 1 to generation 2  (imbalance 22) is most likely that at which there has been a mutation in the probability of emergence of inactive cells. This large heterogeneity in the distribution of the imbalances across the lineage is the clearest symptom of inheritance. The entropy $S$ measures exactly this heterogeneity; in this lineage $S_\text{biological}  = 2.17$, which is rather low compared to:
{\bf(b)} 
One instance out of the $10^6$ randomly reshuffled versions of lineage \texttt{20230503\_13} (RR ensemble - see text); the number of inactive cells and their generation is the same as in the biological lineage, but their radial positions across the tree have been randomly scrambled, hence severing all potential hereditary relationships. In this tree the imbalance is more homogeneously distributed across the nodes than in the biological tree, thus giving a much lower inheritance signal: the entropy is indeed significantly larger than the biological one, $S_\mathrm{scrambled}  = 3.21$. If we repeat the scrambling $10^6$ times, in only 7 cases we find $S_\mathrm{scrambled}  \leq S_\text{biological}$, proving that the evidence of inheritance is very strong in this clone.
}
\end{figure*}
%%%%%%%%%%%%%%%%%%%%%%%%%%%%%%%%%%%%%%%%%%%

We then define the {\it inactivity imbalance} of mitosis $m$ as the difference between the number of missing $k_7$ cells within its left and right branches (Fig.2),
\begin{equation}
  I_m = \lvert M_m^\mathrm{left} - M_m^\mathrm{right} \rvert \ ,
  \label{dupina}
\end{equation}
where we use the absolute value because the actual left-right labelling of the branches after a division is arbitrary.

The definition of imbalance can be generalized to any value of $k_\mathrm{max}$. Moreover, the imbalance can be defined also for trees where all cells become inactive before $k_\mathrm{max}$: in this case, one can define an effective $k_\mathrm{max}^\mathrm{eff}$, as the generation of the latest inactive cell; then, the definition proceeds as before, with $k_\mathrm{max} \to k_\mathrm{max}^\mathrm{eff}$. Such incomplete trees are not present in our experimental dataset, but they may be generated by the null models that we will consider in the next Section.

The interesting thing about the inactivity imbalance is that it provides a sort of `mutation' map, giving a specific information about what happens at each mitosis along the tree (see Fig.3, left).  A large imbalance on a certain node means that the two branches generated by that mitosis are {\it ultimately} very different from each other, suggesting that a mutation in the probability of generating inactive cells occurred there.  From a topological point of view, the mutation is most easily attributed to the node, i.e. to the  mitosis, whose large imbalance marks the difference between the two sub-branches born out of that mitosis; from a biological point of view, though, it is more appropriate to attribute the mutation to the one of the two sister cells born out of that mitosis, whose progeny is riddled with inactive descendants (this is also the most natural interpretation in the context of hereditary models).
The main point is that the inactivity imbalance provides information about a possible change in the propensity to inactivity {\it before} -- namely upstream in the tree --  the inactivity phenotype is actually expressed.  The imbalance is therefore a promising observable in the search of an hereditary test of cell-cycle exit.

\subsection{Entropy}

At variance with other measures of tree imbalance, such as the Colless index \cite{colless1982phylogenetics,fischer2023tree}, we do not consolidate the inactivity imbalance \eqref{dupina} into a sum over all nodes of the tree. This is a crucial point: when trying to characterize inheritance, a large inactivity imbalance at a few nodes is \emph{not} the same as a buildup of small inactivity imbalances scattered over many nodes.

Instead, inheritance is signalled by an exceptional event, namely a large imbalance concentrated on one or a few nodes, revealing a difference between the two cells generated at that mitosis, a difference that is passed on to their progenies (Fig.3, left). Therefore, we need a quantity able to reveal when there are atypically large values of the inactivity imbalance in the lineage, and when --- on the other hand --- small imbalances are scattered all over the tree. The most promising tool to do this is the Shannon entropy. From the imbalance, we first derive the normalized weight of mitosis $m$,
\beq
w_m = \frac{I_m}{\sum_n I_n} \ ,
\label{pesa}
\eeq
where the sum is extended over all mitosis in the lineage. Then, using the weights $w_m$, we define the lineage entropy as,
\begin{equation}
S = - \sum_m w_m \log w_m \ .
\label{stacce}
\end{equation}
The entropy $S$ is high when the inactivity imbalances $I_m$ all have approximately the same value, forming a random pattern across the tree, so that we receive no particular information about the occurrence of an inherited change at any mitosis in the lineage; on the other hand, entropy is low when there are few large imbalances concentrated on a handful of mitosis, as in that case we have reliable information that an inherited change occurred at those mitosis (Fig.3, left). In short --- and very coarsely --- high entropy means that inheritance is unlikely, while low entropy means positive evidence of inheritance. As we shall see in the next Section, though,``high" and ``low" entropy will only be significant in a comparative way.

We stress that the entropy in \eqref{stacce} is {\it not} the entropy associated to the probability distribution of the inactivity imbalance,  because the weight $w_m$ is not the probability of a certain value of $I_m$, but is proportional to the imbalance itself; indeed, the sums do not run over all possible values of the imbalance, but over all possible nodes (i.e.~mitosis) over which the imbalance is defined. For the same reason, the normalized weights $w_m$  do not have an obvious frequentist interpretation; therefore, we should not blindly rely on the standard properties of $S$ in the context of Shannon information theory.  
It is also worth noting that similar concepts have been employed in different contexts: in astronomy an analogous quantity has been defined from the intensity field of galaxy radioastronomical data for image reconstruction \cite{gull1978image}; in economics, the entropy in \eqref{stacce} is related to Theil's index of economic inequality \cite{theil1967economics}. In all these contexts, the key idea is to pinpoint the anomalous values of some observable (imbalance, intensity, inequality) across a certain ``space" (lineage, image, country), rather than across a certain probability distribution. This is what the entropy defined in \eqref{stacce} does.

\subsection{Inheritance test}
% inheritance P-value
The entropy of a lineage is designed to depend on the specific arrangement of the inactive cells, i.e. on their hereditary correlations. But $S$ also depends on the total number of inactive and missing cells, irrespective of their hereditary distribution, and on the overall size of the tree. Therefore, it is hard to establish the absolute degree of inheritance of a tree by simply measuring its entropy $S$. Instead, the correct way to proceed is to compare the entropy of a specific biological lineage to that of an ensemble of {\it non-hereditary} trees.
 
We proceed as follows: the inheritance entropy is measured for a large number $\cal R$ of trees belonging to the non-hereditary (or null) ensemble, thus producing a non-hereditary probability distribution of $S$; the hereditary significance of the biological tree, namely its P-value, is given then by the fraction of non-hereditary trees that happen to have $S \leq S_\text{biological}$, namely by the probability that the null ensemble produces a tree with evidence of inheritance stronger than or equal to the biological one. The lowest P-value (highest significance) is $P< 1/\cal R$, which happens when none of the $\cal R$ non-hereditary trees has entropy smaller than the biological one.

Such inheritance test requires the definition of an appropriate null ensemble of non-hereditary lineages. One way to proceed is to generate this ensemble by means of a non-hereditary null-model; given a biological lineage we want to test, the null model must be fitted to that lineage, which is generally feasible by using the maximum likelihood method, considering that the null model will not be overly complicated (otherwise it would hardly be a {\it null} model). This apparently straightforward procedure, though, is less trivial than it might seem: as we shall see in the next Section, the choice of the null model will not be unequivocal. Moreover, we will discover that the null ensemble that is most effective to assess inheritance, i.e. the one with the highest sensitivity, is not generated by any specific null model.

%%%%%%%%%%%%%%%%%%%%%%%%%%%%%%%%%%%%%%%%%%%%%%%%%

\section{Non-Hereditary ensembles and test validation}
In this Section we will discuss three different types of null ensembles that can be used to implement the inheritance test. Two of them are generated by their relative null models, while the third null ensemble requires no generating model, as it simply derives from the data.
In the end, after a careful benchmarking of the three methods, only one inheritance test will be selected
(called Random Reshuffling -- RR). The reader not interested in this benchmarking tour-de-force can skip this Section.

\subsection{The Galton-Watson ensemble (GW)}
The simplest non-hereditary model of inactivity is one in which each cell is inactive with probability $q$. This is equivalent to a Galton-Watson branching process (GW --  \cite{harris1963theory}), with probabilities to have either zero or two children given by $p_0=q$ and $p_2=1-q$. Fitting GW to a biological tree is simple: the parameter $q$ which maximizes the likelihood is  the empirical value of $q$, i.e. the total number of inactive cells divided by the total number of cells of that tree (see SM-1).

Note that some trees of the null ensemble may not arrive at $k_\mathrm{max}=7$, because all cells become inactive before; this is the reason why in Section III we have defined the imbalance also for incomplete trees. Moreover, a very small fraction of trees of the ensemble have imbalance zero at each site, in which case the entropy is not defined; we discard these trees from the ensemble.

Even within a non-hereditary tree the imbalance has a mean trend with the generations: early generations expect a higher imbalance than late ones, just because the randomness in the longer subtrees adds up. When using a null model it is important to normalize the imbalance of the tree to be tested and of the trees in the null ensemble, by the expected imbalance of the null model, which can be calculated exactly for GW (see SM-1). Once the imbalance has been thus normalized, the calculation of the entropy proceeds as in \eqref{pesa} and \eqref{stacce}. Finally, the P-value of a certain lineage, $P_\mathrm{GW}$, is given by the fraction of GW trees with entropy smaller than or equal to the entropy of that lineage.

Despite the clarity and elegance of the GW model, it is possible to foresee a problem: this simple version of the GW ensemble {\it does not} encompass all possible non-hereditary trees; the probability to be inactive could very well be dependent on the generation, $q=q_k$, producing trees that are still non-hereditary, but unlikely to be produced by a GW model with constant $q$, thus exposing the method to the risk of creating false positives. The response to that is to use a more complete null ensemble, which can be done by introducing a GW model where the probability of inactivity depends on generation.

\subsection{The Non-Homogeneous Galton-Watson \\ ensemble (NHGW)}
The Non-Homogeneous Galton-Watson  model (NHGW) is an uncorrelated branching process where the probability to be inactive depends on the generation $k$, so that the model is defined by a vector $q_k$ \cite{d1992supercritical,kersting2020unifying}. There are sound biological reasons for using NHGW instead of GW, irrespective of any consideration of inheritance: there is plenty of evidence \cite{hayflick1965limited, robbins1970morphologic,rubin1997cell} showing that the fraction of senescent or dead cells can change significantly with the age of the cell culture, and therefore with the generations. Hence, it is unwise to commit to a null model that excludes this possibility.

As in the case of GW, also for NHGW the maximum likelihood method prescribes that the correct vector $q_k$ to fit a specific tree to the model is simply given by the empirical set of inactivity probabilities, i.e. the total number of inactive cells divided by the total number of cells, both calculated at fixed generation $k$ for that tree (see SM-2). Moreover, even though the calculation of the expected NHGW imbalance is significantly more complicated than in GW, it can still be done exactly (see SM-2), so we can normalize the imbalance and calculate the entropies, finally producing a P-value of this ensemble, $P_\mathrm{NHGW}$.

\subsection{The Randomly Reshuffled ensemble (RR)}
The third non-hereditary ensemble that we consider does not rely on a null model, but it operates directly on the lineage tree to be probed. The idea is to produce an ensemble of trees that have as much as possible the same structure in terms of inactive cells and missing cells as the original tree, but where all potential inheritance relations between mother and daughter have been severed. This ensemble is easily achieved by performing a Random Reshuffling (RR) of the input  tree: at each generation $k$ all cells are scrambled {\it within that same generation}; when two cells are exchanged, so are their progenies; then we move to generation $k+1$ and we scramble again, and so on (see SM-3 for a detailed description of the reshuffling procedure). Notice that in this way all potential hereditary connections are randomized and yet the number of inactive cells at each generation remains the same (see Fig.3, right). We shall see that this structural stability gives to the RR ensemble a decisive advantage over the other two ensembles. The P-value of the RR method, $P_\mathrm{RR}$, is finally given by the fraction of scrambled trees that happen to have entropy lower than or equal to the one of the lineage we are probing.

\subsection{Benchmarking the methods}
To evaluate the performance of the three entropy-based inheritance tests we must compare their output to some ground truth, i.e. to run the tests on a large set of lineage trees of known inheritance properties. This can be done by using a testbed of trees generated by a non-hereditary model (in which case the test should be negative) and by an hereditary model (in which case the test should be positive). We must be careful, though: at {\it finite size}, a tree generated by a non-hereditary model has a finite probability to have an hereditary structure, and {\it vice-versa}; any stochastic model can produce these cross-cases when the trees' size is small enough. Therefore, evaluating the methods only on trees of one size (for example, $k_\mathrm{max}=7$, as in our experimental dataset) is unwise. It is only in the {\it infinite size} limit, $k_\mathrm{max} \to \infty$, that the hereditary character of the generating model must emerge. Hence, the correct way to proceed is to generate a testbed of lineages using both non-hereditary and hereditary models, submit them to the different inheritance tests, and check how the results scale with the size of the trees.

We first benchmark the methods on a testbed of homogeneous non-hereditary trees, which we generate through the GW model with $q=0.1$.
For each method we measure the fraction $\Phi$ of trees that pass the inheritance test, i.e. such that $P<0.05$; we then plot $\Phi$ as a function of the trees' size, $k_\mathrm{max}$. Results are reported in Fig.4a: a sound inheritance test must produce a small fraction of trees passing the test; more precisely, because we are using $0.05$ as a significance threshold for the P-value, any correct method should give $\Phi \leq 0.05$. The results show that all three methods perform fine on homogeneous non-hereditary trees.

Things start to change when we benchmark the methods on a set of non-homogeneous non-hereditary trees, which we generate by using the NHGW model with a generation-dependent probability of inactivity, $q_k= q_0\exp(-k/k_0)$, with $q_0=0.8$ and $k_0=1.1$. Because this is still a non-hereditary set of trees, we expect all methods to give $\Phi\leq 0.05$; however, we see from Fig.4b that while this is true for the NHGW and RR methods, the inheritance test based on the GW null ensemble creates a very large number of false positives (around $60\%$). This is what we anticipated: many non-hereditary but non-homogeneous trees (as the stump tree in the inset of Fig.4b) are very rare in the homogeneous GW ensemble, thus giving a low P-value, and therefore a false positive. Other trees of this type, giving false positives for the inheritance test based on the GW ensemble, can be found in the tree zoology reported in SM-Section 9.

Finally, we benchmark the methods on a set of hereditary trees, which we generate through a simple model: 
a wild cell has probability $0$ to be inactive and probability $p$ to mutate;  
a mutated cell {\it and all its descendants} have probability $q$ to be inactive. 
As we shall see in Section VI, this model captures several features of our biological dataset, hence we generate the trees 
by using the mean parameters fitted to the biological lineages, $p=0.1$ and $q=0.46$ (see Section VI). Results are reported in Fig.4c. Both the NHGW and RR methods give a fraction of hereditary trees which increases with size; however, $\Phi_\mathrm{RR}$ grows dramatically faster than $\Phi_\mathrm{NHGW}$. More importantly, we can compare these curves to the ground truth: the asymptotic limit of $\Phi$ for $k_\mathrm{max}\to \infty$ can be calculated exactly for this hereditary model (see SM-5.6); for these parameters we have $\Phi_\infty=0.985$. Fig.4c shows that already at $k_\mathrm{max}=14$ the RR method is remarkably close to this exact value, $\Phi_\mathrm{RR}(14)=0.965$, a $2\%$ error with respect to the ground truth. On the contrary, the data show that the NHGW method plateaus at a value much smaller than the exact limit, $\Phi_\mathrm{NHGW}(14)=0.214$. 
We have repeated this benchmarking at different values of the parameters of the hereditary model ($p=0.08$, $q=0.4$) and found a very similar scenario: $\Phi_\mathrm{RR}$ converges to the ground truth with a $0.3\%$ error, while $\Phi_\mathrm{NHGW}$ remains off by more than $50\%$.
Finally, the GW method always underachieves dreadfully.

%%%%%%%%%%%%%%%%%%%%%%%%%%%%%%%%%%%%%%%%%%%%%%%%%
\begin{figure}
\centering
\includegraphics[width=0.47 \textwidth]{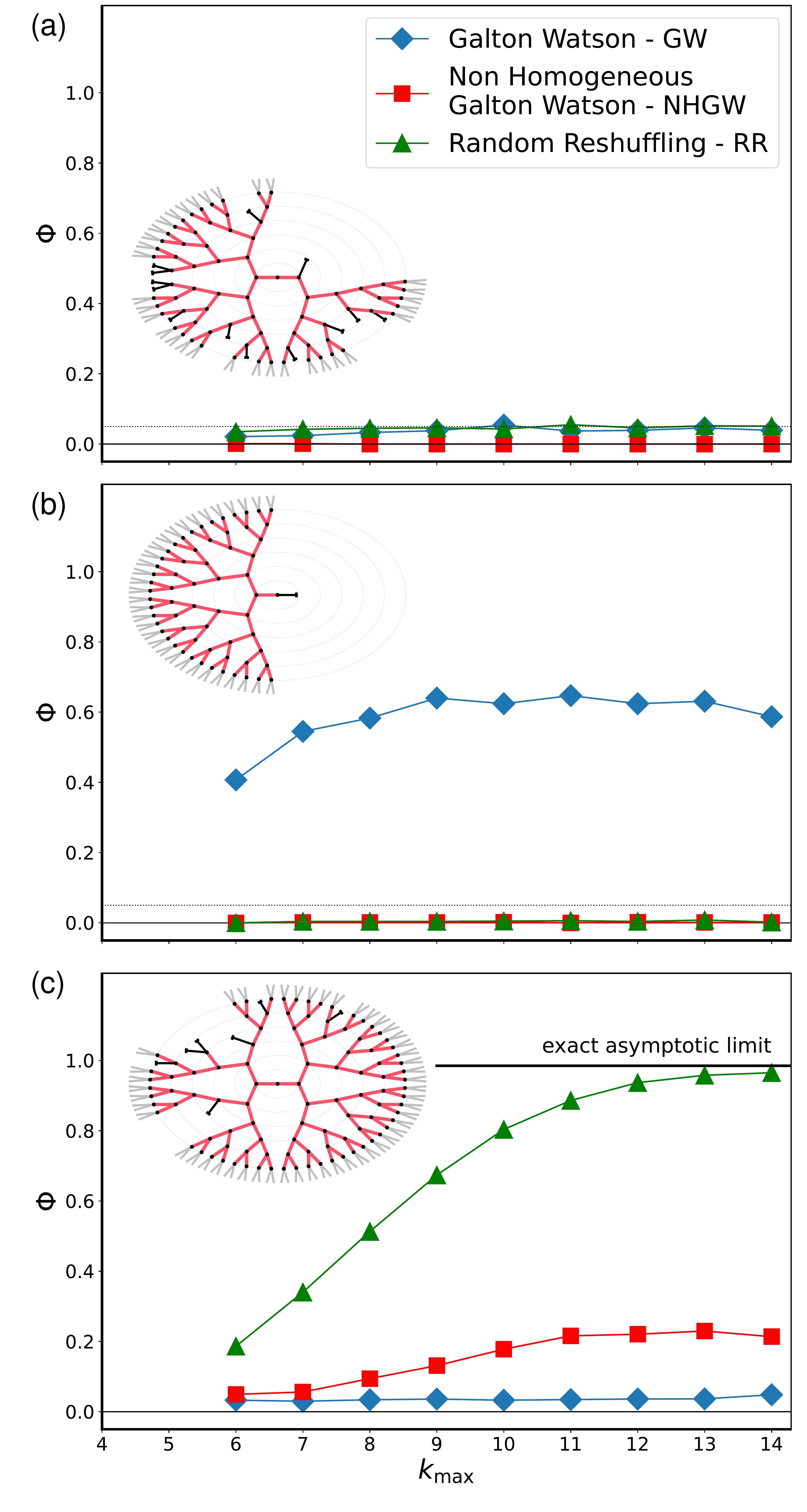}
\caption{
{\bf Benchmarking the inheritance tests.}
We plot the fraction $\Phi$ of trees that pass the inheritance test ($P< 0.05$) as a function of the size $k_\mathrm{max}$, for lineages of different hereditary nature.
{\bf a)} The three methods are benchmarked on a set of homogeneous non-hereditary trees. All inheritance tests give a satisfactory result in this case, i.e. a very small fraction of trees passing the test (below $0.05$, dotted line); we report a sample tree in the inset.
{\bf b)} On the set of non-homogeneous non-hereditary trees the GW inheritance test produces a large number of false positives, $\Phi_\mathrm{GW}\approx 0.6$; the tree in the inset is a vivid example: it has a probability of inactivity that decays rapidly with the generations, which GW is unable to account for, thus giving a low P-value.
Instead, both NHGW and RR perform well on this set.
{\bf c)} On the set of  hereditary trees, the RR method decisively outperforms the other two: $\Phi_\mathrm{RR}$ reaches the exact asymptotic value quite rapidly, while $\Phi_\mathrm{NHGW}$ saturates at a much smaller value and $\Phi_\mathrm{GW}$ never grows. 
For each tree the P-value is calculated from ${\cal R}=10^3$ extractions; for sets (a) and (b) we generate $10^3$ trees; for (c) -- given its relevance - we generate $10^4$ trees. In compliance with the experimental dataset, only trees in which at least two cells arrive at the last generation have been used to benchmark the methods.
}
\end{figure}
%%%%%%%%%%%%%%%%%%%%%%%%%%%%%%%%%%%%%%%%%%%%%%%

%%%%%%%%%%%%%%%%%%%%%%%%%%%%%%%%%%%%%%%%%%%%%%%%%
\begin{figure*}
\centering
\includegraphics[width=0.9 \textwidth]{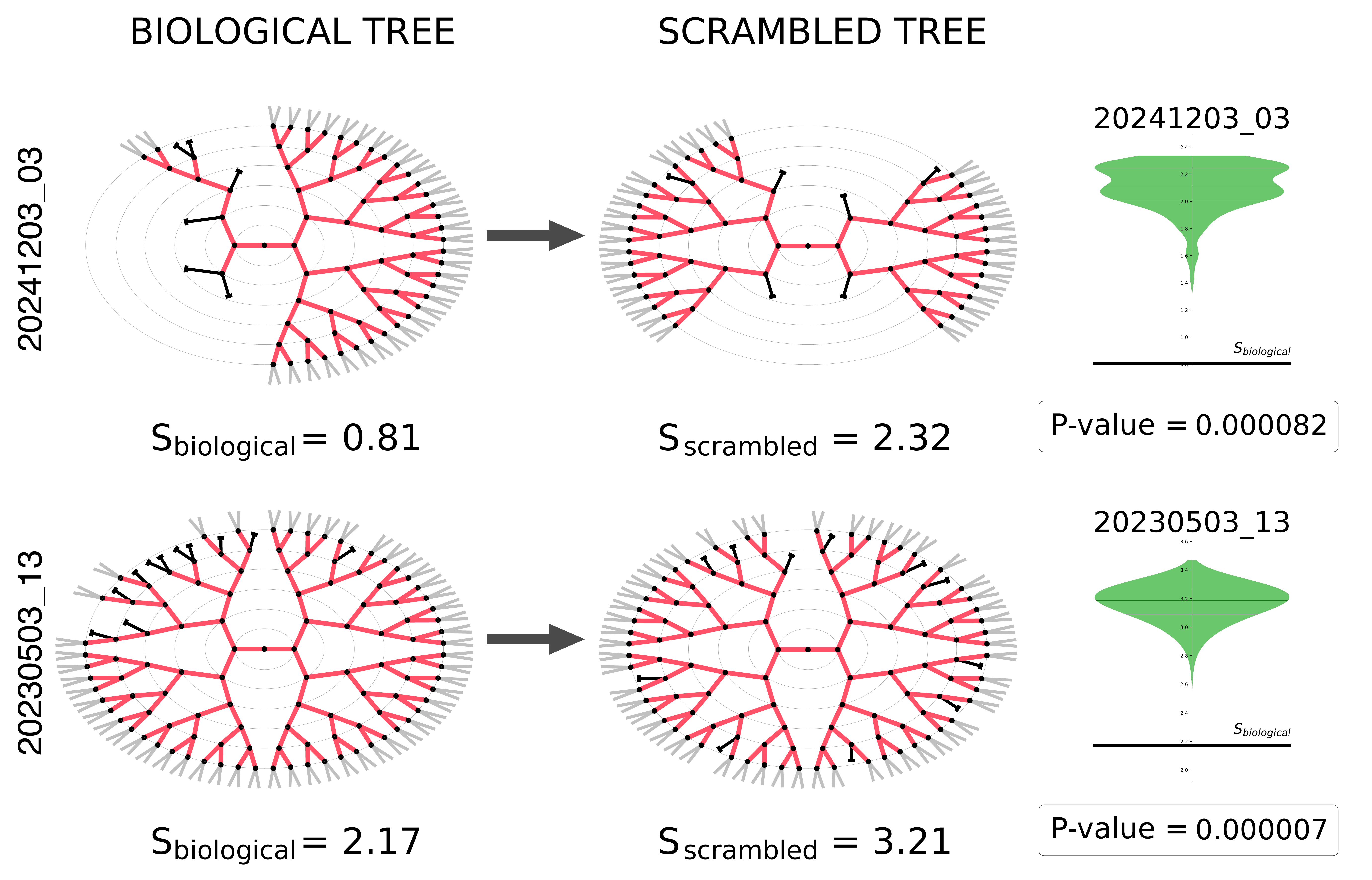}
\caption{{\bf Reshuffling BMSC lineages.}
Two examples of how the hereditary structure of the inactive cells is erased by the Random Reshuffling procedure (RR) and how this impacts on the inheritance entropy.
{\bf Top:} Lineage \texttt{20241203\_03} (left) has a strong concentration of inactive cells, and therefore of missing cells, in the west wing of the lineage; in this tree the imbalance is generated at the very first mitosis (the central node).  Instead, a randomly scrambled version of the biological tree (center) has inactive cells distributed evenly across all branches. The entropy of the biological tree is $S_\text{biological}=0.81$, while the  entropy of one its scrambled versions is $S_\mathrm{scrambled}=2.32$.  When we produce $10^6$ scrambled trees of \texttt{20241203\_03}, we very rarely find $S_\mathrm{scrambled} \leq S_\text{biological}$; this means that the probability that the entropy of the biological case is so small by pure chance is extremely low. The distribution of the randomly reshuffled entropies (right) is shown as a violin-plot on the right: it is evident that the biological value of the entropy is way below the bulk of the distribution of the reshuffled entropies. The P-value is simply the integral of this distribution below $S_\text{biological}$, which gives P$=8.2 \times 10^{-5}$, making the inheritance test for colony \texttt{20241203\_03} highly significant. 
{\bf Bottom:} A similar situation holds for lineage \texttt{20230503\_13}, which also gives a very strong inheritance signal.
}
\end{figure*}
%%%%%%%%%%%%%%%%%%%%%%%%%%%%%%%%%%%%%%%%%%%%%%%

The reasons why GW and NHGW perform poorly on the set of  hereditary trees are discussed and illustrated at length in SM-Section 4; here we just give a succinct version of the argument. Irrespective of the potential hereditary correlations between inactive cells, the entropy $S$ grows with their total number, $N_\mathrm{inact}$. While sampling their respective ensembles,  the GW and NHGW models explores many trees with numbers of inactive cells significantly different from that of the input hereditary tree they have been fitted to; such large fluctuations of $N_\mathrm{inact}$  produce in turn large fluctuations of $S$, introducing many small entropies in the ensemble, hence biasing to the left the probability distribution $P(S)$, ultimately raising the P-value. Instead, all trees in the RR ensemble have the same number of inactive cells at each generation as the input hereditary tree, hence the entropy $S$ changes only due to the {\it bona fide} changes in their hereditary correlations. An extensive zoology of trees generated with all models, together with their relative P-values, can be found in SM-Section 9.

In conclusion, the RR method emerges as the most effective and sensitive entropy-based test to assess the degree of inheritance in a lineage tree, irrespective of the type of tree. Luckily, RR is also by far the simplest to implement and the fastest to run of the three methods we scrutinized: it does not rely on any model, it does not require any calculation, it simply employs a large number of random reshufflings of the lineage tree that needs to be tested. Therefore, in the following we will use the RR entropy-based inheritance test to analyse the biological database of BMSC clonal colonies.

%%%%%%%%%%%%%%%%%%%%%%%%%%%%%%%%%%%%%%%%%%%%%%%%%
\begin{figure*}
\centering
\includegraphics[width=1.0 \textwidth]{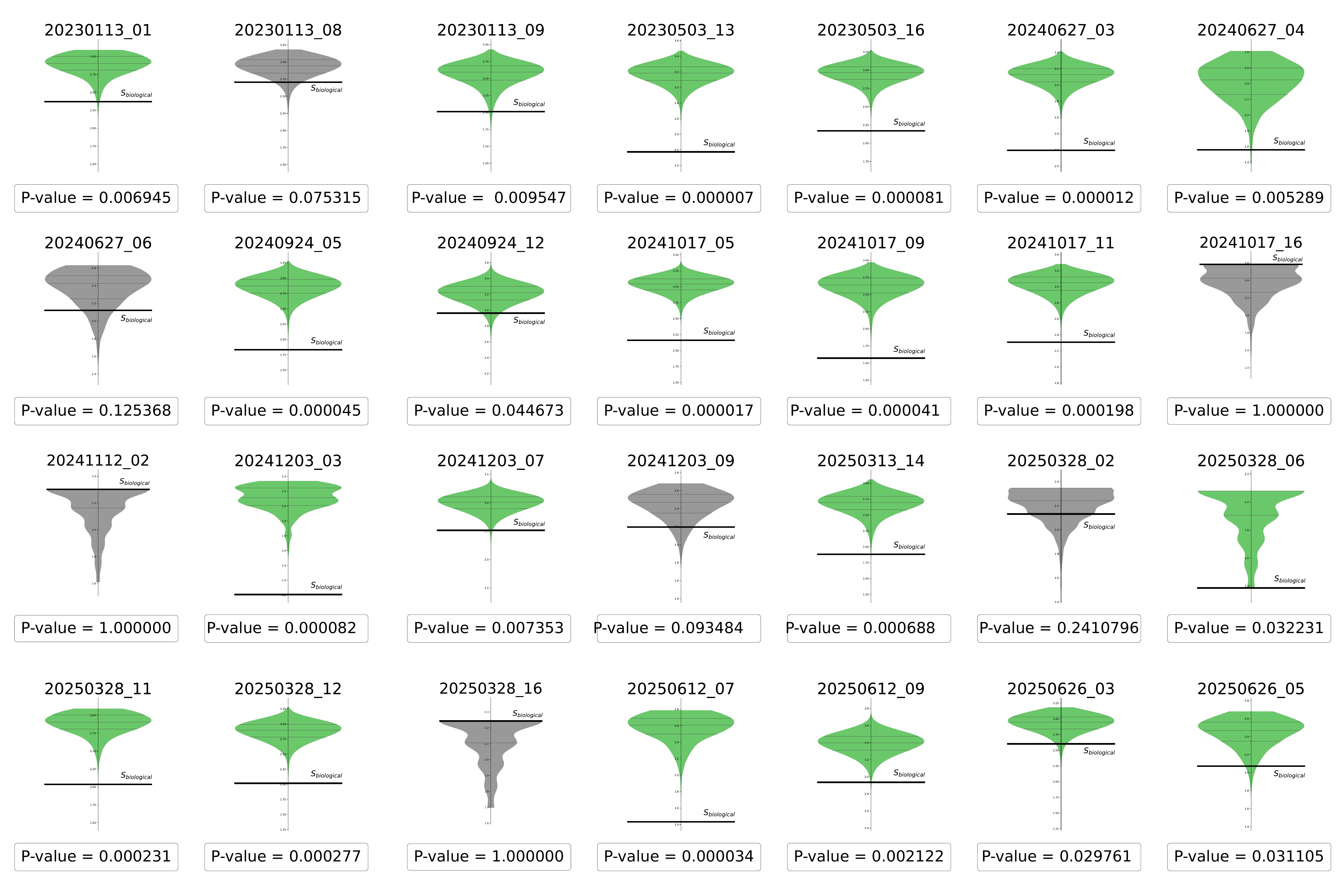}
\caption{{\bf Inheritance of the BMSC colonies.} Result of the entropy-based inheritance test, using the RR null ensemble, for all colonies in the dataset for which the test is non-trivial (i.e.\ for lineages with more than one inactive cell). For  each colony, the violin plot represents the probability distribution of the entropy for the set of ${\cal R}=10^6$ randomly reshuffled trees, while the black line is the value of the entropy of the original experimental tree; the P-value is the integral of the probability below the black line. Green: significant result of the inheritance test (P-value $< 0.05$). Gray: non-significant result of the inheritance test (P-value $\geq 0.05$). 
}
\end{figure*}
%%%%%%%%%%%%%%%%%%%%%%%%%%%%%%%%%%%%%%%%%%%%%%%

\section{Results}

% example
We first  illustrate how the inheritance test works on the two experimental BMSC colonies in Fig.5: the random reshuffling clearly destroys all correlations between inactive cells, yet keeping fixed their number at each generation, as well as the number of missing cells. The bulk of the reshuffled distribution of the entropy in these two trees is way above the actual biological value, hence giving a strong inheritance signal in both cases.

\subsection{Inheritance in the experimental BMSC lineages}
% total results
The outcome of the inheritance test on our entire BMSC dataset is reported in Fig.6. We have 32 colonies in total; 4 colonies have either zero or one inactive cell, hence the scrambling procedure always produces the same tree, trivially giving P=1. Of the 28 colonies for which we can fruitfully run the test, 21 (75\%) have lineages with a significant hereditary character (P $< 0.05$), while in 7 colonies (25\%) the inheritance test gives a non-significant result (P $\geq 0.05$). The non-significant cases are mostly given by colonies with a very small number of inactive cells, for which the inheritance test -- which is based on scrambling in many {\it different} ways the positions of the inactive cells -- is not expected to be very effective: the four colonies with P$> 0.1$ have either 2 (\texttt{20241112\_02}, \texttt{20250328\_16}) or 3  (\texttt{20241017\_16}, \texttt{20250328\_02}) inactive cells; two colonies with 5 (\texttt{20241203\_09}) and 7 (\texttt{20230113\_08}) inactive cells have $0.05 <$ P $< 0.1$; in just one colony in our dataset (\texttt{20240627\_06}) the inheritance test gives a non-significant result (P$=0.12$) despite having 22 inactive cells.  In conclusion, the overall result of the inheritance test indicates that the topology of the largest majority of human BMSC colonies cannot be attributed to random variations in the positions of inactive cells across different branches, thus proving that the processes responsible for regulating the probability of emergence of inactive cells have a strong hereditary character (in SM-Section 7 we demonstrate that the structure of inactive cells cannot be due to physical proximity). Given the short timescales of the experiment and the low number of generations, it is highly unlikely that such structure is determined by genetic mutations \cite{rennerfeldt2016concise, chang2008transcriptome}; instead, our results are most likely due to the hereditary propagation of non-genetic, that is epigenetic, factors regulating cell-cycle exit.

%%%%%%%%%%%%%%%%%%%%%%%%%%%%%%%%%%%%%%%%%%
\begin{figure*}
\centering
\includegraphics[width=0.9 \textwidth]{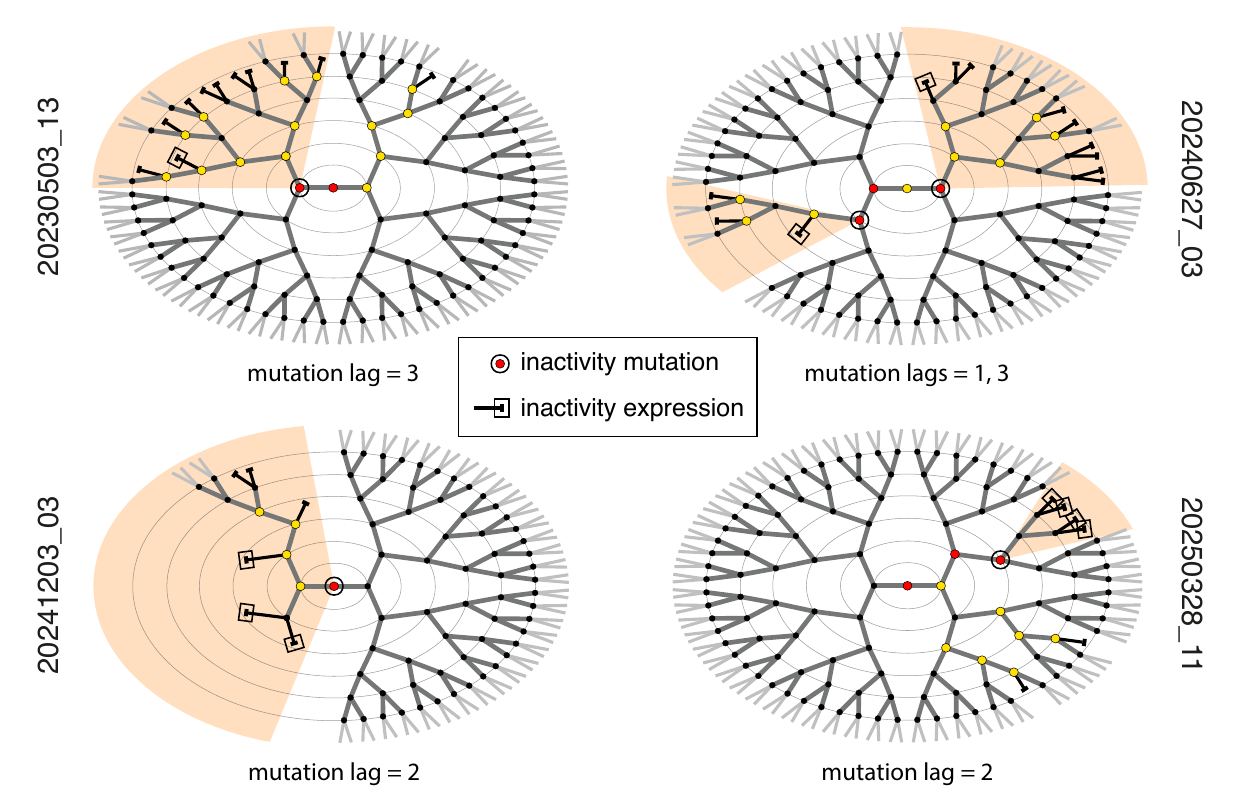}
\caption{
{\bf Mutation lag in BMSC lineages.} Each mitosis $m$ is coloured according to its normalized imbalance: $w_m=0$ -- black; $0 < w_m\leq 0.15$ -- yellow; $w_m > 0.15$ -- red. 
The mitosis whose anomalous imbalance signals a mutation in the probability of inactivity is marked by an extra external black circle and
the shaded area highlights the branch impacted upon by that mutation; the first (most upstream) link in this branch corresponds to the actually mutated cell.
On the other hand, the first inactive cell in the mutated branch emerges a few generations downstream and it is marked by an extra external black rectangle. 
The mutation lag, $l$, is given by the number of generations -- i.e. of nodes -- separating the mutated cell from of the first inactive cell.
In colony \texttt{20240627\_03} there are two independent mutation events. Colony \texttt{20250328\_11} is a more complex case: it seems that there are two concatenated mutation events, one at the very first mitosis in the lineage, differentiating the left and right wings of the tree, and another (stronger) event separating generation $k=3$ from $k=4$ (the one highlighted in the panel); a more in depth analysis is needed to work out whether these two events are correlated.
}
\end{figure*}
%%%%%%%%%%%%%%%%%%%%%%%%%%%%%%%%%%%%%%%%%%%

% sufficient not necessary - more sophisticated inheritance tests
We stress that a non-significant result of the test {\it does not} imply that there is no inheritance at work, but merely that there is no detectable change in the probability of inactivity in the lineage: the lack of variation in the expression of a character makes inheritance undetectable, not necessarily absent; hence, the test is a sufficient, but not necessary, condition for the presence of an hereditary mechanism regulating inactivity. Were we to consider more complex factors, it could very well turn out that a lineage that did not pass the inheritance test at the bare topological level, still results hereditary according to other factors. But we cannot help noticing that calculating the entropy and its P-value is easy and quick, as one needs nothing but the bare structure of the lineage. Hence, it seems lucky, and interesting, that most BMSC colonies display a sharp hereditary character even in relation to the quite fundamental cell state of inactivity.

\subsection{Navigating the imbalance map}

As we have already remarked, the value of the inactivity imbalance defined on each node of the tree, i.e. on each mitosis of the colony, provides a sort of map of the most likely sites where mutation events occurred across the tree. We have to be careful about one point, though: when there is a chain, crossing different generations, of subsequent nodes all with similar anomalous imbalances, it is the node at the {\it latest} generation (largest value of $k$) that actually corresponds to the mutation event. 

In order to understand this point let us go back to Fig.3. In this highly hereditary lineage, the largest part of the imbalance is concentrated on just two nodes: {\it i)} the central mitosis, which gives rise to the two cells of generation 1, let us call this node M1; {\it ii)} and one of the two mitosis separating generation 1 from generation 2, let us call this node M2. We claim that the actual mutation occurred in M2, while the spike in M1 is simply a byproduct of that in M2. The reason is that the significance of a spike in the imbalance must be assessed going {\it upstream} in the lineage, starting from the terminal cells, i.e. the $k_7$ leaves. If we do this, we see that the first anomalous node that we meet is indeed M2: something clearly happened at this mitosis, because its right branch (no inactive, nor missing cells) is completely different from its left branch (which instead has 10 inactive cells, causing 22 missing cells), and this large left-right asymmetry occurs for the first time at this node, when we navigate the tree upstream. On the other hand, if no other mutation occurs at M1, this node would still display roughly the same large imbalance as M2, up to small random fluctuations, which is exactly what happens in this colony.
Reversing the order of the argument, i.e. starting from the central node and navigating the lineage {\it downstream}, each time we find a mitosis with a very large anomalous imbalance, we can only conclude that in one mitosis of its descendants there is a mutation. 

% It is therefore important to avoid giving to the inactivity imbalance any temporal interpretation: it is a quantity that has a value and a meaning only once the entire colony has been traced up to a certain generation. If, in the colony of Fig.3, we traced one more generation of cells, the imbalance could change on {\it all} nodes, even those at the early generations; this is not surprising, as in the colony traced up to $k=7$ there might have already been some mutations that will only become visible in terms of inactive cells once we trace one or two extra generations beyond the $7^\mathrm{th}$.

\subsection{The inactivity mutation lag}

Once we have established a procedure to identify the mitosis where there is a change in the probability of inactivity, an inspection of the imbalance map in those lineages that pass the inheritance test shows that the mutation ultimately leading to the anomalous accumulation of inactive cells within a branch, typically happens a couple of generations {\it before} these cells start emerging in that branch (Fig.7). Inactive cells may either gradually appear in the mutated branch across different generations (as in \texttt{20230503\_13} or \texttt{20241203\_03}), or they may show up all at the same generation  (\texttt{20250328\_11}); moreover, in some clones there are two independent mutation events (\texttt{20240627\_03}). Due to this complex scenario, there might be more than one way to quantify the inactivity mutation lag, $\mathit{l}$; we stick to the simplest definition and count how many generations (i.e. mitosis, or nodes) separate the mutation node from the {\it first} inactive cell emerging after the mutation (see Fig.7). The average value of the mutation lag over all BMSC clones passing the inheritance test is $l_\mathrm{BMSC}\approx 2$ generations. 

The existence of this mutation lag indicates that what we hypothesized in the Introduction, namely that the epigenetic change responsible for the emergence of inactivity occurs a few generations {\it before} inactivity itself is expressed, is actually true. As we have already noted, this phenomenon is consistent with results reported in the literature about cellular senescence \cite{ogrodnik2021cellular}. Moreover, in a different context, the presence of  generation delays between cell fate decisions and the onset of lineage markers has been found in hematopoietic stem cells \cite{strasser2018lineage}.
We also notice that our method is able to capture the progression to inactivity (and quite
possibly senescence) on the short time scales of the first
proliferation stages of the clonal colony, contrary to most studies,
which rely on mass population analysis across multiple passages and
focus on the long time scales \cite{wagner2008replicative}.

% no lag no party
Finally, we stress once again that it is only thanks to the inactivity mutation lag that the hereditary nature of cell-cycle exit could be established: if the inactive cell was born {\it exactly} at the same mitosis where the epigenetic change responsible for inactivity occurred ($l=0$), it would be impossible to assess inheritance. This seems at odds with the fact that -- lag or not -- if the number of $k_7$ leaves cut by an inactive cell is large, the imbalance will also be large, thus lowering  the entropy, which should give a strong inheritance signal. In fact, the situation is more subtle. 
Imagine a tree in which there is just one mutation and imagine that that mutation occurs at the central node of the tree, the one producing the two $k=1$ cells; imagine also that {\it there is no mutation lag}, $l=0$, hence (say) the right cell immediately becomes inactive, thus cutting all its $64$ potential $k_7$ leaves, while its sister -- the left cell -- generates a full half-tree, ultimately producing all its $64$ $k_7$ leaves (this is exactly the tree represented in Fig.4b). In this case, a large inactivity imbalance, $I=64$, is entirely concentrated on just that central mitosis, which gives a very small entropy, in fact the smallest, $S=0$. However, when we attempt to randomly reshuffle this tree, we immediately recognize that the only move we can make is to switch the two $k=1$ sisters, which produces exactly the same lineage; hence, {\it all} reshuffled trees have $S=0$, so that the P-value with the RR method is equal to $1$ and we have zero inheritance signal. 
This is a general phenomenon: the absence of a mutation lag would imply that the locations of inactive cells coincide with the locations of the mutation events; if these events are uncorrelated from each other (and there are no strong reasons to think otherwise), then reshuffling the inactive cells creates the same statistical lineage and the entropy is not lower in the RR set, on average. 
This argument highlights very vividly that the {\it absolute} value of the entropy can trick us; only its value {\it relative} to the null ensemble of lineages is significant: a lineage is highly hereditary not because its entropy is low, but because it is {\it lower} than that of the corresponding non-hereditary ensemble, in our case RR. In this respect, the reshuffling procedure is as crucial an ingredient of the inheritance test as the entropy itself.

%%%%%%%%%%%%%%%%%%%%%%%%%%%%%%%%%%%%%%%%%%%%%%%%%%%%%%%%%%%

% relation to Commitment theory of cellular aging
\section{A simple hereditary model \\ of the data}
The existence of a lag between the epigenetic change leading to
inactivity and its expression, is very reminiscent of the {\it
  incubation period} introduced by the commitment theory of cellular
aging of Kirkwood and Holliday \cite{kirkwood1975commitment, holliday1977testing,harley1980retesting}, whose goal
was to model the evolution of senescence in human diploid fibroblasts. There is an
important difference with our experimental findings, though: according to
that theory, after the
incubation period, {\it all} the descendants generated by the committed
cell die out, while this is clearly not the case in our data. Instead,
BMSC lineages suggest that at the mutation there is no deterministic
commitment to inactivity, but rather a sudden increase in the
probability to produce inactive cells; this probability is passed-on
in a hereditary way and it acquires a phenotypic impact only a few
generations later, when enough cells have been generated to 
produce some inactive descendants. This state of affairs suggests a simple stochastic generalization of the 
Kirkwood-Holliday hereditary model.

\subsection{The Stochastic Kirkwood-Hollyday model (SKH)}
As in the original Kirkwood-Holliday model \cite{kirkwood1975commitment}, we assume that a wild cell has probability $0$ to be inactive and probability $p$ to mutate; once a cell mutates, all its descendants have probability $q$ to be inactive, including the mutated cell itself (the descendants of a mutated cell can only become inactive, they cannot mutate again). Hence, at the mutation there is a sharp change in the probability of inactivity, from $0$ to $q$, and this change is inherited by the progeny of the mutated cell, whence the hereditary character of the model. In contrast with the original Kirkwood-Holliday model, not only the mutation, but also the emergence of inactivity is a stochastic process, which prevents the unrealistic case of perfect synchronization in the emergence of inactivity. We call this the Stochastic Kirkwood-Holliday model (SKH). 

The readers can easily convince themselves that for $q=1$ the SKH model reduces to the GW model with probability $p$: this is exactly the case of {\it zero} mutation lag that we described at the end of the previous Section; moreover, for $p=1$ SKH reduces to GW with probability $q$: once the root has mutated, all proceeds as in GW. Therefore, although SKH is in general an hereditary model, it contains also some non-hereditary sectors in the space of parameters. In fact, the non-hereditary region of SKH extends far beyond the two GW edges. For each pair $(p,q)$ in a grid, we compute the fraction $\Phi$ of SKH trees that pass the inheritance test, hence drawing an hereditary map of the model (Fig.8): this map shows that the most hereditary region of SKH is an oval-shaped zone around $p = 0.1$ and $q = 0.45$: here, the SKH model produces up to 35\% of trees that pass the entropy-based inheritance test. 

The two GW non-hereditary edges of SKH are symmetric, $(p=1 \ \forall \ q)$ and $(q=1 \ \forall \ p)$; besides, the mean number of inactive cells is also symmetric, $\Nina^\mathrm{SKH}(p,q)=\Nina^\mathrm{SKH}(q,p)$, and the same happens for the mean number of missing cells, $\Nmiss^\mathrm{SKH}$ (see SM-5.3 and 5.4). However, the hereditary map is strongly {\it non}-symmetric, hence suggesting that the two ways to recover GW from SKH are very different from each other. This is indeed the case. The non-hereditary limit $p\to 1$ is smooth: a tree with a very large probability of mutation, $p=1-\epsilon$, has a mutated root almost all the times, hence giving rise to a GW model, which has a very low probability to pass the inheritance test. On the contrary, the other non-hereditary limit, $q\to 1$, is {\it not} smooth, as any tree with $q=1-\epsilon$ is still hereditary in the infinite-size limit, because in a large enough tree sooner or later inactivity will express not right at the mutation, giving rise to an hereditary signal; only for $q=1$ the two models coincide. A finite-size echo of this interesting property shows up in Fig.8, in which the strip $p=0.9$ produces trees significantly less hereditary than the strip $q=0.9$.

%%%%%%%%%%%%%%%%%%%%%%%%%%%%%%%%%%%%%%%%%%
\begin{figure}
\centering
\includegraphics[width=0.48 \textwidth]{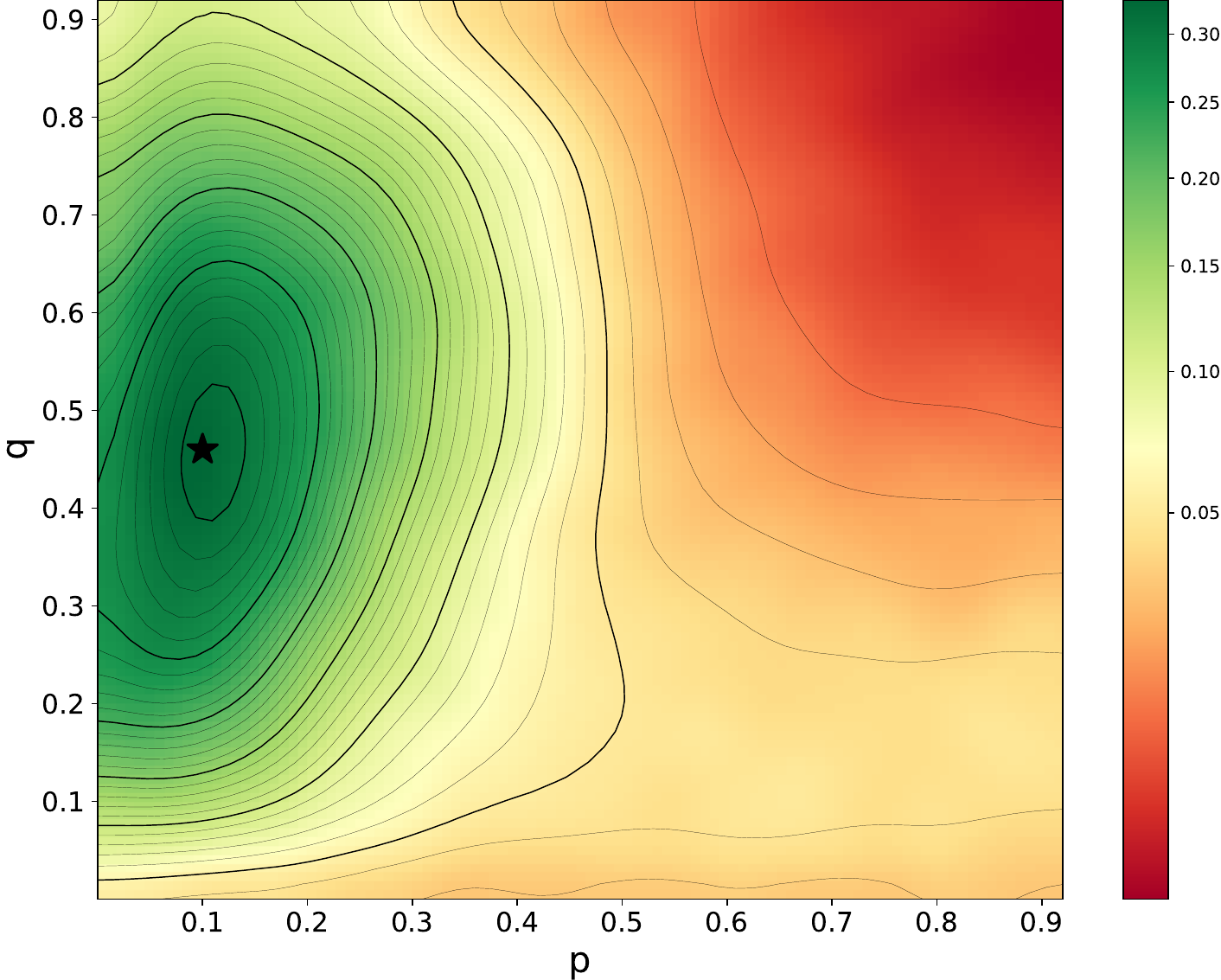}
\caption{{\bf Hereditary map of the SKH model.} For each pair of parameters $(p,q)$ we generate $10^3$ SKH trees with $k_\mathrm{max}=7$; for each tree we run the inheritance test with ${\cal R}=10^3$ reshufflings and measure the fraction $\Phi$ of trees that pass the test; a larger value of $\Phi$ (green) represents a more hereditary region of the model, while a smaller value of $\Phi$ (red) represents a less hereditary one. Contour lines start at $\Phi=0.01$ (up-right) and have spacing $\Delta\Phi=0.01$; their steep density in the green zone shows how sharp is the inheritance sector of SKH. The star marks the position of the mean MLE fit of the SKH model to the experimental colonies. 
}
\end{figure}
%%%%%%%%%%%%%%%%%%%%%%%%%%%%%%%%%%%%%%%%%%%

%%%%%%%%%%%%%%%%%%%%%%%%%%%%%%%%%%%%%%%%%%%%%%%
\begin{figure*}
\centering
\includegraphics[width=0.9 \textwidth]{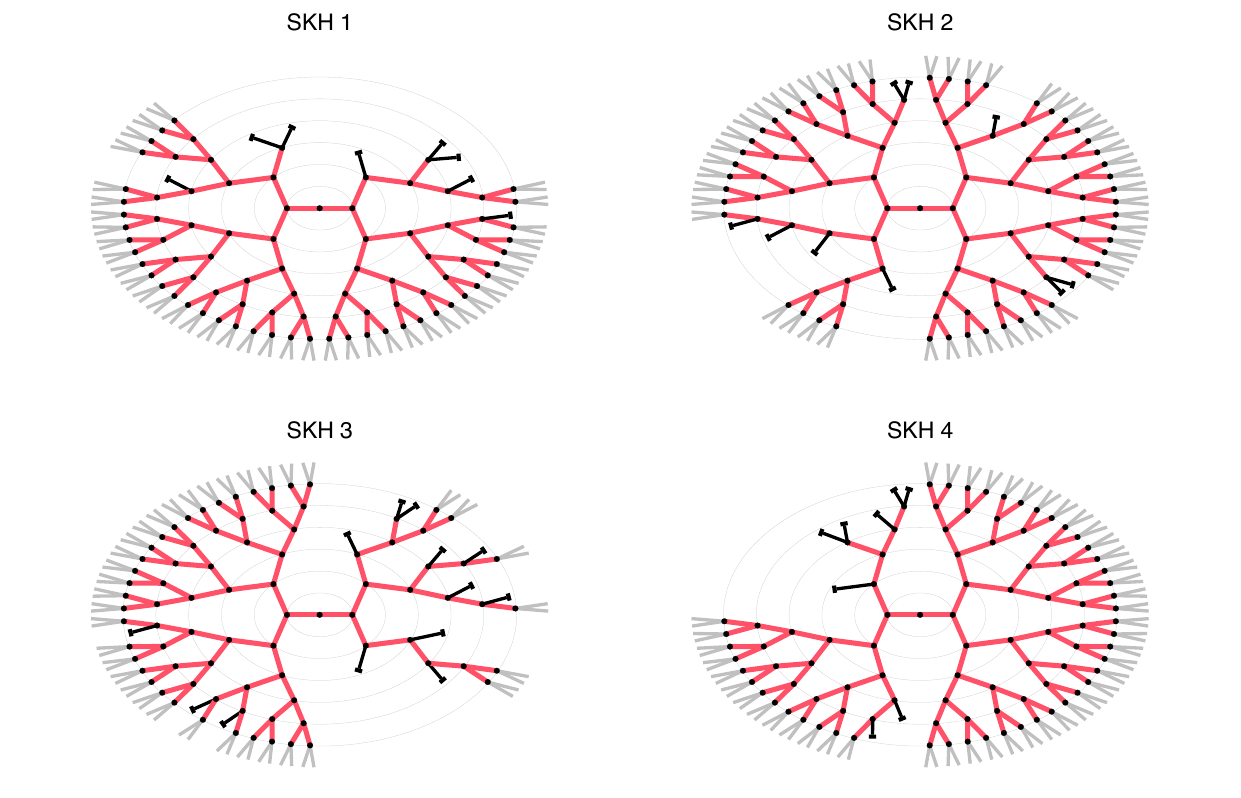}
\caption{
{\bf Sample lineages of the hereditary SKH model.} The parameters $(p,q)$ of each one of these trees are equal to those fitted to the four biological trees in Fig.7, so to make as fair a comparison as possible between the two figures (for the actual parameters $(p,q)$ see SM-Table 1). The hereditary structure of the SKH trees is quite evident even by eye and indeed all four lineages pass the inheritance test ($P^\mathrm{SKH1}=0.0261$, $P^\mathrm{SKH2}=0.0173$, $P^\mathrm{SKH3}=0.0003$, $P^\mathrm{SKH4}<10^{-6}$).
}
\end{figure*}
%%%%%%%%%%%%%%%%%%%%%%%%%%%%%%%%%%%%%%%%%%%%%%%%

\subsection{Fitting the SKH model to the biological lineages}
Given an experimental tree, $\cal T$, we want to find the parameters $(p_\tau, q_\tau)$ that give the best SKH fit to the data; 
we do this in three different ways.
We first use a Maximum Likelihood Estimate (MLE): we calculate the likelihood of the SKH model given the tree and check whether or not it has a well-defined maximum $(p_\tau, q_\tau)$; this happens only for experimental lineages with a large number of inactive cells, which are also those trees that pass the entropy-based inheritance test; BMSC trees with small $\Nina$ have a flat SKH likelihood profile, which makes it hard to accurately locate the maximum (SM-5.7.1). 
The second method uses the inactivity imbalance to locate the mutation within the experimental tree and employs MLE restricted to the mutated sub-tree to calculate $q$; then, it fits $p$ by matching the total number of inactive and missing cells to the analytic SKH expressions. 
The third method fits $q$ by matching the mutation lag of the experimental tree to the analytic SKH expression, while $p$ is fixed as in the second method.
Also these last two methods work only on BMSC colonies that pass the entropy-based inheritance test, the sole for which it is possible to locate the mutations. Full analytic details of the three methods are reported in SM-Section 5.7. Since MLE is the gold standard of model inference,  it is reassuring that the other two fitting methods -- which rely heavily on our definitions of imbalance, mutation and mutation lag -- give results consistent with MLE (see SM-Table 1). 
Finally, if we average the log-likelihood over all BMSC trees, we get $p^\star=0.1$, $q^\star=0.46$, the parameters used in Fig.4c to benchmark the inheritance tests; $(p^\star, q^\star)$ falls right at the centre of the hereditary SKH zone.

% why putting the points on the map is misleading
The reader may wonder why we do not report on the SKH inheritance map the positions of the individual BMSC lineages; if we did, we would find that all experimental colonies comfortably tuck in the hereditary region of SKH, which could be interpreted as further proof of their hereditary nature. In fact, doing this would be a mistake: assessing the inheritance of a tree by fitting it to the SKH model and comparing the fit to the hereditary region of SKH, may lead to wrong conclusions. We show in SM-Section 6 that there are ensembles of non-hereditary trees that are nevertheless located by an MLE fit in proximity of the hereditary sector of SKH. The fact is that, fitting the SKH model to a tree does not necessarily prioritize inheritance as the main fitted trait.
An example of non-hereditary tree that sneaks into the hereditary zone of SKH is the following: consider a {\it non-homogeneous} tree $\Xi$ with inactive cells all concentrated at the last generations (SM-Fig.5). The non-hereditary region of SKH has probabilities of mutation/inactivity quite large and -- {\it because it is an homogeneous model} -- SKH produces in this zone many inactive cells also at the early generations, making it very unlikely for a tree to survive to the last generation. Hence, the likelihood finds it more appropriate to fit $\Xi$ through SKH parameters able to generate complete trees, indifferent to the fact that these parameters fall in the hereditary region.

This is a deeper issue than it seems.
If instead of naively gauging the MLE fit on the hereditary SKH map, we employ the more serious parametric bootstrap likelihood-ratio test \cite{tibshirani1993introduction,davison1997bootstrap,self1987asymptotic} to compare the SKH and GW likelihoods, we would still wrongly conclude that the non-hereditary tree $\Xi$ is instead hereditary (SM-6). This is a stark reminder that any inheritance-assessment method based on comparing the likelihoods of different models, inevitably relies on the arbitrary choice of these models: there might always exist lineages so alien to the structure of the selected models, to fool the likelihood-based methods. Instead, the entropy-based inheritance test is {\it completely independent} from the choice of models, so it correctly classifies $\Xi$ as non-hereditary.

%%%%%%%%%%%%%%%%%%%%%%%%%%%%%%%%%%%%%%%%%%%%%%%%

\subsection{Insights from the SKH Model}
How do SKH trees {\it look} like? In Fig.9 we represent four SKH trees, each one drawn with the MLE parameters $(p,q)$ relatives to the four biological trees of Fig.7, so to make a fair comparison. We see that the SKH trees look indeed quite hereditary, with clustering of inactive cells along the same branches, very much as in the biological dataset. If the reader checks the tree zoology reported in SM-Section 9, they will further appreciate the qualitative similarity between SKH and BMSC trees. To evaluate in a slightly more quantitatively fashion the quality of the SKH fit to the BMSC data, we can calculate the mutation lag correlation: after we fit the SKH model to an experimental lineage $\cal T$ through MLE, we use the fitted parameters to calculate the mutation lag, $l^\tau_\mathrm{SKH}$, for that tree, using the analytic relations for the distribution of the mutation lag in the SKH model that we calculate in SM-Section 5.5; besides, we determine the lag of the experimental tree, $l^\tau_\mathrm{exp}$, by using the peaks of the imbalance, as described in the previous Section. We find that the two sets of lags have decent Spearman correlation, $\rho=0.67$, and very significant P-value, $P=0.001$ ($10^6$ extractions).

Hence, it seems that SKH is a sound starting point to describe inheritance in our data; more specifically, our interpretation of BMSC imbalance maps, namely that sometimes along the lineage there is a mutation of probability to become inactive, seems very reasonable, given the SKH fit to the experimental data. In this respect, it is interesting to note that the mean MLE fit to SKH of the biological lineages indicates that the probability of mutation is quite low, $p \approx 0.1$, but that -- after the mutation occurs -- the probability to become inactive is quite more substantial, $q\approx  0.45$. One could have envisaged an opposite scenario, in which mutations occurred more frequently, but the mutated inactivity probability was lower, $p\approx 0.45$ and $q\approx 0.1$; what the SKH inheritance map shows is that this scenario would have given a far smaller inheritance signature, so small to go probably undetected by the inheritance test.

The SKH model has also limitations, though. The inheritance level of the SKH model is lower than that of BMSC lineages; as we have seen in Fig.8, even in the most hereditary sector of the parameters space, about $35\%$ of SKH trees (with $k_\mathrm{max}=7$) pass the inheritance test, against the  $75\%$ of hereditary trees in the experimental dataset. Moreover, the correlation between the mutation lags (inferred with SKH vs experimental) is good, but not great. In fact, we believe that both these discrepancies could be overcome by introducing more articulated (and complex) versions of the model. Instead of having a constant probability of inactivity $q$ after the mutation, which produces an exponential distribution of the mutation lag (SM-5.5), one can formulate a model in which the inactivity probability depends on the distance ${\Delta k}$ (in generations) from the mutation. In this way one can easily produce a non-exponential probability distribution of the mutation lag, with the desired mean and variance; by lowering the variance one can obtain as large an inheritance signal as desired. Of course, more parameters means more problems to fit the model, so one should balance simplicity vs accuracy, as always. Exploring  such broader class of SKH hereditary models goes beyond the scope of this work, but our results seem already to suggest that SKH is a very reasonable class of models for describing the inheritance of inactivity in cell lineages.

%%%%%%%%%%%%%%%%%%%%%%%%%%%%%%%%%%%%%%%%%%%%%%%%%%%%%%%%%%%

\section{Conclusions}

% the new method
We introduced a new entropy-based method to assess inheritance in a lineage tree. The method is model-agnostic, as it relies solely on the topological structure of the truncated branches, namely of those elements of the tree that have stopped proliferating. In the case of cell lineages, truncation of a branch is due to inactivity or death of a cell, but in general the interruption of a branch may be due to any phenomenon impacting on the growth of the tree, irrespective of the specific origin and context of the lineage. In this respect, our method -- which we proved to be reliable on a great number of different synthetic trees -- seems to have a strong potential for very broad applications, well beyond the assessment of epigenetic inheritance in stem cell colonies.

% what we found plus intra-colony  heterogeneities
The entropy-based inheritance test shows that human BMSC clonal colonies have a strong hereditary structure of their inactive cells: $75\%$ of colonies pass the inheritance test, and those that do not are characterized by such low numbers of inactive cells to make the test void.
It is important to note that such evidence of inheritance -- which we obtain from the inactivity imbalance --  is due to the existence of {\it intra-colony} heterogeneities: it is only thanks to the emergence throughout the lineage of changes in the propensity to become inactive that we can detect the anomalous imbalances that contribute to lowering the entropy; 
on the contrary, colonies with no inactive cells do not give any signal; as we have already noted, inheritance is undetectable in absence of mutations. 
Once the presence of inheritance is established through the intra-colony fluctuations, it is legitimate to infer that the same hereditary mechanism is responsible for the colony-level permanence of epigenetic mutations, thus giving rise to {\it inter-colony} heterogeneities. 
Hence, our results support the hypothesis that the strong colony-level heterogeneities met when dealing with in vivo transplant of BMSCs --- heterogeneities that strongly affects the effectiveness of transplants and thus of skeletal regeneration therapies --- do indeed have an epigenetic hereditary origin.

% imbalance map, mutation, new model
The imbalance map of a colony allows to identify the mutation in the probability of inactivity, a mutation then inherited by the descendants, hence causing the hereditary structure of the lineage. We find that in BMSC colonies the mutation occurs, on average, 2 generations before the expression of inactivity. The observation of this mutation lag prompted the development of a simple stochastic variant of the Kirkwood-Holliday model (SKH), which has an interesting hereditary landscape in the space of parameters. The SKH model produces synthetic trees that -- despite being less hereditary than the BMSC dataset -- have a strong topological similarity with the experimental lineages. An MLE fit of the experimental data to the SKH model suggests that the epigenetic hereditary structure of cell-cycle exit in BMSCs is due to a relatively low probability of mutation, associated to a relatively high probability of inactivity after the mutation. Generalisations of the SKH model seem easy to implement, hence paving the way to even more accurate characterisations of epigenetic inheritance in stem cells lineages.

% inheritance of other traits - correlation topology kinetics - cycle time kinetics
The distribution of inactive cells across a colony determines the inheritance entropy. Active cells, on the other hand, constitute a separate population that enters the entropy-based hereditary test only in a passive way. Experiments, though, shows that the topology of BMSC lineages (determined by inactive cells) is strongly correlated to the statistics of the division times of the active cells \cite{allegrezza2025topology}: colonies with the larger populations of inactive cells are also characterized by the slower populations of active cells \cite{allegrezza2025topology}. This result suggests that not only cell-cycle exit, but also cell-cycle duration might be regulated by epigenetic hereditary mechanisms. To confirm this point, though, it is necessary to study the correlations of the division times. This has been done for bacterial colonies in relation to genetic factors  \cite{powell1955some, staudte1984additive}, and to a scarcer degree for epigenetic factors in eukaryotic cells \cite{sandler2015lineage, hormoz2015inferring}, but never in the context of stem cells. Besides, in all past studies, the assessed correlation was not quite deep enough to trace in a robust way the hereditary ramifications of the division times across the lineage tree. This remains therefore an interesting open experimental and theoretical problem along the path to establishing epigenetic factors as the primary cause of the strong heterogeneity affecting stem cell colonies.

%%%%%%%%%%%%%%%%%%%%%%%%%%%%%%%%%%%%%%%%%%%%%%%%%%%%%%%%%%%%%%%%%

\section{Acknowledgements}

We thank William Bialek, Fabio Cecconi, Edo Kussel, Leonid Mirny, Felix Ritort, Aleksandra Walczak
and two anonymous referees for very insightful advice on the manuscript. ACa and TSG thank Enzo Branchini for discussions within the CC.  ACa acknowledges the support and advice of Francesco Cavagna and of the late Giovanni Cavagna. This work was supported by ERC Grant RG.BIO (Contract n. 785932), MIUR Grant INFO.BIO (Protocol n. R18JNYYMEY), and MIUR Grant PRIN2020 (Contract n. 2020PFCXPE-005).

%%%%%%%%%%%%%%%%%%%%%%%%%%%%%%%%%%%%%%%%%%%%%%%%%%%%%%%%%%%%%%%%%
%%%%%%%%%%%%%%%%%%%%%%%%%%%%%%%%%%%%%%%%%%%%%%%%%%%%%%%%%%%%%%%%%
%%%%%%%%%%%%%%%%%%%%%%%%%%%%%%%%%%%%%%%%%%%%%%%%%%%%%%%%%%%%%%%%%

%%%%%%%%%%%%%%%%%%%%%%%%%%%%%%%%%%%%%%%%%%%%%%%
%%%%%%%%%%%%%%%%%%%%%%%%%%%%%%%%%%%%%%%%%%%%%%%
%\bibliographystyle{apsrev4-1}
\bibliography{bibliography-stem-cells-cobbs}

\begin{thebibliography}{66}%
\makeatletter
\providecommand \@ifxundefined [1]{%
 \@ifx{#1\undefined}
}%
\providecommand \@ifnum [1]{%
 \ifnum #1\expandafter \@firstoftwo
 \else \expandafter \@secondoftwo
 \fi
}%
\providecommand \@ifx [1]{%
 \ifx #1\expandafter \@firstoftwo
 \else \expandafter \@secondoftwo
 \fi
}%
\providecommand \natexlab [1]{#1}%
\providecommand \enquote  [1]{``#1''}%
\providecommand \bibnamefont  [1]{#1}%
\providecommand \bibfnamefont [1]{#1}%
\providecommand \citenamefont [1]{#1}%
\providecommand \href@noop [0]{\@secondoftwo}%
\providecommand \href [0]{\begingroup \@sanitize@url \@href}%
\providecommand \@href[1]{\@@startlink{#1}\@@href}%
\providecommand \@@href[1]{\endgroup#1\@@endlink}%
\providecommand \@sanitize@url [0]{\catcode `\\12\catcode `\$12\catcode `\&12\catcode `\#12\catcode `\^12\catcode `\_12\catcode `\%12\relax}%
\providecommand \@@startlink[1]{}%
\providecommand \@@endlink[0]{}%
\providecommand \url  [0]{\begingroup\@sanitize@url \@url }%
\providecommand \@url [1]{\endgroup\@href {#1}{\urlprefix }}%
\providecommand \urlprefix  [0]{URL }%
\providecommand \Eprint [0]{\href }%
\providecommand \doibase [0]{https://doi.org/}%
\providecommand \selectlanguage [0]{\@gobble}%
\providecommand \bibinfo  [0]{\@secondoftwo}%
\providecommand \bibfield  [0]{\@secondoftwo}%
\providecommand \translation [1]{[#1]}%
\providecommand \BibitemOpen [0]{}%
\providecommand \bibitemStop [0]{}%
\providecommand \bibitemNoStop [0]{.\EOS\space}%
\providecommand \EOS [0]{\spacefactor3000\relax}%
\providecommand \BibitemShut  [1]{\csname bibitem#1\endcsname}%
\let\auto@bib@innerbib\@empty
%</preamble>
\bibitem [{\citenamefont {Ng}\ and\ \citenamefont {Gurdon}(2008)}]{ng2008epigenetic}%
  \BibitemOpen
  \bibfield  {author} {\bibinfo {author} {\bibfnamefont {R.~K.}\ \bibnamefont {Ng}}\ and\ \bibinfo {author} {\bibfnamefont {J.~B.}\ \bibnamefont {Gurdon}},\ }\bibfield  {title} {\bibinfo {title} {Epigenetic inheritance of cell differentiation status},\ }\href@noop {} {\bibfield  {journal} {\bibinfo  {journal} {Cell cycle}\ }\textbf {\bibinfo {volume} {7}},\ \bibinfo {pages} {1173} (\bibinfo {year} {2008})}\BibitemShut {NoStop}%
\bibitem [{\citenamefont {Zion}\ \emph {et~al.}(2020)\citenamefont {Zion}, \citenamefont {Chandrasekhara},\ and\ \citenamefont {Chen}}]{zion2020asymmetric}%
  \BibitemOpen
  \bibfield  {author} {\bibinfo {author} {\bibfnamefont {E.~H.}\ \bibnamefont {Zion}}, \bibinfo {author} {\bibfnamefont {C.}~\bibnamefont {Chandrasekhara}},\ and\ \bibinfo {author} {\bibfnamefont {X.}~\bibnamefont {Chen}},\ }\bibfield  {title} {\bibinfo {title} {Asymmetric inheritance of epigenetic states in asymmetrically dividing stem cells},\ }\href@noop {} {\bibfield  {journal} {\bibinfo  {journal} {Current opinion in cell biology}\ }\textbf {\bibinfo {volume} {67}},\ \bibinfo {pages} {27} (\bibinfo {year} {2020})}\BibitemShut {NoStop}%
\bibitem [{\citenamefont {Sandler}\ \emph {et~al.}(2015)\citenamefont {Sandler}, \citenamefont {Mizrahi}, \citenamefont {Weiss}, \citenamefont {Agam}, \citenamefont {Simon},\ and\ \citenamefont {Balaban}}]{sandler2015lineage}%
  \BibitemOpen
  \bibfield  {author} {\bibinfo {author} {\bibfnamefont {O.}~\bibnamefont {Sandler}}, \bibinfo {author} {\bibfnamefont {S.~P.}\ \bibnamefont {Mizrahi}}, \bibinfo {author} {\bibfnamefont {N.}~\bibnamefont {Weiss}}, \bibinfo {author} {\bibfnamefont {O.}~\bibnamefont {Agam}}, \bibinfo {author} {\bibfnamefont {I.}~\bibnamefont {Simon}},\ and\ \bibinfo {author} {\bibfnamefont {N.~Q.}\ \bibnamefont {Balaban}},\ }\bibfield  {title} {\bibinfo {title} {Lineage correlations of single cell division time as a probe of cell-cycle dynamics},\ }\href@noop {} {\bibfield  {journal} {\bibinfo  {journal} {Nature}\ }\textbf {\bibinfo {volume} {519}},\ \bibinfo {pages} {468} (\bibinfo {year} {2015})}\BibitemShut {NoStop}%
\bibitem [{\citenamefont {Mosheiff}\ \emph {et~al.}(2018)\citenamefont {Mosheiff}, \citenamefont {Martins}, \citenamefont {Pearl-Mizrahi}, \citenamefont {Gr{\"u}nberger}, \citenamefont {Helfrich}, \citenamefont {Mihalcescu}, \citenamefont {Kohlheyer}, \citenamefont {Locke}, \citenamefont {Glass},\ and\ \citenamefont {Balaban}}]{mosheiff2018inheritance}%
  \BibitemOpen
  \bibfield  {author} {\bibinfo {author} {\bibfnamefont {N.}~\bibnamefont {Mosheiff}}, \bibinfo {author} {\bibfnamefont {B.~M.}\ \bibnamefont {Martins}}, \bibinfo {author} {\bibfnamefont {S.}~\bibnamefont {Pearl-Mizrahi}}, \bibinfo {author} {\bibfnamefont {A.}~\bibnamefont {Gr{\"u}nberger}}, \bibinfo {author} {\bibfnamefont {S.}~\bibnamefont {Helfrich}}, \bibinfo {author} {\bibfnamefont {I.}~\bibnamefont {Mihalcescu}}, \bibinfo {author} {\bibfnamefont {D.}~\bibnamefont {Kohlheyer}}, \bibinfo {author} {\bibfnamefont {J.~C.}\ \bibnamefont {Locke}}, \bibinfo {author} {\bibfnamefont {L.}~\bibnamefont {Glass}},\ and\ \bibinfo {author} {\bibfnamefont {N.~Q.}\ \bibnamefont {Balaban}},\ }\bibfield  {title} {\bibinfo {title} {Inheritance of cell-cycle duration in the presence of periodic forcing},\ }\href@noop {} {\bibfield  {journal} {\bibinfo  {journal} {Physical Review X}\ }\textbf {\bibinfo {volume} {8}},\ \bibinfo {pages} {021035} (\bibinfo {year} {2018})}\BibitemShut {NoStop}%
\bibitem [{\citenamefont {Hormoz}\ \emph {et~al.}(2016)\citenamefont {Hormoz}, \citenamefont {Singer}, \citenamefont {Linton}, \citenamefont {Antebi}, \citenamefont {Shraiman},\ and\ \citenamefont {Elowitz}}]{hormoz2016inferring}%
  \BibitemOpen
  \bibfield  {author} {\bibinfo {author} {\bibfnamefont {S.}~\bibnamefont {Hormoz}}, \bibinfo {author} {\bibfnamefont {Z.~S.}\ \bibnamefont {Singer}}, \bibinfo {author} {\bibfnamefont {J.~M.}\ \bibnamefont {Linton}}, \bibinfo {author} {\bibfnamefont {Y.~E.}\ \bibnamefont {Antebi}}, \bibinfo {author} {\bibfnamefont {B.~I.}\ \bibnamefont {Shraiman}},\ and\ \bibinfo {author} {\bibfnamefont {M.~B.}\ \bibnamefont {Elowitz}},\ }\bibfield  {title} {\bibinfo {title} {Inferring cell-state transition dynamics from lineage trees and endpoint single-cell measurements},\ }\href@noop {} {\bibfield  {journal} {\bibinfo  {journal} {Cell systems}\ }\textbf {\bibinfo {volume} {3}},\ \bibinfo {pages} {419} (\bibinfo {year} {2016})}\BibitemShut {NoStop}%
\bibitem [{\citenamefont {Chang}\ \emph {et~al.}(2008)\citenamefont {Chang}, \citenamefont {Hemberg}, \citenamefont {Barahona}, \citenamefont {Ingber},\ and\ \citenamefont {Huang}}]{chang2008transcriptome}%
  \BibitemOpen
  \bibfield  {author} {\bibinfo {author} {\bibfnamefont {H.~H.}\ \bibnamefont {Chang}}, \bibinfo {author} {\bibfnamefont {M.}~\bibnamefont {Hemberg}}, \bibinfo {author} {\bibfnamefont {M.}~\bibnamefont {Barahona}}, \bibinfo {author} {\bibfnamefont {D.~E.}\ \bibnamefont {Ingber}},\ and\ \bibinfo {author} {\bibfnamefont {S.}~\bibnamefont {Huang}},\ }\bibfield  {title} {\bibinfo {title} {Transcriptome-wide noise controls lineage choice in mammalian progenitor cells},\ }\href@noop {} {\bibfield  {journal} {\bibinfo  {journal} {Nature}\ }\textbf {\bibinfo {volume} {453}},\ \bibinfo {pages} {544} (\bibinfo {year} {2008})}\BibitemShut {NoStop}%
\bibitem [{\citenamefont {Plambeck}\ \emph {et~al.}(2022)\citenamefont {Plambeck}, \citenamefont {Kazeroonian}, \citenamefont {Loeffler}, \citenamefont {Kretschmer}, \citenamefont {Salinno}, \citenamefont {Schroeder}, \citenamefont {Busch}, \citenamefont {Flossdorf},\ and\ \citenamefont {Buchholz}}]{plambeck2022heritable}%
  \BibitemOpen
  \bibfield  {author} {\bibinfo {author} {\bibfnamefont {M.}~\bibnamefont {Plambeck}}, \bibinfo {author} {\bibfnamefont {A.}~\bibnamefont {Kazeroonian}}, \bibinfo {author} {\bibfnamefont {D.}~\bibnamefont {Loeffler}}, \bibinfo {author} {\bibfnamefont {L.}~\bibnamefont {Kretschmer}}, \bibinfo {author} {\bibfnamefont {C.}~\bibnamefont {Salinno}}, \bibinfo {author} {\bibfnamefont {T.}~\bibnamefont {Schroeder}}, \bibinfo {author} {\bibfnamefont {D.~H.}\ \bibnamefont {Busch}}, \bibinfo {author} {\bibfnamefont {M.}~\bibnamefont {Flossdorf}},\ and\ \bibinfo {author} {\bibfnamefont {V.~R.}\ \bibnamefont {Buchholz}},\ }\bibfield  {title} {\bibinfo {title} {Heritable changes in division speed accompany the diversification of single t cell fate},\ }\href@noop {} {\bibfield  {journal} {\bibinfo  {journal} {Proceedings of the National Academy of Sciences}\ }\textbf {\bibinfo {volume} {119}},\ \bibinfo {pages} {e2116260119} (\bibinfo {year} {2022})}\BibitemShut {NoStop}%
\bibitem [{\citenamefont {Yampolskaya}\ \emph {et~al.}(2025)\citenamefont {Yampolskaya}, \citenamefont {Ikonomou},\ and\ \citenamefont {Mehta}}]{yampolskaya2025}%
  \BibitemOpen
  \bibfield  {author} {\bibinfo {author} {\bibfnamefont {M.}~\bibnamefont {Yampolskaya}}, \bibinfo {author} {\bibfnamefont {L.}~\bibnamefont {Ikonomou}},\ and\ \bibinfo {author} {\bibfnamefont {P.}~\bibnamefont {Mehta}},\ }\bibfield  {title} {\bibinfo {title} {Finding signatures of low-dimensional geometric landscapes in high-dimensional cell fate transitions},\ }\href@noop {} {\bibfield  {journal} {\bibinfo  {journal} {arXiv preprint 2506.04219}\ } (\bibinfo {year} {2025})}\BibitemShut {NoStop}%
\bibitem [{\citenamefont {Wagner}\ and\ \citenamefont {Klein}(2020)}]{wagner2020lineage}%
  \BibitemOpen
  \bibfield  {author} {\bibinfo {author} {\bibfnamefont {D.~E.}\ \bibnamefont {Wagner}}\ and\ \bibinfo {author} {\bibfnamefont {A.~M.}\ \bibnamefont {Klein}},\ }\bibfield  {title} {\bibinfo {title} {Lineage tracing meets single-cell omics: opportunities and challenges},\ }\href@noop {} {\bibfield  {journal} {\bibinfo  {journal} {Nature Reviews Genetics}\ }\textbf {\bibinfo {volume} {21}},\ \bibinfo {pages} {410} (\bibinfo {year} {2020})}\BibitemShut {NoStop}%
\bibitem [{\citenamefont {Baron}\ and\ \citenamefont {van Oudenaarden}(2019)}]{baron2019unravelling}%
  \BibitemOpen
  \bibfield  {author} {\bibinfo {author} {\bibfnamefont {C.~S.}\ \bibnamefont {Baron}}\ and\ \bibinfo {author} {\bibfnamefont {A.}~\bibnamefont {van Oudenaarden}},\ }\bibfield  {title} {\bibinfo {title} {Unravelling cellular relationships during development and regeneration using genetic lineage tracing},\ }\href@noop {} {\bibfield  {journal} {\bibinfo  {journal} {Nature reviews molecular cell biology}\ }\textbf {\bibinfo {volume} {20}},\ \bibinfo {pages} {753} (\bibinfo {year} {2019})}\BibitemShut {NoStop}%
\bibitem [{\citenamefont {Weinreb}\ \emph {et~al.}(2020)\citenamefont {Weinreb}, \citenamefont {Rodriguez-Fraticelli}, \citenamefont {Camargo},\ and\ \citenamefont {Klein}}]{weinreb2020lineage}%
  \BibitemOpen
  \bibfield  {author} {\bibinfo {author} {\bibfnamefont {C.}~\bibnamefont {Weinreb}}, \bibinfo {author} {\bibfnamefont {A.}~\bibnamefont {Rodriguez-Fraticelli}}, \bibinfo {author} {\bibfnamefont {F.~D.}\ \bibnamefont {Camargo}},\ and\ \bibinfo {author} {\bibfnamefont {A.~M.}\ \bibnamefont {Klein}},\ }\bibfield  {title} {\bibinfo {title} {Lineage tracing on transcriptional landscapes links state to fate during differentiation},\ }\href@noop {} {\bibfield  {journal} {\bibinfo  {journal} {Science}\ }\textbf {\bibinfo {volume} {367}},\ \bibinfo {pages} {eaaw3381} (\bibinfo {year} {2020})}\BibitemShut {NoStop}%
\bibitem [{\citenamefont {Raj}(2025)}]{raj2025single}%
  \BibitemOpen
  \bibfield  {author} {\bibinfo {author} {\bibfnamefont {B.}~\bibnamefont {Raj}},\ }\bibfield  {title} {\bibinfo {title} {Single-cell profiling of lineages and cell types in the vertebrate brain},\ }in\ \href@noop {} {\emph {\bibinfo {booktitle} {Lineage Tracing: Methods and Protocols}}}\ (\bibinfo  {publisher} {Springer},\ \bibinfo {year} {2025})\ pp.\ \bibinfo {pages} {299--310}\BibitemShut {NoStop}%
\bibitem [{\citenamefont {Spanjaard}\ \emph {et~al.}(2018)\citenamefont {Spanjaard}, \citenamefont {Hu}, \citenamefont {Mitic}, \citenamefont {Olivares-Chauvet}, \citenamefont {Janjuha}, \citenamefont {Ninov},\ and\ \citenamefont {Junker}}]{spanjaard2018simultaneous}%
  \BibitemOpen
  \bibfield  {author} {\bibinfo {author} {\bibfnamefont {B.}~\bibnamefont {Spanjaard}}, \bibinfo {author} {\bibfnamefont {B.}~\bibnamefont {Hu}}, \bibinfo {author} {\bibfnamefont {N.}~\bibnamefont {Mitic}}, \bibinfo {author} {\bibfnamefont {P.}~\bibnamefont {Olivares-Chauvet}}, \bibinfo {author} {\bibfnamefont {S.}~\bibnamefont {Janjuha}}, \bibinfo {author} {\bibfnamefont {N.}~\bibnamefont {Ninov}},\ and\ \bibinfo {author} {\bibfnamefont {J.~P.}\ \bibnamefont {Junker}},\ }\bibfield  {title} {\bibinfo {title} {Simultaneous lineage tracing and cell-type identification using crispr--cas9-induced genetic scars},\ }\href@noop {} {\bibfield  {journal} {\bibinfo  {journal} {Nature biotechnology}\ }\textbf {\bibinfo {volume} {36}},\ \bibinfo {pages} {469} (\bibinfo {year} {2018})}\BibitemShut {NoStop}%
\bibitem [{\citenamefont {Ihry}\ \emph {et~al.}(2018)\citenamefont {Ihry}, \citenamefont {Worringer}, \citenamefont {Salick}, \citenamefont {Frias}, \citenamefont {Ho}, \citenamefont {Theriault}, \citenamefont {Kommineni}, \citenamefont {Chen}, \citenamefont {Sondey}, \citenamefont {Ye} \emph {et~al.}}]{ihry2018p53}%
  \BibitemOpen
  \bibfield  {author} {\bibinfo {author} {\bibfnamefont {R.~J.}\ \bibnamefont {Ihry}}, \bibinfo {author} {\bibfnamefont {K.~A.}\ \bibnamefont {Worringer}}, \bibinfo {author} {\bibfnamefont {M.~R.}\ \bibnamefont {Salick}}, \bibinfo {author} {\bibfnamefont {E.}~\bibnamefont {Frias}}, \bibinfo {author} {\bibfnamefont {D.}~\bibnamefont {Ho}}, \bibinfo {author} {\bibfnamefont {K.}~\bibnamefont {Theriault}}, \bibinfo {author} {\bibfnamefont {S.}~\bibnamefont {Kommineni}}, \bibinfo {author} {\bibfnamefont {J.}~\bibnamefont {Chen}}, \bibinfo {author} {\bibfnamefont {M.}~\bibnamefont {Sondey}}, \bibinfo {author} {\bibfnamefont {C.}~\bibnamefont {Ye}}, \emph {et~al.},\ }\bibfield  {title} {\bibinfo {title} {p53 inhibits crispr--cas9 engineering in human pluripotent stem cells},\ }\href@noop {} {\bibfield  {journal} {\bibinfo  {journal} {Nature medicine}\ }\textbf {\bibinfo {volume} {24}},\ \bibinfo {pages} {939} (\bibinfo {year} {2018})}\BibitemShut {NoStop}%
\bibitem [{\citenamefont {Deichmann}(2016)}]{Deichmann2016epigenetics}%
  \BibitemOpen
  \bibfield  {author} {\bibinfo {author} {\bibfnamefont {U.}~\bibnamefont {Deichmann}},\ }\bibfield  {title} {\bibinfo {title} {Epigenetics: The origins and evolution of a fashionable topic},\ }\href@noop {} {\bibfield  {journal} {\bibinfo  {journal} {Developmental biology}\ }\textbf {\bibinfo {volume} {416}},\ \bibinfo {pages} {249} (\bibinfo {year} {2016})}\BibitemShut {NoStop}%
\bibitem [{\citenamefont {Jablonka}\ and\ \citenamefont {Lamb}(2002)}]{Jablonka2002epigeneticsterm}%
  \BibitemOpen
  \bibfield  {author} {\bibinfo {author} {\bibfnamefont {E.}~\bibnamefont {Jablonka}}\ and\ \bibinfo {author} {\bibfnamefont {M.~J.}\ \bibnamefont {Lamb}},\ }\bibfield  {title} {\bibinfo {title} {The changing concept of epigenetics},\ }\href@noop {} {\bibfield  {journal} {\bibinfo  {journal} {Annals of the New York Academy of Sciences}\ }\textbf {\bibinfo {volume} {981}},\ \bibinfo {pages} {82} (\bibinfo {year} {2002})}\BibitemShut {NoStop}%
\bibitem [{\citenamefont {Lunyak}\ and\ \citenamefont {Rosenfeld}(2008)}]{Lunyak2008epigenetic}%
  \BibitemOpen
  \bibfield  {author} {\bibinfo {author} {\bibfnamefont {V.~V.}\ \bibnamefont {Lunyak}}\ and\ \bibinfo {author} {\bibfnamefont {M.~G.}\ \bibnamefont {Rosenfeld}},\ }\bibfield  {title} {\bibinfo {title} {Epigenetic regulation of stem cell fate},\ }\href@noop {} {\bibfield  {journal} {\bibinfo  {journal} {Human molecular genetics}\ }\textbf {\bibinfo {volume} {17}},\ \bibinfo {pages} {R28} (\bibinfo {year} {2008})}\BibitemShut {NoStop}%
\bibitem [{\citenamefont {Meissner}(2010)}]{Meissner2010epigeneticstemcells}%
  \BibitemOpen
  \bibfield  {author} {\bibinfo {author} {\bibfnamefont {A.}~\bibnamefont {Meissner}},\ }\bibfield  {title} {\bibinfo {title} {Epigenetic modifications in pluripotent and differentiated cells},\ }\href@noop {} {\bibfield  {journal} {\bibinfo  {journal} {Nature biotechnology}\ }\textbf {\bibinfo {volume} {28}},\ \bibinfo {pages} {1079} (\bibinfo {year} {2010})}\BibitemShut {NoStop}%
\bibitem [{\citenamefont {Ogrodnik}(2021)}]{ogrodnik2021cellular}%
  \BibitemOpen
  \bibfield  {author} {\bibinfo {author} {\bibfnamefont {M.}~\bibnamefont {Ogrodnik}},\ }\bibfield  {title} {\bibinfo {title} {Cellular aging beyond cellular senescence: Markers of senescence prior to cell cycle arrest in vitro and in vivo},\ }\href@noop {} {\bibfield  {journal} {\bibinfo  {journal} {Aging cell}\ }\textbf {\bibinfo {volume} {20}},\ \bibinfo {pages} {e13338} (\bibinfo {year} {2021})}\BibitemShut {NoStop}%
\bibitem [{\citenamefont {Koch}\ \emph {et~al.}(2012)\citenamefont {Koch}, \citenamefont {Joussen}, \citenamefont {Schellenberg}, \citenamefont {Lin}, \citenamefont {Zenke},\ and\ \citenamefont {Wagner}}]{koch2012monitoring}%
  \BibitemOpen
  \bibfield  {author} {\bibinfo {author} {\bibfnamefont {C.~M.}\ \bibnamefont {Koch}}, \bibinfo {author} {\bibfnamefont {S.}~\bibnamefont {Joussen}}, \bibinfo {author} {\bibfnamefont {A.}~\bibnamefont {Schellenberg}}, \bibinfo {author} {\bibfnamefont {Q.}~\bibnamefont {Lin}}, \bibinfo {author} {\bibfnamefont {M.}~\bibnamefont {Zenke}},\ and\ \bibinfo {author} {\bibfnamefont {W.}~\bibnamefont {Wagner}},\ }\bibfield  {title} {\bibinfo {title} {Monitoring of cellular senescence by dna-methylation at specific cpg sites},\ }\href@noop {} {\bibfield  {journal} {\bibinfo  {journal} {Aging cell}\ }\textbf {\bibinfo {volume} {11}},\ \bibinfo {pages} {366} (\bibinfo {year} {2012})}\BibitemShut {NoStop}%
\bibitem [{\citenamefont {Franzen}\ \emph {et~al.}(2017)\citenamefont {Franzen}, \citenamefont {Zirkel}, \citenamefont {Blake}, \citenamefont {Rath}, \citenamefont {Benes}, \citenamefont {Papantonis},\ and\ \citenamefont {Wagner}}]{franzen2017senescence}%
  \BibitemOpen
  \bibfield  {author} {\bibinfo {author} {\bibfnamefont {J.}~\bibnamefont {Franzen}}, \bibinfo {author} {\bibfnamefont {A.}~\bibnamefont {Zirkel}}, \bibinfo {author} {\bibfnamefont {J.}~\bibnamefont {Blake}}, \bibinfo {author} {\bibfnamefont {B.}~\bibnamefont {Rath}}, \bibinfo {author} {\bibfnamefont {V.}~\bibnamefont {Benes}}, \bibinfo {author} {\bibfnamefont {A.}~\bibnamefont {Papantonis}},\ and\ \bibinfo {author} {\bibfnamefont {W.}~\bibnamefont {Wagner}},\ }\bibfield  {title} {\bibinfo {title} {Senescence-associated dna methylation is stochastically acquired in subpopulations of mesenchymal stem cells},\ }\href@noop {} {\bibfield  {journal} {\bibinfo  {journal} {Aging cell}\ }\textbf {\bibinfo {volume} {16}},\ \bibinfo {pages} {183} (\bibinfo {year} {2017})}\BibitemShut {NoStop}%
\bibitem [{\citenamefont {Franzen}\ \emph {et~al.}(2021)\citenamefont {Franzen}, \citenamefont {Georgomanolis}, \citenamefont {Selich}, \citenamefont {Kuo}, \citenamefont {St{\"o}ger}, \citenamefont {Brant}, \citenamefont {Mulabdi{\'c}}, \citenamefont {Fernandez-Rebollo}, \citenamefont {Grezella}, \citenamefont {Ostrowska} \emph {et~al.}}]{franzen2021dna}%
  \BibitemOpen
  \bibfield  {author} {\bibinfo {author} {\bibfnamefont {J.}~\bibnamefont {Franzen}}, \bibinfo {author} {\bibfnamefont {T.}~\bibnamefont {Georgomanolis}}, \bibinfo {author} {\bibfnamefont {A.}~\bibnamefont {Selich}}, \bibinfo {author} {\bibfnamefont {C.-C.}\ \bibnamefont {Kuo}}, \bibinfo {author} {\bibfnamefont {R.}~\bibnamefont {St{\"o}ger}}, \bibinfo {author} {\bibfnamefont {L.}~\bibnamefont {Brant}}, \bibinfo {author} {\bibfnamefont {M.~S.}\ \bibnamefont {Mulabdi{\'c}}}, \bibinfo {author} {\bibfnamefont {E.}~\bibnamefont {Fernandez-Rebollo}}, \bibinfo {author} {\bibfnamefont {C.}~\bibnamefont {Grezella}}, \bibinfo {author} {\bibfnamefont {A.}~\bibnamefont {Ostrowska}}, \emph {et~al.},\ }\bibfield  {title} {\bibinfo {title} {Dna methylation changes during long-term in vitro cell culture are caused by epigenetic drift},\ }\href@noop {} {\bibfield  {journal} {\bibinfo  {journal} {Communications biology}\ }\textbf {\bibinfo {volume} {4}},\ \bibinfo {pages} {598} (\bibinfo {year} {2021})}\BibitemShut {NoStop}%
\bibitem [{\citenamefont {Nachtwey}\ and\ \citenamefont {Cameron}(1969)}]{nachtwey1969cell}%
  \BibitemOpen
  \bibfield  {author} {\bibinfo {author} {\bibfnamefont {D.}~\bibnamefont {Nachtwey}}\ and\ \bibinfo {author} {\bibfnamefont {I.}~\bibnamefont {Cameron}},\ }\bibfield  {title} {\bibinfo {title} {Cell cycle analysis},\ }in\ \href@noop {} {\emph {\bibinfo {booktitle} {Methods in Cell Biology}}},\ Vol.~\bibinfo {volume} {3}\ (\bibinfo  {publisher} {Elsevier},\ \bibinfo {year} {1969})\ pp.\ \bibinfo {pages} {213--259}\BibitemShut {NoStop}%
\bibitem [{\citenamefont {Cheung}\ and\ \citenamefont {Rando}(2013)}]{cheung2013molecular}%
  \BibitemOpen
  \bibfield  {author} {\bibinfo {author} {\bibfnamefont {T.~H.}\ \bibnamefont {Cheung}}\ and\ \bibinfo {author} {\bibfnamefont {T.~A.}\ \bibnamefont {Rando}},\ }\bibfield  {title} {\bibinfo {title} {Molecular regulation of stem cell quiescence},\ }\href@noop {} {\bibfield  {journal} {\bibinfo  {journal} {Nature reviews Molecular cell biology}\ }\textbf {\bibinfo {volume} {14}},\ \bibinfo {pages} {329} (\bibinfo {year} {2013})}\BibitemShut {NoStop}%
\bibitem [{\citenamefont {Behl}\ and\ \citenamefont {Ziegler}(2013)}]{behl2013cell}%
  \BibitemOpen
  \bibfield  {author} {\bibinfo {author} {\bibfnamefont {C.}~\bibnamefont {Behl}}\ and\ \bibinfo {author} {\bibfnamefont {C.}~\bibnamefont {Ziegler}},\ }\bibfield  {title} {\bibinfo {title} {Cell cycle: The life cycle of a cell},\ }in\ \href@noop {} {\emph {\bibinfo {booktitle} {Cell aging: Molecular mechanisms and implications for disease}}}\ (\bibinfo  {publisher} {Springer},\ \bibinfo {year} {2013})\ pp.\ \bibinfo {pages} {9--19}\BibitemShut {NoStop}%
\bibitem [{\citenamefont {Mens}\ and\ \citenamefont {Ghanbari}(2018)}]{mens2018cell}%
  \BibitemOpen
  \bibfield  {author} {\bibinfo {author} {\bibfnamefont {M.~M.}\ \bibnamefont {Mens}}\ and\ \bibinfo {author} {\bibfnamefont {M.}~\bibnamefont {Ghanbari}},\ }\bibfield  {title} {\bibinfo {title} {Cell cycle regulation of stem cells by micrornas},\ }\href@noop {} {\bibfield  {journal} {\bibinfo  {journal} {Stem cell reviews and reports}\ }\textbf {\bibinfo {volume} {14}},\ \bibinfo {pages} {309} (\bibinfo {year} {2018})}\BibitemShut {NoStop}%
\bibitem [{\citenamefont {Bianco}\ and\ \citenamefont {Robey}(2015)}]{bianco2015skeletal}%
  \BibitemOpen
  \bibfield  {author} {\bibinfo {author} {\bibfnamefont {P.}~\bibnamefont {Bianco}}\ and\ \bibinfo {author} {\bibfnamefont {P.~G.}\ \bibnamefont {Robey}},\ }\bibfield  {title} {\bibinfo {title} {Skeletal stem cells},\ }\href@noop {} {\bibfield  {journal} {\bibinfo  {journal} {Development}\ }\textbf {\bibinfo {volume} {142}},\ \bibinfo {pages} {1023} (\bibinfo {year} {2015})}\BibitemShut {NoStop}%
\bibitem [{\citenamefont {Bianco}\ \emph {et~al.}(2013)\citenamefont {Bianco}, \citenamefont {Cao}, \citenamefont {Frenette}, \citenamefont {Mao}, \citenamefont {Robey}, \citenamefont {Simmons},\ and\ \citenamefont {Wang}}]{bianco2013meaning}%
  \BibitemOpen
  \bibfield  {author} {\bibinfo {author} {\bibfnamefont {P.}~\bibnamefont {Bianco}}, \bibinfo {author} {\bibfnamefont {X.}~\bibnamefont {Cao}}, \bibinfo {author} {\bibfnamefont {P.~S.}\ \bibnamefont {Frenette}}, \bibinfo {author} {\bibfnamefont {J.~J.}\ \bibnamefont {Mao}}, \bibinfo {author} {\bibfnamefont {P.~G.}\ \bibnamefont {Robey}}, \bibinfo {author} {\bibfnamefont {P.~J.}\ \bibnamefont {Simmons}},\ and\ \bibinfo {author} {\bibfnamefont {C.-Y.}\ \bibnamefont {Wang}},\ }\bibfield  {title} {\bibinfo {title} {The meaning, the sense and the significance: translating the science of mesenchymal stem cells into medicine},\ }\href@noop {} {\bibfield  {journal} {\bibinfo  {journal} {Nature medicine}\ }\textbf {\bibinfo {volume} {19}},\ \bibinfo {pages} {35} (\bibinfo {year} {2013})}\BibitemShut {NoStop}%
\bibitem [{\citenamefont {Mareddy}\ \emph {et~al.}(2007)\citenamefont {Mareddy}, \citenamefont {Crawford}, \citenamefont {Brooke},\ and\ \citenamefont {Xiao}}]{mareddy2007clonal}%
  \BibitemOpen
  \bibfield  {author} {\bibinfo {author} {\bibfnamefont {S.}~\bibnamefont {Mareddy}}, \bibinfo {author} {\bibfnamefont {R.}~\bibnamefont {Crawford}}, \bibinfo {author} {\bibfnamefont {G.}~\bibnamefont {Brooke}},\ and\ \bibinfo {author} {\bibfnamefont {Y.}~\bibnamefont {Xiao}},\ }\bibfield  {title} {\bibinfo {title} {Clonal isolation and characterization of bone marrow stromal cells from patients with osteoarthritis},\ }\href@noop {} {\bibfield  {journal} {\bibinfo  {journal} {Tissue engineering}\ }\textbf {\bibinfo {volume} {13}},\ \bibinfo {pages} {819} (\bibinfo {year} {2007})}\BibitemShut {NoStop}%
\bibitem [{\citenamefont {Mareddy}\ \emph {et~al.}(2010)\citenamefont {Mareddy}, \citenamefont {Dhaliwal}, \citenamefont {Crawford},\ and\ \citenamefont {Xiao}}]{mareddy2010stem}%
  \BibitemOpen
  \bibfield  {author} {\bibinfo {author} {\bibfnamefont {S.}~\bibnamefont {Mareddy}}, \bibinfo {author} {\bibfnamefont {N.}~\bibnamefont {Dhaliwal}}, \bibinfo {author} {\bibfnamefont {R.}~\bibnamefont {Crawford}},\ and\ \bibinfo {author} {\bibfnamefont {Y.}~\bibnamefont {Xiao}},\ }\bibfield  {title} {\bibinfo {title} {Stem cell--related gene expression in clonal populations of mesenchymal stromal cells from bone marrow},\ }\href@noop {} {\bibfield  {journal} {\bibinfo  {journal} {Tissue Engineering Part A}\ }\textbf {\bibinfo {volume} {16}},\ \bibinfo {pages} {749} (\bibinfo {year} {2010})}\BibitemShut {NoStop}%
\bibitem [{\citenamefont {Kuczek}\ and\ \citenamefont {Axelrod}(1986)}]{kuczek1986importance}%
  \BibitemOpen
  \bibfield  {author} {\bibinfo {author} {\bibfnamefont {T.}~\bibnamefont {Kuczek}}\ and\ \bibinfo {author} {\bibfnamefont {D.~E.}\ \bibnamefont {Axelrod}},\ }\bibfield  {title} {\bibinfo {title} {The importance of clonal heterogeniety and interexperiment variability in modeling the eukaryotic cell cycle},\ }\href@noop {} {\bibfield  {journal} {\bibinfo  {journal} {Mathematical biosciences}\ }\textbf {\bibinfo {volume} {79}},\ \bibinfo {pages} {87} (\bibinfo {year} {1986})}\BibitemShut {NoStop}%
\bibitem [{\citenamefont {Rennerfeldt}\ and\ \citenamefont {Van~Vliet}(2016)}]{rennerfeldt2016concise}%
  \BibitemOpen
  \bibfield  {author} {\bibinfo {author} {\bibfnamefont {D.~A.}\ \bibnamefont {Rennerfeldt}}\ and\ \bibinfo {author} {\bibfnamefont {K.~J.}\ \bibnamefont {Van~Vliet}},\ }\bibfield  {title} {\bibinfo {title} {Concise review: when colonies are not clones: evidence and implications of intracolony heterogeneity in mesenchymal stem cells},\ }\href@noop {} {\bibfield  {journal} {\bibinfo  {journal} {Stem cells}\ }\textbf {\bibinfo {volume} {34}},\ \bibinfo {pages} {1135} (\bibinfo {year} {2016})}\BibitemShut {NoStop}%
\bibitem [{\citenamefont {Powell}(1955)}]{powell1955some}%
  \BibitemOpen
  \bibfield  {author} {\bibinfo {author} {\bibfnamefont {E.}~\bibnamefont {Powell}},\ }\bibfield  {title} {\bibinfo {title} {Some features of the generation times of individual bacteria},\ }\href@noop {} {\bibfield  {journal} {\bibinfo  {journal} {Biometrika}\ }\textbf {\bibinfo {volume} {42}},\ \bibinfo {pages} {16} (\bibinfo {year} {1955})}\BibitemShut {NoStop}%
\bibitem [{\citenamefont {Genthon}\ \emph {et~al.}(2023)\citenamefont {Genthon}, \citenamefont {Nozoe}, \citenamefont {Peliti},\ and\ \citenamefont {Lacoste}}]{genthon2023cell}%
  \BibitemOpen
  \bibfield  {author} {\bibinfo {author} {\bibfnamefont {A.}~\bibnamefont {Genthon}}, \bibinfo {author} {\bibfnamefont {T.}~\bibnamefont {Nozoe}}, \bibinfo {author} {\bibfnamefont {L.}~\bibnamefont {Peliti}},\ and\ \bibinfo {author} {\bibfnamefont {D.}~\bibnamefont {Lacoste}},\ }\bibfield  {title} {\bibinfo {title} {Cell lineage statistics with incomplete population trees},\ }\href@noop {} {\bibfield  {journal} {\bibinfo  {journal} {PRX Life}\ }\textbf {\bibinfo {volume} {1}},\ \bibinfo {pages} {013014} (\bibinfo {year} {2023})}\BibitemShut {NoStop}%
\bibitem [{\citenamefont {Allegrezza}\ \emph {et~al.}(2026)\citenamefont {Allegrezza}, \citenamefont {Beschi}, \citenamefont {Caudo}, \citenamefont {Cavagna}, \citenamefont {Corsi}, \citenamefont {Culla}, \citenamefont {Donsante}, \citenamefont {Giannicola}, \citenamefont {Giardina}, \citenamefont {Gosti} \emph {et~al.}}]{allegrezza2025topology}%
  \BibitemOpen
  \bibfield  {author} {\bibinfo {author} {\bibfnamefont {A.}~\bibnamefont {Allegrezza}}, \bibinfo {author} {\bibfnamefont {R.}~\bibnamefont {Beschi}}, \bibinfo {author} {\bibfnamefont {D.}~\bibnamefont {Caudo}}, \bibinfo {author} {\bibfnamefont {A.}~\bibnamefont {Cavagna}}, \bibinfo {author} {\bibfnamefont {A.}~\bibnamefont {Corsi}}, \bibinfo {author} {\bibfnamefont {A.}~\bibnamefont {Culla}}, \bibinfo {author} {\bibfnamefont {S.}~\bibnamefont {Donsante}}, \bibinfo {author} {\bibfnamefont {G.}~\bibnamefont {Giannicola}}, \bibinfo {author} {\bibfnamefont {I.}~\bibnamefont {Giardina}}, \bibinfo {author} {\bibfnamefont {G.}~\bibnamefont {Gosti}}, \emph {et~al.},\ }\bibfield  {title} {\bibinfo {title} {Lineage topology, replication kinetics and cell cycle synchronization reveal regulated growth dynamics in human bone marrow stromal cell colonies},\ }\bibfield  {journal} {\bibinfo  {journal} {Scientific Reports}\ }\href {https://doi.org/https://doi.org/10.1038/s41598-026-51809-z} {https://doi.org/10.1038/s41598-026-51809-z} (\bibinfo {year} {2026})\BibitemShut {NoStop}%
\bibitem [{SM()}]{SM}%
  \BibitemOpen
  \href@noop {} {}\bibinfo {note} {See Supplemental Material-SM for all technical details about the models and for an exhaustive illustration of the lineage trees. The SM includes Refs.\cite{tran2024lineage, felsenstein1981evolutionary, neher2014predicting, aragon2026learning, dieselhorst2026phylodynamics, wilks1938large, froese1964distribution, kubitschek1962normal}.}\BibitemShut {Stop}%
\bibitem [{\citenamefont {Kirkwood}\ and\ \citenamefont {Holliday}(1975)}]{kirkwood1975commitment}%
  \BibitemOpen
  \bibfield  {author} {\bibinfo {author} {\bibfnamefont {T.}~\bibnamefont {Kirkwood}}\ and\ \bibinfo {author} {\bibfnamefont {R.}~\bibnamefont {Holliday}},\ }\bibfield  {title} {\bibinfo {title} {Commitment to senescence: a model for the finite and infinite growth of diploid and transformed human fibroblasts in culture},\ }\href@noop {} {\bibfield  {journal} {\bibinfo  {journal} {Journal of Theoretical Biology}\ }\textbf {\bibinfo {volume} {53}},\ \bibinfo {pages} {481} (\bibinfo {year} {1975})}\BibitemShut {NoStop}%
\bibitem [{\citenamefont {Holliday}\ \emph {et~al.}(1977)\citenamefont {Holliday}, \citenamefont {Huschtscha}, \citenamefont {Tarrant},\ and\ \citenamefont {Kirkwood}}]{holliday1977testing}%
  \BibitemOpen
  \bibfield  {author} {\bibinfo {author} {\bibfnamefont {R.}~\bibnamefont {Holliday}}, \bibinfo {author} {\bibfnamefont {L.}~\bibnamefont {Huschtscha}}, \bibinfo {author} {\bibfnamefont {G.}~\bibnamefont {Tarrant}},\ and\ \bibinfo {author} {\bibfnamefont {T.}~\bibnamefont {Kirkwood}},\ }\bibfield  {title} {\bibinfo {title} {Testing the commitment theory of cellular aging: The finite lifespan of human fibroblasts may be due to the decline and loss of a subpopulation of immortal cells.},\ }\href@noop {} {\bibfield  {journal} {\bibinfo  {journal} {Science}\ }\textbf {\bibinfo {volume} {198}},\ \bibinfo {pages} {366} (\bibinfo {year} {1977})}\BibitemShut {NoStop}%
\bibitem [{\citenamefont {Harley}\ and\ \citenamefont {Goldstein}(1980)}]{harley1980retesting}%
  \BibitemOpen
  \bibfield  {author} {\bibinfo {author} {\bibfnamefont {C.~B.}\ \bibnamefont {Harley}}\ and\ \bibinfo {author} {\bibfnamefont {S.}~\bibnamefont {Goldstein}},\ }\bibfield  {title} {\bibinfo {title} {Retesting the commitment theory of cellular aging},\ }\href@noop {} {\bibfield  {journal} {\bibinfo  {journal} {Science}\ }\textbf {\bibinfo {volume} {207}},\ \bibinfo {pages} {191} (\bibinfo {year} {1980})}\BibitemShut {NoStop}%
\bibitem [{\citenamefont {Luria}\ and\ \citenamefont {Delbr{\"u}ck}(1943)}]{luria1943mutations}%
  \BibitemOpen
  \bibfield  {author} {\bibinfo {author} {\bibfnamefont {S.~E.}\ \bibnamefont {Luria}}\ and\ \bibinfo {author} {\bibfnamefont {M.}~\bibnamefont {Delbr{\"u}ck}},\ }\bibfield  {title} {\bibinfo {title} {Mutations of bacteria from virus sensitivity to virus resistance},\ }\href@noop {} {\bibfield  {journal} {\bibinfo  {journal} {Genetics}\ }\textbf {\bibinfo {volume} {28}},\ \bibinfo {pages} {491} (\bibinfo {year} {1943})}\BibitemShut {NoStop}%
\bibitem [{\citenamefont {Rosche}\ and\ \citenamefont {Foster}(2000)}]{rosche2000determining}%
  \BibitemOpen
  \bibfield  {author} {\bibinfo {author} {\bibfnamefont {W.~A.}\ \bibnamefont {Rosche}}\ and\ \bibinfo {author} {\bibfnamefont {P.~L.}\ \bibnamefont {Foster}},\ }\bibfield  {title} {\bibinfo {title} {Determining mutation rates in bacterial populations},\ }\href@noop {} {\bibfield  {journal} {\bibinfo  {journal} {Methods}\ }\textbf {\bibinfo {volume} {20}},\ \bibinfo {pages} {4} (\bibinfo {year} {2000})}\BibitemShut {NoStop}%
\bibitem [{\citenamefont {Colless}(1982)}]{colless1982phylogenetics}%
  \BibitemOpen
  \bibfield  {author} {\bibinfo {author} {\bibfnamefont {D.~H.}\ \bibnamefont {Colless}},\ }\bibfield  {title} {\bibinfo {title} {Phylogenetics: the theory and practice of phylogenetic systematics},\ }\href@noop {} {\bibfield  {journal} {\bibinfo  {journal} {Systematic Zoology}\ }\textbf {\bibinfo {volume} {31}},\ \bibinfo {pages} {100} (\bibinfo {year} {1982})}\BibitemShut {NoStop}%
\bibitem [{\citenamefont {Fischer}\ \emph {et~al.}(2023)\citenamefont {Fischer}, \citenamefont {Herbst}, \citenamefont {Kersting}, \citenamefont {K{\"u}hn},\ and\ \citenamefont {Wicke}}]{fischer2023tree}%
  \BibitemOpen
  \bibfield  {author} {\bibinfo {author} {\bibfnamefont {M.}~\bibnamefont {Fischer}}, \bibinfo {author} {\bibfnamefont {L.}~\bibnamefont {Herbst}}, \bibinfo {author} {\bibfnamefont {S.}~\bibnamefont {Kersting}}, \bibinfo {author} {\bibfnamefont {A.~L.}\ \bibnamefont {K{\"u}hn}},\ and\ \bibinfo {author} {\bibfnamefont {K.}~\bibnamefont {Wicke}},\ }\href@noop {} {\emph {\bibinfo {title} {Tree balance indices: A comprehensive survey}}}\ (\bibinfo  {publisher} {Springer Nature},\ \bibinfo {year} {2023})\BibitemShut {NoStop}%
\bibitem [{\citenamefont {Gull}\ and\ \citenamefont {Daniell}(1978)}]{gull1978image}%
  \BibitemOpen
  \bibfield  {author} {\bibinfo {author} {\bibfnamefont {S.~F.}\ \bibnamefont {Gull}}\ and\ \bibinfo {author} {\bibfnamefont {G.~J.}\ \bibnamefont {Daniell}},\ }\bibfield  {title} {\bibinfo {title} {Image reconstruction from incomplete and noisy data},\ }\href@noop {} {\bibfield  {journal} {\bibinfo  {journal} {Nature}\ }\textbf {\bibinfo {volume} {272}},\ \bibinfo {pages} {686} (\bibinfo {year} {1978})}\BibitemShut {NoStop}%
\bibitem [{\citenamefont {Theil}(1967)}]{theil1967economics}%
  \BibitemOpen
  \bibfield  {author} {\bibinfo {author} {\bibfnamefont {H.}~\bibnamefont {Theil}},\ }\href@noop {} {\emph {\bibinfo {title} {Economics and information theory}}}\ (\bibinfo  {publisher} {Amsterdam: North-Holland,},\ \bibinfo {year} {1967})\BibitemShut {NoStop}%
\bibitem [{\citenamefont {Harris}\ \emph {et~al.}(1963)\citenamefont {Harris} \emph {et~al.}}]{harris1963theory}%
  \BibitemOpen
  \bibfield  {author} {\bibinfo {author} {\bibfnamefont {T.~E.}\ \bibnamefont {Harris}} \emph {et~al.},\ }\href@noop {} {\emph {\bibinfo {title} {The theory of branching processes}}},\ Vol.~\bibinfo {volume} {6}\ (\bibinfo  {publisher} {Springer Berlin},\ \bibinfo {year} {1963})\BibitemShut {NoStop}%
\bibitem [{\citenamefont {d'Souza}\ and\ \citenamefont {Biggins}(1992)}]{d1992supercritical}%
  \BibitemOpen
  \bibfield  {author} {\bibinfo {author} {\bibfnamefont {J.}~\bibnamefont {d'Souza}}\ and\ \bibinfo {author} {\bibfnamefont {J.}~\bibnamefont {Biggins}},\ }\bibfield  {title} {\bibinfo {title} {The supercritical galton-watson process in varying environments},\ }\href@noop {} {\bibfield  {journal} {\bibinfo  {journal} {Stochastic processes and their applications}\ }\textbf {\bibinfo {volume} {42}},\ \bibinfo {pages} {39} (\bibinfo {year} {1992})}\BibitemShut {NoStop}%
\bibitem [{\citenamefont {Kersting}(2020)}]{kersting2020unifying}%
  \BibitemOpen
  \bibfield  {author} {\bibinfo {author} {\bibfnamefont {G.}~\bibnamefont {Kersting}},\ }\bibfield  {title} {\bibinfo {title} {A unifying approach to branching processes in a varying environment},\ }\href@noop {} {\bibfield  {journal} {\bibinfo  {journal} {Journal of Applied Probability}\ }\textbf {\bibinfo {volume} {57}},\ \bibinfo {pages} {196} (\bibinfo {year} {2020})}\BibitemShut {NoStop}%
\bibitem [{\citenamefont {Hayflick}(1965)}]{hayflick1965limited}%
  \BibitemOpen
  \bibfield  {author} {\bibinfo {author} {\bibfnamefont {L.}~\bibnamefont {Hayflick}},\ }\bibfield  {title} {\bibinfo {title} {The limited in vitro lifetime of human diploid cell strains},\ }\href@noop {} {\bibfield  {journal} {\bibinfo  {journal} {Experimental cell research}\ }\textbf {\bibinfo {volume} {37}},\ \bibinfo {pages} {614} (\bibinfo {year} {1965})}\BibitemShut {NoStop}%
\bibitem [{\citenamefont {Robbins}\ \emph {et~al.}(1970)\citenamefont {Robbins}, \citenamefont {Levine},\ and\ \citenamefont {Eagle}}]{robbins1970morphologic}%
  \BibitemOpen
  \bibfield  {author} {\bibinfo {author} {\bibfnamefont {E.}~\bibnamefont {Robbins}}, \bibinfo {author} {\bibfnamefont {E.}~\bibnamefont {Levine}},\ and\ \bibinfo {author} {\bibfnamefont {H.}~\bibnamefont {Eagle}},\ }\bibfield  {title} {\bibinfo {title} {Morphologic changes accompanying senescence of cultured human diploid cells},\ }\href@noop {} {\bibfield  {journal} {\bibinfo  {journal} {The Journal of experimental medicine}\ }\textbf {\bibinfo {volume} {131}},\ \bibinfo {pages} {1211} (\bibinfo {year} {1970})}\BibitemShut {NoStop}%
\bibitem [{\citenamefont {Rubin}(1997)}]{rubin1997cell}%
  \BibitemOpen
  \bibfield  {author} {\bibinfo {author} {\bibfnamefont {H.}~\bibnamefont {Rubin}},\ }\bibfield  {title} {\bibinfo {title} {Cell aging in vivo and in vitro},\ }\href@noop {} {\bibfield  {journal} {\bibinfo  {journal} {Mechanisms of ageing and development}\ }\textbf {\bibinfo {volume} {98}},\ \bibinfo {pages} {1} (\bibinfo {year} {1997})}\BibitemShut {NoStop}%
\bibitem [{\citenamefont {Strasser}\ \emph {et~al.}(2018)\citenamefont {Strasser}, \citenamefont {Hoppe}, \citenamefont {Loeffler}, \citenamefont {Kokkaliaris}, \citenamefont {Schroeder}, \citenamefont {Theis},\ and\ \citenamefont {Marr}}]{strasser2018lineage}%
  \BibitemOpen
  \bibfield  {author} {\bibinfo {author} {\bibfnamefont {M.~K.}\ \bibnamefont {Strasser}}, \bibinfo {author} {\bibfnamefont {P.~S.}\ \bibnamefont {Hoppe}}, \bibinfo {author} {\bibfnamefont {D.}~\bibnamefont {Loeffler}}, \bibinfo {author} {\bibfnamefont {K.~D.}\ \bibnamefont {Kokkaliaris}}, \bibinfo {author} {\bibfnamefont {T.}~\bibnamefont {Schroeder}}, \bibinfo {author} {\bibfnamefont {F.~J.}\ \bibnamefont {Theis}},\ and\ \bibinfo {author} {\bibfnamefont {C.}~\bibnamefont {Marr}},\ }\bibfield  {title} {\bibinfo {title} {Lineage marker synchrony in hematopoietic genealogies refutes the pu. 1/gata1 toggle switch paradigm},\ }\href@noop {} {\bibfield  {journal} {\bibinfo  {journal} {Nature communications}\ }\textbf {\bibinfo {volume} {9}},\ \bibinfo {pages} {2697} (\bibinfo {year} {2018})}\BibitemShut {NoStop}%
\bibitem [{\citenamefont {Wagner}\ \emph {et~al.}(2008)\citenamefont {Wagner}, \citenamefont {Horn}, \citenamefont {Castoldi}, \citenamefont {Diehlmann}, \citenamefont {Bork}, \citenamefont {Saffrich}, \citenamefont {Benes}, \citenamefont {Blake}, \citenamefont {Pfister}, \citenamefont {Eckstein} \emph {et~al.}}]{wagner2008replicative}%
  \BibitemOpen
  \bibfield  {author} {\bibinfo {author} {\bibfnamefont {W.}~\bibnamefont {Wagner}}, \bibinfo {author} {\bibfnamefont {P.}~\bibnamefont {Horn}}, \bibinfo {author} {\bibfnamefont {M.}~\bibnamefont {Castoldi}}, \bibinfo {author} {\bibfnamefont {A.}~\bibnamefont {Diehlmann}}, \bibinfo {author} {\bibfnamefont {S.}~\bibnamefont {Bork}}, \bibinfo {author} {\bibfnamefont {R.}~\bibnamefont {Saffrich}}, \bibinfo {author} {\bibfnamefont {V.}~\bibnamefont {Benes}}, \bibinfo {author} {\bibfnamefont {J.}~\bibnamefont {Blake}}, \bibinfo {author} {\bibfnamefont {S.}~\bibnamefont {Pfister}}, \bibinfo {author} {\bibfnamefont {V.}~\bibnamefont {Eckstein}}, \emph {et~al.},\ }\bibfield  {title} {\bibinfo {title} {Replicative senescence of mesenchymal stem cells: a continuous and organized process},\ }\href@noop {} {\bibfield  {journal} {\bibinfo  {journal} {PloS one}\ }\textbf {\bibinfo {volume} {3}},\ \bibinfo {pages} {e2213} (\bibinfo {year} {2008})}\BibitemShut {NoStop}%
\bibitem [{\citenamefont {Tibshirani}\ and\ \citenamefont {Efron}(1993)}]{tibshirani1993introduction}%
  \BibitemOpen
  \bibfield  {author} {\bibinfo {author} {\bibfnamefont {R.~J.}\ \bibnamefont {Tibshirani}}\ and\ \bibinfo {author} {\bibfnamefont {B.}~\bibnamefont {Efron}},\ }\href@noop {} {\emph {\bibinfo {title} {An introduction to the bootstrap}}}\ (\bibinfo  {publisher} {Chapman $\&$ Hall, New York, London},\ \bibinfo {year} {1993})\BibitemShut {NoStop}%
\bibitem [{\citenamefont {Davison}\ and\ \citenamefont {Hinkley}(1997)}]{davison1997bootstrap}%
  \BibitemOpen
  \bibfield  {author} {\bibinfo {author} {\bibfnamefont {A.~C.}\ \bibnamefont {Davison}}\ and\ \bibinfo {author} {\bibfnamefont {D.~V.}\ \bibnamefont {Hinkley}},\ }\href@noop {} {\emph {\bibinfo {title} {Bootstrap methods and their application}}},\ \bibinfo {number} {1}\ (\bibinfo  {publisher} {Cambridge university press},\ \bibinfo {year} {1997})\BibitemShut {NoStop}%
\bibitem [{\citenamefont {Self}\ and\ \citenamefont {Liang}(1987)}]{self1987asymptotic}%
  \BibitemOpen
  \bibfield  {author} {\bibinfo {author} {\bibfnamefont {S.~G.}\ \bibnamefont {Self}}\ and\ \bibinfo {author} {\bibfnamefont {K.-Y.}\ \bibnamefont {Liang}},\ }\bibfield  {title} {\bibinfo {title} {Asymptotic properties of maximum likelihood estimators and likelihood ratio tests under nonstandard conditions},\ }\href@noop {} {\bibfield  {journal} {\bibinfo  {journal} {Journal of the American Statistical Association}\ }\textbf {\bibinfo {volume} {82}},\ \bibinfo {pages} {605} (\bibinfo {year} {1987})}\BibitemShut {NoStop}%
\bibitem [{\citenamefont {Staudte}\ \emph {et~al.}(1984)\citenamefont {Staudte}, \citenamefont {Guiguet},\ and\ \citenamefont {d'Hooghe}}]{staudte1984additive}%
  \BibitemOpen
  \bibfield  {author} {\bibinfo {author} {\bibfnamefont {R.}~\bibnamefont {Staudte}}, \bibinfo {author} {\bibfnamefont {M.}~\bibnamefont {Guiguet}},\ and\ \bibinfo {author} {\bibfnamefont {M.~C.}\ \bibnamefont {d'Hooghe}},\ }\bibfield  {title} {\bibinfo {title} {Additive models for dependent cell populations},\ }\href@noop {} {\bibfield  {journal} {\bibinfo  {journal} {Journal of theoretical biology}\ }\textbf {\bibinfo {volume} {109}},\ \bibinfo {pages} {127} (\bibinfo {year} {1984})}\BibitemShut {NoStop}%
\bibitem [{\citenamefont {Hormoz}\ \emph {et~al.}(2015)\citenamefont {Hormoz}, \citenamefont {Desprat},\ and\ \citenamefont {Shraiman}}]{hormoz2015inferring}%
  \BibitemOpen
  \bibfield  {author} {\bibinfo {author} {\bibfnamefont {S.}~\bibnamefont {Hormoz}}, \bibinfo {author} {\bibfnamefont {N.}~\bibnamefont {Desprat}},\ and\ \bibinfo {author} {\bibfnamefont {B.~I.}\ \bibnamefont {Shraiman}},\ }\bibfield  {title} {\bibinfo {title} {Inferring epigenetic dynamics from kin correlations},\ }\href@noop {} {\bibfield  {journal} {\bibinfo  {journal} {Proceedings of the National Academy of Sciences}\ }\textbf {\bibinfo {volume} {112}},\ \bibinfo {pages} {E2281} (\bibinfo {year} {2015})}\BibitemShut {NoStop}%
\bibitem [{\citenamefont {Tran}\ \emph {et~al.}(2024)\citenamefont {Tran}, \citenamefont {Askary},\ and\ \citenamefont {Elowitz}}]{tran2024lineage}%
  \BibitemOpen
  \bibfield  {author} {\bibinfo {author} {\bibfnamefont {M.}~\bibnamefont {Tran}}, \bibinfo {author} {\bibfnamefont {A.}~\bibnamefont {Askary}},\ and\ \bibinfo {author} {\bibfnamefont {M.~B.}\ \bibnamefont {Elowitz}},\ }\bibfield  {title} {\bibinfo {title} {Lineage motifs as developmental modules for control of cell type proportions},\ }\href@noop {} {\bibfield  {journal} {\bibinfo  {journal} {Developmental Cell}\ }\textbf {\bibinfo {volume} {59}},\ \bibinfo {pages} {812} (\bibinfo {year} {2024})}\BibitemShut {NoStop}%
\bibitem [{\citenamefont {Felsenstein}(1981)}]{felsenstein1981evolutionary}%
  \BibitemOpen
  \bibfield  {author} {\bibinfo {author} {\bibfnamefont {J.}~\bibnamefont {Felsenstein}},\ }\bibfield  {title} {\bibinfo {title} {Evolutionary trees from dna sequences: a maximum likelihood approach},\ }\href@noop {} {\bibfield  {journal} {\bibinfo  {journal} {Journal of molecular evolution}\ }\textbf {\bibinfo {volume} {17}},\ \bibinfo {pages} {368} (\bibinfo {year} {1981})}\BibitemShut {NoStop}%
\bibitem [{\citenamefont {Neher}\ \emph {et~al.}(2014)\citenamefont {Neher}, \citenamefont {Russell},\ and\ \citenamefont {Shraiman}}]{neher2014predicting}%
  \BibitemOpen
  \bibfield  {author} {\bibinfo {author} {\bibfnamefont {R.~A.}\ \bibnamefont {Neher}}, \bibinfo {author} {\bibfnamefont {C.~A.}\ \bibnamefont {Russell}},\ and\ \bibinfo {author} {\bibfnamefont {B.~I.}\ \bibnamefont {Shraiman}},\ }\bibfield  {title} {\bibinfo {title} {Predicting evolution from the shape of genealogical trees},\ }\href@noop {} {\bibfield  {journal} {\bibinfo  {journal} {elife}\ }\textbf {\bibinfo {volume} {3}},\ \bibinfo {pages} {e03568} (\bibinfo {year} {2014})}\BibitemShut {NoStop}%
\bibitem [{\citenamefont {Aragon}\ \emph {et~al.}(2026)\citenamefont {Aragon}, \citenamefont {Lambert}, \citenamefont {Mora},\ and\ \citenamefont {Walczak}}]{aragon2026learning}%
  \BibitemOpen
  \bibfield  {author} {\bibinfo {author} {\bibfnamefont {A.}~\bibnamefont {Aragon}}, \bibinfo {author} {\bibfnamefont {A.}~\bibnamefont {Lambert}}, \bibinfo {author} {\bibfnamefont {T.}~\bibnamefont {Mora}},\ and\ \bibinfo {author} {\bibfnamefont {A.~M.}\ \bibnamefont {Walczak}},\ }\bibfield  {title} {\bibinfo {title} {Learning evolutionary parameters from genealogies using allelic trees},\ }\href@noop {} {\bibfield  {journal} {\bibinfo  {journal} {Genetics}\ }\textbf {\bibinfo {volume} {232}},\ \bibinfo {pages} {iyaf112} (\bibinfo {year} {2026})}\BibitemShut {NoStop}%
\bibitem [{\citenamefont {Dieselhorst}\ and\ \citenamefont {Berg}(2026)}]{dieselhorst2026phylodynamics}%
  \BibitemOpen
  \bibfield  {author} {\bibinfo {author} {\bibfnamefont {T.}~\bibnamefont {Dieselhorst}}\ and\ \bibinfo {author} {\bibfnamefont {J.}~\bibnamefont {Berg}},\ }\bibfield  {title} {\bibinfo {title} {Phylodynamics of somatic evolution: A likelihood-based approach for cellular reproduction},\ }\href@noop {} {\bibfield  {journal} {\bibinfo  {journal} {Molecular Biology and Evolution}\ }\textbf {\bibinfo {volume} {43}},\ \bibinfo {pages} {msag002} (\bibinfo {year} {2026})}\BibitemShut {NoStop}%
\bibitem [{\citenamefont {Wilks}(1938)}]{wilks1938large}%
  \BibitemOpen
  \bibfield  {author} {\bibinfo {author} {\bibfnamefont {S.~S.}\ \bibnamefont {Wilks}},\ }\bibfield  {title} {\bibinfo {title} {The large-sample distribution of the likelihood ratio for testing composite hypotheses},\ }\href@noop {} {\bibfield  {journal} {\bibinfo  {journal} {The annals of mathematical statistics}\ }\textbf {\bibinfo {volume} {9}},\ \bibinfo {pages} {60} (\bibinfo {year} {1938})}\BibitemShut {NoStop}%
\bibitem [{\citenamefont {Froese}(1964)}]{froese1964distribution}%
  \BibitemOpen
  \bibfield  {author} {\bibinfo {author} {\bibfnamefont {G.}~\bibnamefont {Froese}},\ }\bibfield  {title} {\bibinfo {title} {The distribution and interdependence of generation times of hela cells},\ }\href@noop {} {\bibfield  {journal} {\bibinfo  {journal} {Experimental cell research}\ }\textbf {\bibinfo {volume} {35}},\ \bibinfo {pages} {415} (\bibinfo {year} {1964})}\BibitemShut {NoStop}%
\bibitem [{\citenamefont {Kubitschek}(1962)}]{kubitschek1962normal}%
  \BibitemOpen
  \bibfield  {author} {\bibinfo {author} {\bibfnamefont {H.}~\bibnamefont {Kubitschek}},\ }\bibfield  {title} {\bibinfo {title} {Normal distribution of cell generation rate},\ }\href@noop {} {\bibfield  {journal} {\bibinfo  {journal} {Experimental cell research}\ }\textbf {\bibinfo {volume} {26}},\ \bibinfo {pages} {439} (\bibinfo {year} {1962})}\BibitemShut {NoStop}%
\end{thebibliography}%


\begin{thebibliography}{15}%
\makeatletter
\providecommand \@ifxundefined [1]{%
 \@ifx{#1\undefined}
}%
\providecommand \@ifnum [1]{%
 \ifnum #1\expandafter \@firstoftwo
 \else \expandafter \@secondoftwo
 \fi
}%
\providecommand \@ifx [1]{%
 \ifx #1\expandafter \@firstoftwo
 \else \expandafter \@secondoftwo
 \fi
}%
\providecommand \natexlab [1]{#1}%
\providecommand \enquote  [1]{``#1''}%
\providecommand \bibnamefont  [1]{#1}%
\providecommand \bibfnamefont [1]{#1}%
\providecommand \citenamefont [1]{#1}%
\providecommand \href@noop [0]{\@secondoftwo}%
\providecommand \href [0]{\begingroup \@sanitize@url \@href}%
\providecommand \@href[1]{\@@startlink{#1}\@@href}%
\providecommand \@@href[1]{\endgroup#1\@@endlink}%
\providecommand \@sanitize@url [0]{\catcode `\\12\catcode `\$12\catcode `\&12\catcode `\#12\catcode `\^12\catcode `\_12\catcode `\%12\relax}%
\providecommand \@@startlink[1]{}%
\providecommand \@@endlink[0]{}%
\providecommand \url  [0]{\begingroup\@sanitize@url \@url }%
\providecommand \@url [1]{\endgroup\@href {#1}{\urlprefix }}%
\providecommand \urlprefix  [0]{URL }%
\providecommand \Eprint [0]{\href }%
\providecommand \doibase [0]{http://dx.doi.org/}%
\providecommand \selectlanguage [0]{\@gobble}%
\providecommand \bibinfo  [0]{\@secondoftwo}%
\providecommand \bibfield  [0]{\@secondoftwo}%
\providecommand \translation [1]{[#1]}%
\providecommand \BibitemOpen [0]{}%
\providecommand \bibitemStop [0]{}%
\providecommand \bibitemNoStop [0]{.\EOS\space}%
\providecommand \EOS [0]{\spacefactor3000\relax}%
\providecommand \BibitemShut  [1]{\csname bibitem#1\endcsname}%
\let\auto@bib@innerbib\@empty
%</preamble>
\bibitem [{\citenamefont {Harris}\ \emph {et~al.}(1963)\citenamefont {Harris} \emph {et~al.}}]{harris1963theory}%
  \BibitemOpen
  \bibfield  {author} {\bibinfo {author} {\bibfnamefont {T.~E.}\ \bibnamefont {Harris}} \emph {et~al.},\ }\href@noop {} {\emph {\bibinfo {title} {The theory of branching processes}}},\ Vol.~\bibinfo {volume} {6}\ (\bibinfo  {publisher} {Springer Berlin},\ \bibinfo {year} {1963})\BibitemShut {NoStop}%
\bibitem [{\citenamefont {d'Souza}\ and\ \citenamefont {Biggins}(1992)}]{d1992supercritical}%
  \BibitemOpen
  \bibfield  {author} {\bibinfo {author} {\bibfnamefont {J.}~\bibnamefont {d'Souza}}\ and\ \bibinfo {author} {\bibfnamefont {J.}~\bibnamefont {Biggins}},\ }\href@noop {} {\bibfield  {journal} {\bibinfo  {journal} {Stochastic processes and their applications}\ }\textbf {\bibinfo {volume} {42}},\ \bibinfo {pages} {39} (\bibinfo {year} {1992})}\BibitemShut {NoStop}%
\bibitem [{\citenamefont {Kersting}(2020)}]{kersting2020unifying}%
  \BibitemOpen
  \bibfield  {author} {\bibinfo {author} {\bibfnamefont {G.}~\bibnamefont {Kersting}},\ }\href@noop {} {\bibfield  {journal} {\bibinfo  {journal} {Journal of Applied Probability}\ }\textbf {\bibinfo {volume} {57}},\ \bibinfo {pages} {196} (\bibinfo {year} {2020})}\BibitemShut {NoStop}%
\bibitem [{\citenamefont {Tran}\ \emph {et~al.}(2024)\citenamefont {Tran}, \citenamefont {Askary},\ and\ \citenamefont {Elowitz}}]{tran2024lineage}%
  \BibitemOpen
  \bibfield  {author} {\bibinfo {author} {\bibfnamefont {M.}~\bibnamefont {Tran}}, \bibinfo {author} {\bibfnamefont {A.}~\bibnamefont {Askary}}, \ and\ \bibinfo {author} {\bibfnamefont {M.~B.}\ \bibnamefont {Elowitz}},\ }\href@noop {} {\bibfield  {journal} {\bibinfo  {journal} {Developmental Cell}\ }\textbf {\bibinfo {volume} {59}},\ \bibinfo {pages} {812} (\bibinfo {year} {2024})}\BibitemShut {NoStop}%
\bibitem [{\citenamefont {Felsenstein}(1981)}]{felsenstein1981evolutionary}%
  \BibitemOpen
  \bibfield  {author} {\bibinfo {author} {\bibfnamefont {J.}~\bibnamefont {Felsenstein}},\ }\href@noop {} {\bibfield  {journal} {\bibinfo  {journal} {Journal of molecular evolution}\ }\textbf {\bibinfo {volume} {17}},\ \bibinfo {pages} {368} (\bibinfo {year} {1981})}\BibitemShut {NoStop}%
\bibitem [{\citenamefont {Neher}\ \emph {et~al.}(2014)\citenamefont {Neher}, \citenamefont {Russell},\ and\ \citenamefont {Shraiman}}]{neher2014predicting}%
  \BibitemOpen
  \bibfield  {author} {\bibinfo {author} {\bibfnamefont {R.~A.}\ \bibnamefont {Neher}}, \bibinfo {author} {\bibfnamefont {C.~A.}\ \bibnamefont {Russell}}, \ and\ \bibinfo {author} {\bibfnamefont {B.~I.}\ \bibnamefont {Shraiman}},\ }\href@noop {} {\bibfield  {journal} {\bibinfo  {journal} {elife}\ }\textbf {\bibinfo {volume} {3}},\ \bibinfo {pages} {e03568} (\bibinfo {year} {2014})}\BibitemShut {NoStop}%
\bibitem [{\citenamefont {Aragon}\ \emph {et~al.}(2026)\citenamefont {Aragon}, \citenamefont {Lambert}, \citenamefont {Mora},\ and\ \citenamefont {Walczak}}]{aragon2026learning}%
  \BibitemOpen
  \bibfield  {author} {\bibinfo {author} {\bibfnamefont {A.}~\bibnamefont {Aragon}}, \bibinfo {author} {\bibfnamefont {A.}~\bibnamefont {Lambert}}, \bibinfo {author} {\bibfnamefont {T.}~\bibnamefont {Mora}}, \ and\ \bibinfo {author} {\bibfnamefont {A.~M.}\ \bibnamefont {Walczak}},\ }\href@noop {} {\bibfield  {journal} {\bibinfo  {journal} {Genetics}\ }\textbf {\bibinfo {volume} {232}},\ \bibinfo {pages} {iyaf112} (\bibinfo {year} {2026})}\BibitemShut {NoStop}%
\bibitem [{\citenamefont {Dieselhorst}\ and\ \citenamefont {Berg}(2026)}]{dieselhorst2026phylodynamics}%
  \BibitemOpen
  \bibfield  {author} {\bibinfo {author} {\bibfnamefont {T.}~\bibnamefont {Dieselhorst}}\ and\ \bibinfo {author} {\bibfnamefont {J.}~\bibnamefont {Berg}},\ }\href@noop {} {\bibfield  {journal} {\bibinfo  {journal} {Molecular Biology and Evolution}\ }\textbf {\bibinfo {volume} {43}},\ \bibinfo {pages} {msag002} (\bibinfo {year} {2026})}\BibitemShut {NoStop}%
\bibitem [{\citenamefont {Wilks}(1938)}]{wilks1938large}%
  \BibitemOpen
  \bibfield  {author} {\bibinfo {author} {\bibfnamefont {S.~S.}\ \bibnamefont {Wilks}},\ }\href@noop {} {\bibfield  {journal} {\bibinfo  {journal} {The annals of mathematical statistics}\ }\textbf {\bibinfo {volume} {9}},\ \bibinfo {pages} {60} (\bibinfo {year} {1938})}\BibitemShut {NoStop}%
\bibitem [{\citenamefont {Tibshirani}\ and\ \citenamefont {Efron}(1993)}]{tibshirani1993introduction}%
  \BibitemOpen
  \bibfield  {author} {\bibinfo {author} {\bibfnamefont {R.~J.}\ \bibnamefont {Tibshirani}}\ and\ \bibinfo {author} {\bibfnamefont {B.}~\bibnamefont {Efron}},\ }\href@noop {} {\emph {\bibinfo {title} {An introduction to the bootstrap}}}\ (\bibinfo  {publisher} {Chapman $\&$ Hall, New York, London},\ \bibinfo {year} {1993})\BibitemShut {NoStop}%
\bibitem [{\citenamefont {Self}\ and\ \citenamefont {Liang}(1987)}]{self1987asymptotic}%
  \BibitemOpen
  \bibfield  {author} {\bibinfo {author} {\bibfnamefont {S.~G.}\ \bibnamefont {Self}}\ and\ \bibinfo {author} {\bibfnamefont {K.-Y.}\ \bibnamefont {Liang}},\ }\href@noop {} {\bibfield  {journal} {\bibinfo  {journal} {Journal of the American Statistical Association}\ }\textbf {\bibinfo {volume} {82}},\ \bibinfo {pages} {605} (\bibinfo {year} {1987})}\BibitemShut {NoStop}%
\bibitem [{\citenamefont {Davison}\ and\ \citenamefont {Hinkley}(1997)}]{davison1997bootstrap}%
  \BibitemOpen
  \bibfield  {author} {\bibinfo {author} {\bibfnamefont {A.~C.}\ \bibnamefont {Davison}}\ and\ \bibinfo {author} {\bibfnamefont {D.~V.}\ \bibnamefont {Hinkley}},\ }\href@noop {} {\emph {\bibinfo {title} {Bootstrap methods and their application}}},\ \bibinfo {number} {1}\ (\bibinfo  {publisher} {Cambridge university press},\ \bibinfo {year} {1997})\BibitemShut {NoStop}%
\bibitem [{\citenamefont {Froese}(1964)}]{froese1964distribution}%
  \BibitemOpen
  \bibfield  {author} {\bibinfo {author} {\bibfnamefont {G.}~\bibnamefont {Froese}},\ }\href@noop {} {\bibfield  {journal} {\bibinfo  {journal} {Experimental cell research}\ }\textbf {\bibinfo {volume} {35}},\ \bibinfo {pages} {415} (\bibinfo {year} {1964})}\BibitemShut {NoStop}%
\bibitem [{\citenamefont {Powell}(1955)}]{powell1955some}%
  \BibitemOpen
  \bibfield  {author} {\bibinfo {author} {\bibfnamefont {E.}~\bibnamefont {Powell}},\ }\href@noop {} {\bibfield  {journal} {\bibinfo  {journal} {Biometrika}\ }\textbf {\bibinfo {volume} {42}},\ \bibinfo {pages} {16} (\bibinfo {year} {1955})}\BibitemShut {NoStop}%
\bibitem [{\citenamefont {Kubitschek}(1962)}]{kubitschek1962normal}%
  \BibitemOpen
  \bibfield  {author} {\bibinfo {author} {\bibfnamefont {H.}~\bibnamefont {Kubitschek}},\ }\href@noop {} {\bibfield  {journal} {\bibinfo  {journal} {Experimental cell research}\ }\textbf {\bibinfo {volume} {26}},\ \bibinfo {pages} {439} (\bibinfo {year} {1962})}\BibitemShut {NoStop}%
\end{thebibliography}%
%%%%%%%%%%%%%%%%%%%%%%%%%%%%%%%%%%%%%%%%%%%%%%%
%%%%%%%%%%%%%%%%%%%%%%%%%%%%%%%%%%%%%%%%%%%%%%%

\end{document}

%\title{Inheritance entropy: A model-independent method to probe the hereditary structure \\ of cell lineage trees \\ SUPPLEMENTAL MATERIAL}

\begin{center}
    \large \textbf{Inheritance entropy: A model-independent method to probe the hereditary structure \\ of cell lineage trees} \\
    \vskip 0.5 truecm
    \large \textbf{SUPPLEMENTAL MATERIAL}
\end{center}

\author{Alessandro Allegrezza}

\author{Riccardo Beschi}

\author{Domenico Caudo}

\author{Andrea Cavagna}

\author{Alessandro Corsi}

\author{Antonio Culla}

\author{Samantha Donsante}

\author{Giuseppe Giannicola}

\author{Irene Giardina}

\author{Giorgio Gosti}

\author{Tom\'as S. Grigera}
    
\author{Stefania Melillo}

\author{Biagio Palmisano}

\author{Leonardo Parisi}

\author{Lorena Postiglione}

\author{Mara Riminucci}

\author{Francesco Saverio Rotondi}

\maketitle

\tableofcontents

\newpage

%%%%%%%%%%%%%%%%%%%%%%%%%%%%%%%%%%%%%%%%%%%%%%%%%%%%%%%%%%%%%%%

%%%%%%%%%%%%%%%%%%%%%%%%%%%%%%%%%%%%%%%%%%%%%%%%%%%%%%%%%%%%%%%%%
\section{Details of the GW model}
\subsection{Model description and theoretical relations}

The Galton-Watson model is a simple model of population growth proposed more than 150 years ago and belonging to the category of branching processes \cite{harris1963theory}.  Population is tracked across generations (i.e.\ individual birth times are disregarded): a set of individuals comprising generation $k$ is replaced by their offspring, defining the next generation, $k+1$.  In our case individuals are cells, so we only allow zero or two offspring for each (the tree is binary).  There is no interaction or correlation among individuals; all cells in the lineage give rise to (sub-)lineages that are statistically identical.

A cell has probability $q$ of being inactive (i.e.\ having zero descendants) and probability $1-q$ of being active (two descendants).  Calling $Z_k$ the number of cells at generation $k$ and starting with a single cell ($Z_0=1$), the mean number of cells at generation $k$ and its variance are given by \cite[I.5]{harris1963theory}
\begin{align}
  \langle Z_k \rangle & = 2^k(1-q)^k, \\
  \text{Var}_Z(k) & =
                    \begin{cases}
                      \frac{ 2q 2^k(1-q)^k [ 2^k(1-q)^k-1]}{1-2q}, & q\neq 1/2 \\
                      k & q=1/2.
                    \end{cases}
                    \label{eq:variance-GW}
\end{align}

It is also possible to compute the asymptotic extinction probability $p_E$ (i.e.\ the probability that $Z_k=0$ for some $k$), which we shall need later on.  It is given by the fixed point equation \cite[I.6]{harris1963theory}
\begin{equation}
  p_E = f(p_E), \qquad f(s) = q + (1-q) s^2,
\end{equation}
where $f(s)$ is the probability generation function, and the form quoted above is appropriate for our case with 0 or 2 descendants.  $p_E=1$ is always a solution, and is the extinction probability unless there is another solution in the interval $[0,1)$.  For our case,
\begin{equation}
  \label{eq:extinction-prob}
  p_E =
  \begin{cases}
    q / (1-q), & 0\le q<1/2, \\
    1,          & 1/2 \le q\le 1.
  \end{cases}
\end{equation}

A mitosis $m$ between generations $k-1$ and $k$ gives rise to left and right sub-lineages, each of which is a Galton-Watson tree.  Calling $Z^L_{k+d,k}$ the number of cells of the left sub-tree at generation $k+d$ (and similarly for the right sub-tree), the imbalance at that mitosis is
\begin{equation}
  I_m(k-1,k) = \lvert M_m^\text{left} - M_m^\text{right} \rvert =
  \left\lvert [ 2^{\kmax-k} - Z^L_{\kmax,k} ] - [ 2^{\kmax-k} - Z^R_{\kmax,k} ] \right\rvert.
  \label{eq:imbalance-general}
\end{equation}
In this model the mean $\langle Z_{k+d,k}\rangle$ (as well as the higher moments) depends only on $d$, but we employ this more general notation which is necessary for the NHGW model below.  We can now write the mean square imbalance,
\begin{equation}
  \langle I^2(k-1,k) \rangle = 2 \langle Z^2_{\kmax,k} \rangle - 2\langle Z_{\kmax,k} \rangle^2 = 2\text{Var}_Z(\kmax-k),
\end{equation}
with the variance given by~\eqref{eq:variance-GW}.

\subsection{Likelihood}
\label{sec:GW-like}

Let us now evaluate the likelihood that a specific tree ${\cal T}$, as specified by its topology and the active/inactive state of the cells at all generations, is generated by the GW model with inactivity parameter $q$.  To do so, we need to compute the occurrence probability of the tree within the GW ensemble.  An observation is helpful: in a GW process a cell is active or inactive independently of the others, and the inactivity probability is the same for all cells. Besides, since the process is Markovian, the number of cells at a given generation is uniquely determined by number of active cells at the previous one. As a consequence, the probability that at generation $k$ there are $N_k$ active cells and $2N_{k-1}-N_k$  inactive ones, as observed in the data, is given by $(1-q)^{N_k} q^{2N_{k-1}-N_k}$. Considering all generations in the tree, we therefore have
\begin{equation}
{\cal L}({\cal T}|q)  = P^{\rm GW}({\cal T})  
 =  \prod_{k=1}^{k_\mathrm{max}} (1-q)^{N_k} q^{2N_{k-1}-N_k}  (1-q)^{N_0} q^{1-N_0} 
 \label{GW_like}
\end{equation}
where we assumed for the root the same inactivity probability as in its descendants, and $N_0$ --- the number of active cells at $k=0$ --- can be either $0$ or $1$. 

From Eq.~(\ref{GW_like}), we can find explicitly the value of $q$ that maximizes the likelihood (or, equivalently the log-likelihood). Indeed, we have
\begin{equation}
\frac{\partial \log {\cal L}({\cal T}|q) }{\partial q}  = 0 \Leftrightarrow  \frac{\partial \log(P^{\rm GW})}{\partial q} =0
\end{equation}
Using Eq.~(\ref{GW_like}), this condition becomes
\begin{equation}
\frac{\partial }{\partial q} \left [ \log(1-q) \left ( \sum_{k=0}^{k_\mathrm{max}} N_k  \right )+ \log(q) \left (1-N_0+\sum_{k=1}^{k_\mathrm{max}}(2N_{k-1}-N_k) \right )  \right ] =0 \ .
\end{equation}
The maximum likelihood estimator for $q$ is then
\begin{equation}
q^{\rm MLE} = \frac{1-N_0+\sum_{k=1}^{k_\mathrm{max}}(2N_{k-1}-N_k)}{1 + \sum_{k=1}^{k_\mathrm{max}} 2N_{k-1}} = \frac{ \sum_{k=0}^{\kmax} Z_{k} - \sum_{k=0}^{\kmax} N_{k} }{\sum_{k=0}^{\kmax} Z_{k}} = \frac{\Nina}{\Nina +\Nacta}  ,
\end{equation}
where we called $\Nacta$ and $\Nina$ the global number of, respectively, active and inactive cells in the whole tree.  This estimation is based on two global quantities, and it is very easily performed for experimental or synthetic trees.

Before concluding this Section, two further remarks. Eq.~(\ref{GW_like}) represents the unconditional likelihood, where the root can either be active or inactive. The case where the root is assumed to be active, can be easily derived by dividing the likelihood by the probability that $N_0=1$, i.e. by $1-q$. This conditional likelihood then reads
\begin{equation}
P^{\rm GW}_c =  \prod_{k=1}^{k_\mathrm{max}} (1-q)^{N_k} q^{2N_{k-1}-N_k}  ,
\label{GW_cond}
\end{equation}
which will be useful later.

Finally, we note that for a GW tree, all the trees that have the same number of active cells $N_k$ per generation, occur with the same probability (the one appearing in the likelihood). By taking into account this combinatorial factor, we can easily write the distribution probability for the number of active cells in the tree
\begin{equation}
 P(N_0,N_1,\cdots N_{k_\mathrm{max}}) =  \prod_{k=1}^{k_\mathrm{max}} P(N_{k}|N_{k-1}) P(N_0)
 =  \prod_{k=1}^{k_\mathrm{max}} \binom{2N_{k-1}}{N_k} (1-q)^{N_k} q^{2N_{k-1}-N_k}  (1-q)^{N_0} q^{1-N_0} 
 \label{GW_P}
\end{equation}
This probability can be used to compute expected values of relevant quantities. For example, the average number of active cells at generation $k$ is given by
\begin{align}
\langle N_k \rangle &= \sum_{N_0,N_1,\cdots N_k} N_k \ P(N_0,N_1,\cdots N_k)  = \sum_{N_0,N_1 \cdots N_k} N_k \ P(N_k|N_{k-1}) P(N_{k-1}|N_{k-2}) \cdots P(N_0)  \nonumber \\
& = 2 (1-q) \sum_{N_{0},N_1 \cdots N_{k-1}} N_{k-1} \ P(N_{k-1}|N_{k-2}) \cdots P(N_0)  = \left [ 2 (1-q)\right ]^k (1-q) \ ,
\end{align}
where we exploited the fact that each transition probability is a binomial distribution. The expected global number of cells at generation $k$ is immediately derived as $\langle Z_k \rangle = \langle 2 N_{k-1} \rangle = (2(1-q))^k$. Variances can also be computed following a similar strategy, leading to equation \eqref{eq:variance-GW}.  For this binary case, this is an alternative to the method of the probability generating function used in Ref.~\cite{harris1963theory}.

%%%%%%%%%%%%%%%%%%%%%%%%%%%%%%%%%%%%%%%%%%%%%%%%%%%%%%%%%%%%%%%%%
\section{Details of the NHGW model}

\subsection{Model description and theoretical relations}

The non-homogeneous Galton-Watson model is defined as the Galton-Watson model except that the probability of being inactive depends on the generation \cite{d1992supercritical, kersting2020unifying}.  Cells are still independent, but they are subject to an environmental influence that changes with time, represented through the $k$-dependence of the  inactivity probability, $q_k$.  To compute the average imbalance we can start from~\eqref{eq:imbalance-general}, but we need new expressions for $\langle Z_{k+d,k} \rangle$ and $\langle Z^2_{k+d,k} \rangle$, which will also yield $\langle Z_k\rangle = \langle Z_{k,0}\rangle$.  To compute these we 
introduce the probability that a cell present at generation $k$ has exactly $Z$ descendants at generation $k+d$, $P_{k+d,k}(Z)$.  We have clearly
\begin{align}
  P_{k,k}(Z) & = \delta_{Z,1}, \\
  P_{k+1,k}(Z) & = q_k \delta_{Z,0} +  (1-q_k)  \delta_{Z,2}.
\end{align}
The probabilities for $d=2,3,\ldots$ can be expressed recursively: If there are $Z_{k+d-1,k}$ descendants at generation $k+d-1$, then one can have $Z_{k+d,k}=0,2,\ldots,2Z_{k+d-1,k}$.  For $Z_{k+d,k}$ to be zero, all cells at $k+d-1$ must become inactive, so the probability is $q_{k+d-1}^{Z_{k+d-1,k}} P(Z_{k+d-1,k})$.  Similarly, to have $Z_{k+d,k}=2Z_{k+d-1,k}$ all cells must be active, then the probability is $(1-q_{k+d-1})^{Z_{k+d-1,k}} P(Z_{k+d-1,k})$.  In general to have $Z$ cells one has $q_{k+d-1}^{Z_{k+d-1,K}-Z/2} (1-q_{k+d-1})^{Z/2} P_{k+d-1,k}(Z_{k+d-1,k})$ multiplied by a combinatorial factor accounting for the different ways to pick the cells that reproduce and those that turn inactive.  Summing over all possible values of $Z_{k+d-1,k}$ one gets
\begin{equation}
  P_{k+d,k}(2s) = \sum_{r=\lceil s/2 \rceil}^{2^{d-2}} \binom{2r}{s} P_{k+d-1,k}(2r) (1-q_{k+d-1})^s (q_{k+d-1})^{2r-s} , \qquad d\ge2.
\end{equation}
$P_{k+d,k}(Z)$ is seen to be normalized by induction, because
\begin{equation}
  \sum_{s=0}^{2^{d-1}} P_{k+d,k}(2s) =
  \sum_{r=0}^{2^{d-2}} P_{k+d-1,k}(2r) \sum_{s=0}^{2r} \binom{2r}{s}  (1-q_{k+d-1})^s (q_{k+d-1})^{2r-s} = \sum_{r=0}^{2^{d-2}} P_{k+d-1,k}(2r).
\end{equation}
Recursive expressions for $\langle Z_{k+d,k} \rangle$ and $\langle Z^2_{k+d,k} \rangle$ can be obtained by evaluating the binomial sums:
\begin{align}
\langle Z_{k+d,k} \rangle &= \sum_{s=0}^{2^{d-1}} (2s) P_{k+d,k}(2s) =
                            \sum_{r=0}^{2^{d-2}} P_{k+d-1,k}(2r) \sum_{s=0}^{2r} \binom{2r}{s} (2s) (1-q_{k+d-1})^s (q_{k+d-1})^{2r-s} \\
  &=  2(1-q_{k+d-1}) \sum_{r=0}^{2^{d-2}} P_{k+d-1,k}(2r) (2r) = 2(1-q_{k+d-1}) \langle Z_{k+d-1,k} \rangle, \\
\langle Z^2_{k+d,k} \rangle &= \sum_{s=0}^{2^{d-1}} (2s)^2 P_{k+d,k}(2s) =
                            \sum_{r=0}^{2^{d-2}} P_{k+d-1,k}(2r) \sum_{s=0}^{2r} \binom{2r}{s} (2s)^2 (1-q_{k+d-1})^s (q_{k+d-1})^{2r-s} \\
                          &=  \sum_{r=0}^{2^{d-2}} P_{k+d-1,k}(2r) 4 [2r q_{k+d-1}(1-q_{k+d-1}) + 4r^2 (1-q_{k+d-1})^2] \\
  & = 4q_{k+d-1}(1-q_{k+d-1}) \langle Z_{k+d-1,k}\rangle + 4(1-q_{k+d-1})^2 \langle Z_{k+d-1,k}^2\rangle .
\end{align}
Defining $m_k=2(1-q_k)$ one finally obtains the following recursion relations for the mean and variance:
\begin{align}
  \langle Z_{k+d,k} \rangle &= m_{k+d-1} \langle Z_{k+d-1,k} \rangle, & \langle Z_{k,k}\rangle & =1, \\
  \text{Var}_Z(k+d,k) & = m^2_{k+d-1} \text{Var}_Z(k+d-1,k) + (2-m_{k+d-1})m_{k+d-1} \langle Z_{k+d-1,k}\rangle , & \text{Var}_Z(k,k) & = 0.
\end{align}

With these expressions we can compute the imbalance,
\begin{equation}
\langle I^2(k-1,k)\rangle = 2\text{Var}_Z(\kmax,k)  
\end{equation}

\subsection{Likelihood}
The computation of the likelihood for the NHGW ensemble proceeds along the same lines as for the standard GW discussed previously. The ensemble is now defined by a set of parameters rather than just one, i.e. the $\{q_k\}$ specifying the inactivity probability at each generation. Also in this case, the likelihood of a given tree is expressed in terms of the number of active cells occurring at each generation. Proceeding as for the GW case, we get
\begin{equation}
{\cal L}({\cal T}|q_0,q_1\cdots q_k) = P^{\rm NHGW}({\cal T}) = \prod_{k=1}^{k_\mathrm{max}}  (1-q_k)^{N_k} q_k^{2N_{k-1}-N_k}  (1-q_0)^{N_0} q_0^{1-N_0} \ ,
\label{NHGW_like}
\end{equation}
where the inactivity parameter is now different at each generation $k$. 
The maximum likelihood estimators are obtained by deriving the logarithm of this probability with respect to each inactivity parameter,
\begin{equation}
 \frac{\partial \log(P^{NHGW}) }{\partial q_k} =0 \qquad \forall k,
 \end{equation}
giving
\begin{equation}
q_k^{\rm MLE} = \frac{2N_{k-1}-N_k}{2N_{k-1}} = \frac{N_k^{I}}{N_k+N_k^{I}} = \frac{N_k^{I}}{Z_k}    ,
\end{equation}  
where $N^I_k$ indicates the number of inactive cells at generation $k$.

%%%%%%%%%%%%%%%%%%%%%%%%%%%%%%%%%%%%%%%%%%%%%%%%%%%%%%%%%%%%%%%%%
\section{ Details of the RR ensemble}
%%%%%%%%%%%%%%%%%%%%%%%%%%%%%%%%%%%%%%%%%%%%%%%%%
The random scrambling of a lineage is a key step in the RR inheritance test, hence it has to be performed very carefully. The first important point is that the scrambled tree must have all potentially hereditary mother-daughter links severed, but at the same time it must belong to the same topological class as the original tree, namely it must have the same number of inactive cells {\it per generation} as the biological lineage; if this second requirement is violated we do not obtain a topologically homogeneous ensemble of trees, hence the entropy can drift to any value, which is the problem with the GW and NHGW null ensembles (see Section 4). To achieve these two conditions we proceed as follows (see Fig.1).

%%%%%%%%%%%%%%%%%%%%%%%%%%%%%%%%%%%%%%%%%%%%%%%
\begin{figure*}[h!]
\centering
\includegraphics[width=0.9 \textwidth]{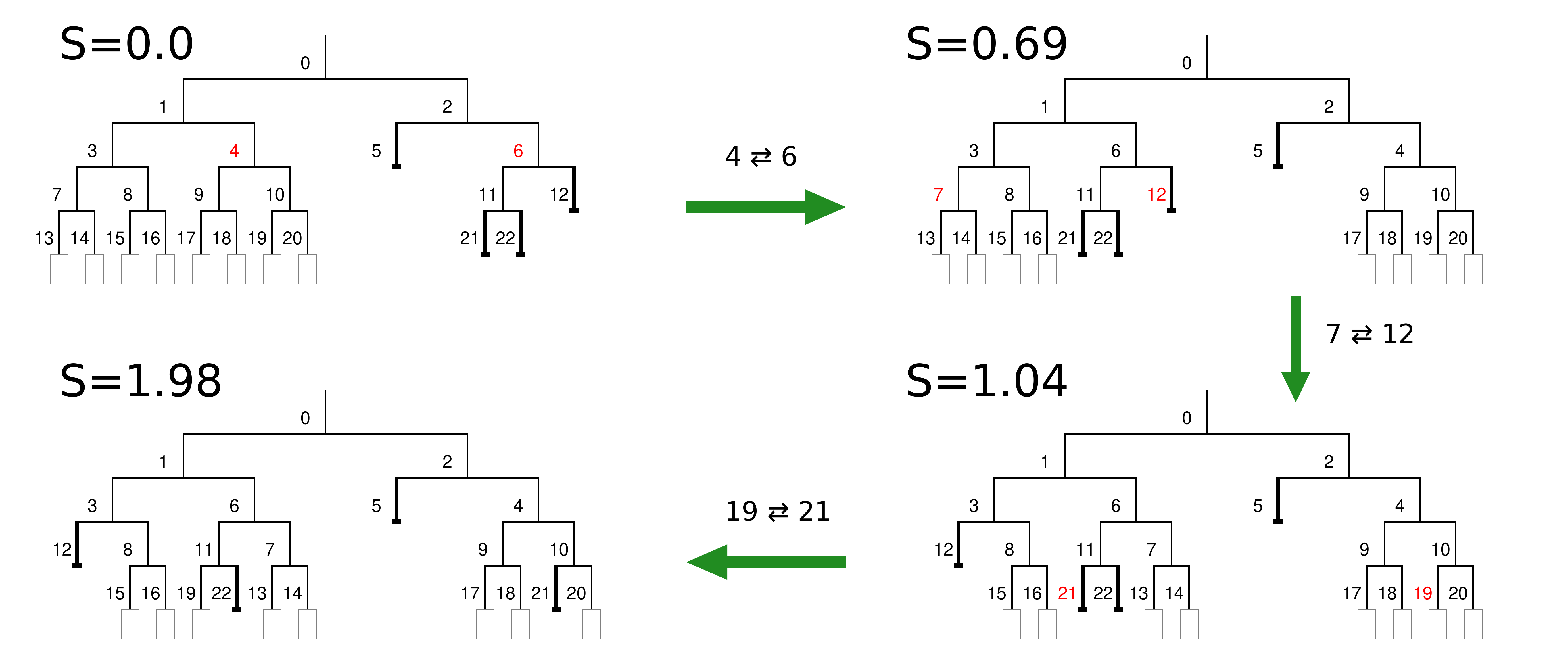}
\caption{Illustration  --- using a fictitious tree --- of the process of lineage scrambling used in the Random Reshuffling (RR) inheritance test. This tree (up-left) is designed to have a clear clustering of inactive cells on the right wing, and therefore a low entropy (in this case the lowest value, $S=0$ --- the reader is invited to check that there is no arrangement of inactive cells that has a more ordered structure than this).  We select randomly two out of the fours cells at generation $k=2$ (those with red label) and we switch them (switching the two cells at generation $k=1$ produces the same lineage); in so doing, we carry with each cell also all its descendants. We then pass to generation $k=3$ and randomly switch two cells and finally do the same at generation $k=4$. With each step of scrambling the entropy grows, and in the end a tree is obtained with a much higher entropy than the original one ($S=1.98$), which is why we can conclude that the inheritance signal of the starting lineage is very significant.}
\end{figure*}
%%%%%%%%%%%%%%%%%%%%%%%%%%%%%%%%%%%%%%%%%%%%%%%

We start the reshuffling at generation $k=2$, since swapping the only two cells of generation $k=1$ leaves the tree topologically unchanged.  Two cells are randomly chosen among the $4$ at $k=2$ (red labels in Fig.1) and these two cells are randomly swapped; when this is done, the descendants of the two cells are also swapped. Then we move to generation $k=3$ and we do the same, i.e. we swap a pair of randomly chosen cells, exchanging also their sub-branches. This procedure is repeated recursively at all generations, each step of the procedure typically increasing the entropy, as any trace of non-random inheritance is deleted. It is crucial to swap only cells at the {\it same} generation, in order to keep the tree within the same topological class as the original biological one.

Even though the procedure that we have just described, and that is illustrated in Fig.1, is the most intuitive one, exchanging just {\it one} pair of cells per generation makes the randomisation of the tree quite slow. To proceed more effectively in severing all potentially heritable connections in the tree, instead of just swapping two cells per generation, we perform a whole random {\it permutation} of all cells at each generation. As in the simple pair swap, also in the case of a permutation, whenever and wherever a cell is moved, it brings along with it all its descendants. The reshuffling procedure of one tree is complete once a random permutation of all cells of each generation is performed.

We notice that a similar randomization procedure, used to investigate the relevance of motifs in cell lineages, has been recently employed in \cite{tran2024lineage}.

%%%%%%%%%%%%%%%%%%%%%%%%%%%%%%%%%%%%%%%%%%%%%%%%%%%%%%%%%%%%%%%%%
%%%%%%%%%%%%%%%%%%%%%%%%%%%%%%%%%%%%%%%%%%%%%%%%%%%%%%%%%%%%%%%%%

%%%%%%%%%%%%%%%%%%%%%%%%%%%%%%%%%%%%%%%%%%%%%%%%%%%%%%%%%%%%%%%%%
\section{ Why GW and NHGW underperform compared to RR?}

As we have seen, the inheritance tests based on the GW and NHGW ensembles perform poorly on hereditary trees, compared to RR. Why is that?
The first thing we must realise is that, for any type of tree, the entropy $S$ grows very significantly -- although not quite linearly --- with the total number of inactive cells of the tree, $N_\mathrm{inact}$ (see Fig.2). This means that the entropy depends not only on the hereditary correlations between inactive cells: its scale depends also on the total number of inactive cells, irrespective of their correlations. Both the GW and the NHGW ensembles (at variance with the RR ensemble) generate trees with fluctuating values of $N_\mathrm{inact}$, in particular often smaller than the number of inactive cells of the hereditary tree to be probed, thus changing radically the original scale of the entropy. This fact biases to the left the probability distribution of the entropy $S$ and therefore the P-value (Fig.2). 

%%%%%%%%%%%%%%%%%%%%%%%%%%%%%%%%%%%%%%%%%%%%%%%
\begin{figure}[h!]
\centering
\includegraphics[width=0.6 \textwidth]{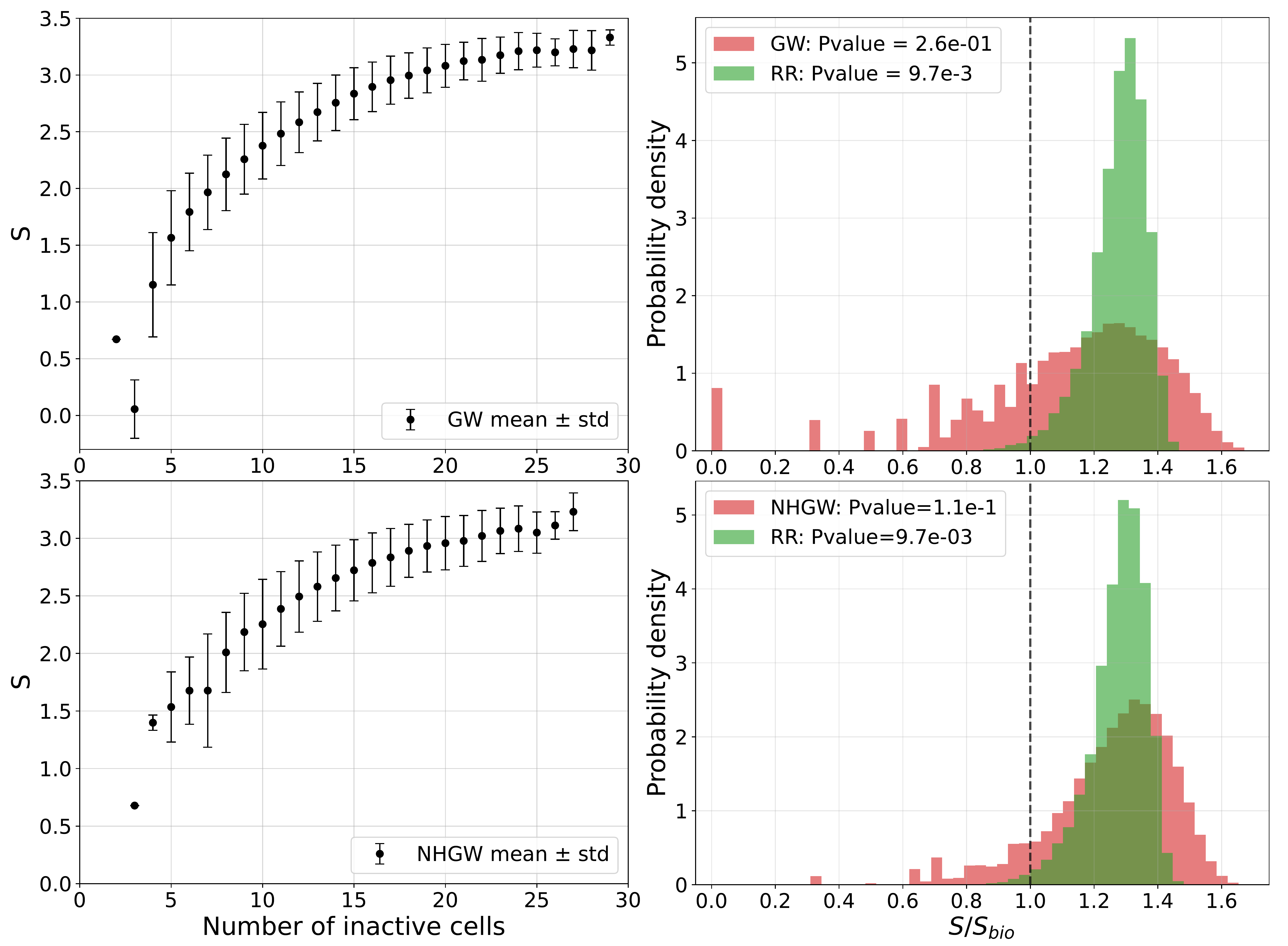}
\caption{
Entropy vs inactivity. Top: GW vs RR ensembles. Left: We take lineage \texttt{20230113-09} in our biological database (14 inactive cells), we fit the GW parameter $q$ to it, and we generate $10^3$ GW trees with that value of $q$.  The entropy of these trees is plotted vs.\ their number of inactive cells (error bars are std deviation): clearly, the GW ensemble explores values of $\Nina$ much smaller than the biological one, and in so doing sampling smaller values of the entropy.  Right: This is reflected in the wider probability distribution of the entropy of the GW ensemble (red) compared to that of the RR ensemble for the same lineage (green), which only explores non-hereditary trees with the same number of inactive cells as the biological lineage.  The respective P-values, $P_\mathrm{GW}$ and $P_\mathrm{RR}$ are -- as a  consequence -- very different: according to GW this is not an hereditary tree, while according to RR it is.  Bottom: same plots for NHGW vs RR. Although the effect is less pronounced, the probability distribution of the entropy in NHGW is still quite wider than that of RR.}
\end{figure}
%%%%%%%%%%%%%%%%%%%%%%%%%%%%%%%%%%%%%%%%%%%%%%%%

To get an idea of what is happening, imagine comparing entropies of two physical systems with different sizes, without normalizing the entropy by the system's size: the results of course would be meaningless.  This example would suggest to normalize $S$ by $N_\mathrm{inact}$ from the outset, but unfortunately this does not work either: the quasi-linear relation between $S$ and $N_\mathrm{inact}$ seems to only holds at low $S$, while over the whole range the relation is definitely not linear. 
Hence, the only way to meaningfully compare entropies of different trees (even with the same overall size $k_\mathrm{max}$) is to work at a fixed value of $N_\mathrm{inact}$ at each generation, which is exactly what the RR method does: within RR the entropy $S$ changes only when there is an actual change in the hereditary structure of the inactive cells. 

Finally, in the case of the GW ensemble, there is something worse going on: as we have already noted, the GW ensemble does not cover all non-hereditary trees, because in general the probability of inactivity may depend on the generation $k$. The fact that NHGW performs better than GW on the numerical set of hereditary trees generated with the SKH model suggests that the closest non-hereditary tree to an hereditary one is a {\it non-homogeneous} tree, hence giving NHGW an advantage over GW.  This makes sense: an hereditary change in the probability of inactivity (which is what SKH does) produces a non-homogeneous empirical vector $q_k$, which is therefore better reproduced by NHGW than GW.

%%%%%%%%%%%%%%%%%%%%%%%%%%%%%%%%%%%%%%%%%%%%%%%%%%%%%%%%%%%%%%%%%

\section{Details of the SKH model}

\subsection{The model}

The Stochastic Kirkwood-Holliday model (SKH) we present here is a two-parameter model which generates an ensemble of binary trees.  The model assumes the existence of three different types of cells:
\begin{itemize}
    \item \textbf{Type W} (wild): each cell of this kind can, at the beginning of its life, mutate into a type M cell with probability $p$ or maintain its original type with probability $1-p$.  At the end of its life a type W cell always divides into two type W daughter cells.
    \item \textbf{Type M} (mutated): each cell of this kind, either at the beginning of its life or immediately after mutation from a wild type, can express the character of inactivity (no progeny) with probability $q$, becoming a type I cell.  With probability $1-q$ the cell does not express the inactivity character, remaining of type M.  At the end of its life a type M cell always divides into two type M daughter cells.
    \item \textbf{Type I} (inactive): each cell of this kind has irreversibly expressed the inactivity character and, therefore, has no progeny.
\end{itemize}
As a result of this formulation the model only requires two parameters: the mutation probability $p$ and the inactivity probability $q$.  If we classify cells with the type they have at the end of their life cycle, each cell can be found in three possible states: W, M, I. The probabilities of the states of the daughter cells, depend on the state of the mother. In particular,
 $P(W\to W)=(1-p)$, $P(W\to M)= p (1-q)$, $P(W\to I)= pq$, $P(M\to M)= (1-q)$, $P(M\to I)= p$, $P(M\to W)= 0$.

We can now compute some useful quantities that help us better describe the ensemble of trees generated by the model: the average number of cells at a given generation $k$, the average number of inactive cells given the maximum generation of a tree $k_\mathrm{max}$, and the average number of missing cells at $k_\mathrm{max}$.

\subsection{Average number of cells at fixed generation}

Let us define $g = 2(1-p)$ and $m = 2(1-q)$, which will be useful in the following.  Since a type W cell always generates two cells in the next generation, each of which will maintain the type W with probability $1-p$, it is easy to see that
\begin{equation}
    \langle N_{k+1}^{W}\rangle = g \langle N_{k}^{W}\rangle
\end{equation}
and since $\langle N_{0}^{W}\rangle = 1-p$, we find
\begin{equation}
    \langle N_{k}^{W}\rangle = (1-p)g^k \Theta(k+1) = \frac{g^{k+1}}{2} \Theta(k+1),
\end{equation}
where the Heaviside theta function enforces $k\geq 0$, under the convention $\Theta(0)=0$.  This is somewhat seem pedantic at the moment, but it will prove to be useful in some of the computations that follow.

The next step is finding the average number of type M cells. We can easily prove the recursive equation
\begin{equation}
    \langle N_{k+1}^{M}\rangle = \langle N_{k}^{W}\rangle pm + \langle N_{k}^{M}\rangle m,
\end{equation}
by observing that a type M cell must originate either from a type M cell in the previous generation or from a type W cell in the previous generation that produced a daughter cell which subsequently mutated into a type M cell. Now, since $\langle N_{0}^{M}\rangle = p(1-q)$, with some algebra we get:

\begin{align}
\langle N_{k+1}^{M}\rangle 
&= \langle N_{k}^{W}\rangle\ pm + \langle N_{k}^{M}\rangle m\\
&= \langle N_{k}^{W}\rangle pm 
 + \left[\langle N_{k-1}^{W}\rangle pm
 + \langle N_{k-1}^{M}\rangle m\right] m \notag\\
&= \langle N_{k}^{W}\rangle pm +...+ \langle N_{0}^{W}\rangle pm^{k+1} + \langle N_{0}^{M}\rangle m^{k+1}\notag\\
&= \sum_{i=0}^{k}  \langle N_{k-i}^{W}\rangle pm^{i+1} + \langle N_{0}^{M}\rangle m^{k+1}\notag\\
&= \sum_{i=0}^{k}  \frac{g^{k-i+1}}{2}\Theta(k-i+1) pm^{i+1} + p \frac{m^{k+2}}{2}\notag
\end{align}

Now we have two possible cases which are: \textbf{1)} $g=0$ and therefore $p=1$, and \textbf{2)} $g\neq 0$ and therefore $p \neq 1$. If case \textbf{1)} applies then we recover the standard result for the average number of active cells in the Galton-Watson model
\begin{equation}
    \langle N_{k+1}^{M}\rangle  = \frac{m^{k+2}}{2}
\end{equation}
but if case \textbf{2)} applies, then we get
\begin{equation}
\langle N_{k+1}^{M}\rangle = pm \frac{g^{k+1}}{2} \sum_{i=0}^{k}  \left(\frac{m}{g}\right)^i + p\frac{m^{k+2}}{2}
\end{equation}
Now we are faced once again with two different cases: \textbf{2a)} $m=g$ and therefore $q=p$, or \textbf{2b)} $m\neq g$ and therefore $q \neq p$. If case \textbf{2a)} applies, then $m/g=1$, and the sum can be computed explicitly:
\begin{align}
\langle N_{k+1}^{M}\rangle 
&= pm \frac{g^{k+1}}{2} (k+1) + p\frac{m^{k+2}}{2}\\
&= p\frac{g^{k+2}}{2}(k+2)\notag
\end{align}
On the other hand if case \textbf{2b)} applies, we can use the known expressions for the geometric series and we obtain:
\begin{align}
\langle N_{k+1}^{M}\rangle 
&= \Theta(k+1) \frac{pmg^{k+1}}{2} \frac{1-\left(\frac{m}{g}\right)^{k+1}}{1-\left(\frac{m}{g}\right)} + p\frac{m^{k+2}}{2}
\end{align}
So the final result is:
\begin{equation}
    \langle N_{k}^{M}\rangle =
    \begin{dcases}
    \frac{m^{k+1}}{2}\Theta(k+1)      & \quad \text{if  } p=1 \\
    p\frac{g^{k+1}}{2}(k+1)\Theta(k+1)      &\quad \text{if } p=q \\
    \frac{pmg^{k}}{2} \frac{1-\left(\frac{m}{g}\right)^{k}}{1-\left(\frac{m}{g}\right)}\Theta(k) + p\frac{m^{k+1}}{2}\Theta(k+1)     &\quad \text{if } p\neq 1, p\neq q
    \end{dcases}
\end{equation}
From now on we shall ignore case \textbf{1)}, as it would be just like computing the standard Galton-Watson quantities all over again.

We can now compute the average number of inactive cells at generation $k$ using the recursive equation
\begin{equation}
    \langle N_{k+1}^{I}\rangle = \langle N_{k}^{W}\rangle 2pq + \langle N_{k}^{M}\rangle 2q,
\end{equation}
which can be readily derived by observing that an inactive cell originate either from a previous-generation type W cell that produces a daughter which mutates and expresses the inactive phenotype in the same generation, or from a previous-generation type M cell whose daughter cell expresses the inactive phenotype.  By substituting the results  we have already found for case \textbf{2b)} we get
\begin{equation}
    \langle N_{k}^{I}\rangle = \frac{g^{k}}{2} \Theta(k) 2pq + \frac{pmg^{k-1}}{2} \frac{1-\left(\frac{m}{g}\right)^{k-1}}{1-\left(\frac{m}{g}\right)}\Theta(k-1)2q
    + p\frac{m^{k}}{2}\Theta(k) 2q .
\end{equation}
We know that we must have $\langle N_{0}^{I}\rangle = pq$, so we insert this initial condition by simply modifying the Heaviside theta function of the first term:
\begin{equation}
    \langle N_{k}^{I}\rangle = g^{k}pq \Theta(k+1) + pqmg^{k-1} \frac{1-\left(\frac{m}{g}\right)^{k-1}}{1-\left(\frac{m}{g}\right)}\Theta(k-1) + pqm^{k}\Theta(k)
\end{equation}
This way, when $k=0$, we will have $\langle N_{0}^{I}\rangle = g^{0}pq \Theta(1) = pq$. By doing the same for case \textbf{2a)} we get the complete result:
\begin{equation}
    \langle N_{k}^{I}\rangle =
    \begin{dcases}
    g^{k}p^2 \Theta(k+1) + p^2g^k k\Theta(k) & \quad \text{if } p=q \\
    g^{k}pq \Theta(k+1) + pqmg^{k-1} \frac{1-\left(\frac{m}{g}\right)^{k-1}}{1-\left(\frac{m}{g}\right)}\Theta(k-1) + pqm^{k}\Theta(k)     & \quad \text{if } p\neq q
    \end{dcases}
\end{equation}
The average number of total cells is obtained by simply adding up the terms we have obtained thus far,
\begin{equation}
    \langle Z_{k}\rangle  = \langle N_{k}^{W}\rangle + \langle N_{k}^{M}\rangle + \langle N_{k}^{I}\rangle,
\end{equation}
and so the final result is
\begin{equation}
    \langle Z_{k}\rangle =
    \begin{dcases}
    \left[\frac{g^{k+1}}{2}  + g^{k}p^2 + p\frac{g^{k+1}}{2}(k+1)\right] \Theta(k+1) + p^2g^k k\Theta(k) & \text{if } p=q\\
    \left[\frac{g^{k+1}}{2}  + \frac{pm^{k+1}}{2}+ g^{k}pq \right]\Theta(k+1)+ \left[\frac{pmg^{k}}{2} \frac{1-\left(\frac{m}{g}\right)^{k}}{1-\left(\frac{m}{g}\right)}+ pqm^{k}\right]\Theta(k) + &\\
    \quad \quad \quad + pqmg^{k-1} \frac{1-\left(\frac{m}{g}\right)^{k-1}}{1-\left(\frac{m}{g}\right)}\Theta(k-1)   & \text{if } p\neq q
    \end{dcases}
\end{equation}

\subsection{Average number of inactive cells}
\label{sec:SKH-total-inactive}
The next step is to compute the average total number of inactive cells, assuming that the maximum generation $k_\mathrm{max}$ is fixed. This is simply
\begin{equation}
    \langle \Nina \rangle = \sum_{k=0}^{k_\mathrm{max} -1} \langle N_{k}^{I}\rangle
\end{equation}
To solve this we need to treat cases \textbf{2a)} and \textbf{2b)} separately, as the expressions we obtain would otherwise get messy and unclear. Let us begin with case \textbf{2a)}, $p=q$.   By substituting the correct expression for $\langle N_{k}^{I}\rangle$ we get:
\begin{align}
\langle \Nina \rangle 
&= \sum_{k=0}^{k_\mathrm{max} -1} g^kp^2\Theta(k+1) + \sum_{k=0}^{k_\mathrm{max} -1} g^kp^2 k\Theta(k)\notag\\
&= p^2 \sum_{k=0}^{k_\mathrm{max} -1} g^k + p^2 \sum_{k=1}^{k_\mathrm{max} -1} k g^k
\end{align}
We need to distinguish between two additional cases, which are:
\begin{enumerate}
    \item $q = p = 1/2$
    \item $q=p \neq 1/2$
\end{enumerate}
If $q = p = 1/2$ (case \textbf{2a.1}) then $g=1$ and the sums have a simple expression:
\begin{align}
    \langle \Nina\rangle 
    &= \left[p^2k_\mathrm{max} + p^2\frac{k_\mathrm{max}(k_\mathrm{max} - 1)}{2}\right]\Theta(k_\mathrm{max})\\
    &=p^2\frac{k_\mathrm{max}(k_\mathrm{max} + 1)}{2} \Theta(k_\mathrm{max})\notag
\end{align}
On the other hand if $q=p \neq 1/2$ (case \textbf{2a.2}) then $g \neq 1$, and it means that we can use the known expressions for the geometric series to obtain the result:
\begin{equation}
    \langle \Nina\rangle = p^2\frac{1-g^{k_\mathrm{max}}}{1-g}\Theta(k_\mathrm{max})+ p^2g\frac{1-k_\mathrm{max}g^{k_\mathrm{max} -1} + (k_\mathrm{max} - 1)g^{k_\mathrm{max}}}{(1-g)^2}\Theta(k_\mathrm{max} -1)
\end{equation}

Now let us treat case \textbf{2b)}, assuming therefore $p\neq q$.
In this case we get:
\begin{align}
\langle \Nina\rangle = \sum_{k=0}^{k_\mathrm{max} -1} g^kpq\Theta(k+1) + \sum_{k=0}^{k_\mathrm{max} -1} pqmg^{k-1} \frac{1-\left(\frac{m}{g}\right)^{k-1}}{1-\left(\frac{m}{g}\right)}\Theta(k-1) + \sum_{k=0}^{k_\mathrm{max} -1} pqm^k\Theta(k)
\end{align}
By moving some quantities around and making the theta functions explicit, we manage to find the shape of a geometric series once more:
\begin{align}
\langle \Nina\rangle 
&= pq\sum_{k=0}^{k_\mathrm{max} -1} g^k + \frac{pqm}{1-\left(\frac{m}{g}\right)}\left[\sum_{k=2}^{k_\mathrm{max} -1}  g^{k-1}-\sum_{k=2}^{k_\mathrm{max}-1} m^{k-1}\right] + pq\sum_{k=1}^{k_\mathrm{max} -1} m^k 
\end{align}
Again we are forced to distinguish between 3 separate cases within the more general case \textbf{2b)}, and these cases are:
\begin{enumerate}
    \item $p,q\neq 1/2$, which yields $g,m \neq 1$ 
    \item $p=1/2$ and $q\neq 1/2$, which yields $g=1$ and $m\neq 1$
    \item $p \neq 1/2$ and $q=1/2$, which yields $g\neq1$ and $m= 1$
\end{enumerate}
We first consider the case that holds for most pairs $p,q \in [0,1)^2$, which is case \textbf{2b.1}. In this case we can simply use the known expressions for the geometric series without any problems:
\begin{align}
\langle \Nina\rangle 
&= pq\frac{1-g^{k_\mathrm{max}}}{1-g}\Theta(k_\mathrm{max}) + \frac{pqm}{1-\left(\frac{m}{g}\right)}\left[\frac{g-g^{k_\mathrm{max}-1}}{1-g} - \frac{m-m^{k_\mathrm{max}-1}}{1-m}\right]\Theta(k_\mathrm{max} -2)  + pq\frac{m-m^{k_\mathrm{max}}}{1-m}\Theta(k_\mathrm{max}-1)
\end{align}

In cases \textbf{2b.2} and \textbf{2b.3} we can use the fact that either $g=1$ or $m=1$ to solve some of the sums. For case \textbf{2b.2}, since $g=1$ and $m\neq 1$, we have:
\begin{align}
\langle \Nina\rangle 
&= pqk_\mathrm{max}\Theta(k_\mathrm{max}) + \frac{pqm}{1-\left(\frac{m}{g}\right)}\left[(k_\mathrm{max} - 2) - \frac{m-m^{k_\mathrm{max}-1}}{1-m}\right]\Theta(k_\mathrm{max} -2) + pq\frac{m-m^{k_\mathrm{max}}}{1-m}\Theta(k_\mathrm{max}-1)
\end{align}
and for case \textbf{2b.3}, where $g\neq 1$ and $m = 1$ we get:
\begin{align}
\langle \Nina\rangle 
&= pq\frac{1-g^{k_\mathrm{max}}}{1-g}\Theta(k_\mathrm{max}) + \frac{pqm}{1-\left(\frac{m}{g}\right)}\left[\frac{g-g^{k_\mathrm{max}-1}}{1-g} - (k_\mathrm{max} - 2)\right]\Theta(k_\mathrm{max} -2) + pq(k_\mathrm{max} - 1)\Theta(k_\mathrm{max}-1)
\end{align}

Putting together the five different expressions for $\langle
\Nina\rangle$ we get a continuous function in the $p,q \in [0,1)^2$ domain.

\subsection{Average number of missing cells}
\label{sec:SKH-missing}
Using the expression for the total number of cells at a fixed generation $k$ we can  compute the number of missing cells at generation $k\leq k_\mathrm{max}$:
\begin{equation}
\langle N_{k}^\mathrm{miss}\rangle =
    \begin{dcases}
    2^{k} - \left[\frac{g^{k+1}}{2}+ g^{k}p^2 + p\frac{g^{k+1}}{2}(k+1)\right]  \Theta(k+1) - p^2g^{k} k\Theta(k)  & \text{if } p=q \\
    2^k - \left[\frac{g^{k+1}}{2}+ \frac{pm^{k+1}}{2} + g^{k}pq \right]\Theta(k+1)-\left[\frac{pmg^{k}}{2} \frac{1-\left(\frac{m}{g}\right)^{k}}{1-\left(\frac{m}{g}\right)} + pqm^{k}\right]\Theta(k) + &\\
    \quad\quad\quad   - pqmg^{k-1} \frac{1-\left(\frac{m}{g}\right)^{k-1}}{1-\left(\frac{m}{g}\right)}\Theta(k-1)   &\text{if } p\neq q
    \end{dcases}
\end{equation}
Since we are particularly interested in the missing cells at the maximum generation we always refer to $\langle N_{k_\mathrm{max}}^\mathrm{miss}\rangle$ as the missing cells of a specific binary tree. So, just like with the average number of total inactive cells, we have managed to find a continuous function in the $p,q \in [0,1)^2$ domain which describes the number of missing cells.

\subsection{Mutation lag distribution}
\label{sec:SKH-mutatation-lag}
We define the mutation lag $l$ as the number of generations that separate the first mutated ancestor from an inactive cell.  In the SKH model it is possible to compute the probability distribution of the mutation lag as follows.  If we define the probability of a type M cell becoming inactive after exactly $l$ generations since the first mutant ancestor, $P(l)$, then the rate $\gamma(l)$ of that same process is the conditional probability for a cell to become inactive after $l$ generations given that none of its ancestors became inactive,
\begin{equation}
    \gamma(l) = \frac{P(l)}{1-\sum_{l^{\prime}=0}^{l-1}P(l^{\prime})},
\end{equation}
But we also know that by the definition of the rate,
\begin{equation}
    1-\sum_{l^{\prime}=0}^{l-1}P(l^{\prime}) = \prod_{l^{\prime}=0}^{l-1}[1-\gamma(l^{\prime})].
\end{equation}
Therefore,
\begin{equation}
    P(l) = \gamma(l)\prod_{l^{\prime}=0}^{l-1}[1-\gamma(l^{\prime})].
\end{equation}
In the specific case of the SKH model we have $\gamma(l) = q$, so
\begin{equation}
    P(l) = q(1-q)^l = q e^{-l/\bar l}
\end{equation}
where the mutation lag scale is
\begin{equation}
  \label{eq:mutation-lag-scale}
  \bar l = \frac{1}{\log\left(\frac{1}{1-q}\right)}  .
\end{equation}
% Now by simply imposing normalization we get the correct shape of the probability density function of the mutation lag $l$:
% \begin{equation}
%     P(l) = \frac{e^{-l/\bar l}}{\bar l}
% \end{equation}
% Using this result we can easily give an estimate of the parameter $q$ by simply measuring the lag scale $\bar l$ experimentally, and then using
% \begin{equation}
%     q = 1 - e^{-1/\bar l}.
% \end{equation}

%%%%%%%%%%%%%%%%%%%%%%%%%%%%%%%%%%

\subsection{Calculation of $\pis$}

Even an infinite SKH tree ($k_\mathrm{max}=\infty$) has a finite probability to be statistically identical to a GW tree and therefore to give a negative inheritance signature. This implies that the asymptotic value of the fraction of infinite SKH trees that pass the test, $\pis$, will {\it not} be equal to 1 (Fig.~4c of the main text).

This can happen for example if the root ($k=0) $ of the tree mutates: then the rest of the tree is identical to a GW tree with inactivity probability $q$; this happens with probability $p$.  But this is not the only case: if the two cells at generation $k=1$ both mutate, the tree is GW from $k=1$ onward, an event that occurs with probability $p^2(1-p)$; and so on. To calculate the probability that SKH generates a GW tree we must sum all these contributions.

Given that in the benchmarking test we are only using trees that arrive to the last generation, $k_\mathrm{max}$, let us be more precise and define $\pino(p,q)$ as the probability that the SKH model with parameters $(p,q)$ generate a GW tree that arrives at the last generation, for $k_\mathrm{max}\to\infty$.  Also, define $\wino(r)$ as the probability that the GW model with inactivity probability $r$ generates a tree that arrives at the last generation, for $k_\mathrm{max}\to\infty$. We can then write,
\beq
\pino(p,q)=[p+p^2(1-p)+p^4(1-p)^3\dots] \ \wino(q) = \wino(q) \sum_{k=0}^\infty p^{2^k}(1-p)^{2^k-1} \ .
\label{zonda}
\eeq
There is however a problem with this formula in one limiting case.  We know that the SKH model is identical to GW for $p=1$ and for $q=1$.  The first case is correctly handled by \eqref{zonda}: $\pino(1,q)=\wino(q)$, as expected. On the other hand, for $q=1$ equation \eqref{zonda} gives $\pino(p,1)=0$, because the GW survival probability for $q=1$ is zero, $\wino(1)=0$.  But this is wrong, as for $q=1$ we should obtain $\pino(p,1)=\wino(p)$.  The glitch is the following: for any $q\neq 1$, no matter how close to $1$, the SKH ensemble is different from the GW ensemble; it is only exactly at $q=1$ that SKH is equivalent to GW. In other words, for $k_\mathrm{max}=\infty$, we have that
\beq
\pino(p,1)\neq \lim_{q\to 1}\pino(p,q) \ .
\eeq 
On the other hand, the limit $p\to 1$ is smooth, thus showing that the two different ways to obtain GW from SKH ($p=1$ and $q=1$) are conceptually very different from each other. In the light of this discussion, we can correct equation \eqref{zonda},
\beq
\pino(p,q) = \wino(q) \sum_{k=0}^\infty p^{2^k}(1-p)^{2^k-1} + \wino(p) \delta_{q,1}    \ , 
\label{zonda2}
\eeq
where we hope the reader will forgive our abuse of the Kronecker delta.  In all our cases $q\neq 1$, hence the extra term does not make any numerical difference in the following.

We can now write an expression for the asymptotic fraction of trees that pass the inheritance test in SKH. We have,
\beq
\pis = \pig \ \pino(p,q) + \Phi_\infty^\mathrm{HRD}(1-\pino(p,q)) \ .
\label{zondissima}
\eeq
In this expression $\pig$ is the asymptotic fraction of GW trees that pass the test, which we know --- and verified in Fig.~4a of the main text --- to be $\pig=0.05$.  On the other hand, $\Phi_\infty^\mathrm{HRD}$ is the fraction of {\it bona fide} hereditary trees that pass the test in the infinite-size limit, hence $\Phi_\infty^\mathrm{HRD}=1$. We can now assemble all pieces in~\eqref{zonda2} and~\eqref{zondissima} and evaluate them at $p=0.1$ and $q=0.46$, the parameters used in Fig.~4c of the main text to benchmark the inheritance tests on the hereditary set.  The value of $\wino(q)$ is obtained from the asymptotic extinction probability, $p_E(q)$ of GW (equation \eqref{eq:extinction-prob}).  We obtain,
\begin{equation*}
  \wino(q) = 1-p_E(q) \approx 0.148 , \qquad
  \sum_{k=0}^\infty p^{2^k}(1-p)^{2^k-1} \approx  0.109, \qquad
  \pino \approx 0.016 ,
\end{equation*}
finally giving
\beq
\pis \approx 0.985  \ ,
\label{millivanilli}
\eeq
which corresponds to the value reported in Fig.4c of the main text as ``exact asymptotic value''.

\subsection{Fitting the SKH model to the biological lineages}

In this Section we describe three different methods to fit the SKH model to the experimental BMSC lineages. The relative values of the fitted parameters $(p,q)$ are reported in SM-Table 1.

\subsubsection{Maximum Likelihood Estimate --- MLE} \label{sec:like_SKH}

Let us now evaluate the likelihood that a given observed tree $\cal T$, as specified by its topology and the active/inactive phenotype of its cells, is generated by a SKH ensemble with mutation probability $p$ and inactivity probability $q$. In the SKH ensemble each cell can be wild (W), mutated active (M) or mutated inactive (I). However, in experiments we only have access to the active/inactive condition.  The state of a cell is therefore a hidden variable, and when computing the likelihood we need to sum over all the realizations of such variables, consistent with the observed phenotype.  If we call $X^k_i$  the state of cell $i$ at generation $k$, and $X=\{X^k_i\}$ the set of all cell states, we then have
\begin{equation}
{\cal L}({\cal T}|q,p) = \sum_{X} P({\cal T} | X) P(X)  \ .
\label{SKH_like}
\end{equation}

The SKH process is Markovian only in terms of the state variables, and correlations are present between inactive cells. As a consequence, contrary to the GW case, the above likelihood cannot be expressed merely in terms of the total number of active and inactive cells. The likelihood can be computed following an iterative procedure, similarly to what done in phylogenetic inference approaches  \cite{felsenstein1981evolutionary,neher2014predicting,aragon2026learning,dieselhorst2026phylodynamics}.
Luckily, the structure of the model allows to compute the recursive equations explicitly, in a relatively simple way.
To see this, it is convenient to introduce some additional notation. Let us call ${\cal T}^{k}_i$ the observed subtree originating at cell $i$ in generation $k$. Starting from the root,  ${\cal T}^0_0$ therefore identifies the whole tree $\cal T$ and, calling $X_0$ the root state variable, we have
 \begin{align}
&\quad {\cal L}({\cal T}|q,p) = P({\cal T}^0_0)  = \sum_{X_0}  P({\cal T}^0_0| X_0) P(X_0) \nonumber \\
&= \left [ P({\cal T}^0_0| W_0) P(W_0) + P({\cal T}^0_0| M_0) P(M_0) \right ] (1-\delta_{{\cal T}^0_0;I}) + \delta_{{\cal T}^0_0;I} P(I) \nonumber\\
\nonumber \\
& = \left [ P({\cal T}^0_0| W_0) P(W_0) + P^{\rm GW}_{q}({\cal T}^0_0) P(M_0) \right ] (1-\delta_{{\cal T}^0_0;I}) + \delta_{{\cal T}^0_0;I} P(I)
\label{SKH_like1}
\end{align}
where we have also considered the case where the observed tree is made by only one inactive cell. This possibility does not occur in our data, but we included it for the sake of generality. In  Eq.~\eqref{SKH_like1}, the state probabilities for the root are known, 
\beqa
P(W)&&=(1-p) \nonumber\\ 
P(M)&&=p(1-q)\nonumber \\
P(I)&&=pq \nonumber \ .
\eeqa
Besides, in the last line, we exploited the fact that, if the root is mutated, the whole tree belongs to a GW ensemble with parameter $q$, and we indicated with $P^{\rm GW}_{q}({\cal T}^0_0)$ the GW probability of the tree (conditioned to an active root). Also for this probability we know the explicit expression (see Section 1, Eq.(\ref{GW_cond})). Therefore, the only quantity in Eq.~\eqref{SKH_like1} that remains to be determined is $P({\cal T}^0_0| W_0)$, which  we wish to derive in a recursive way. 

A useful observation is that, given the root, the tree departing from it is built by the two subtrees ${\cal T}^1_1$ and ${\cal T}^1_2$ originating at the two daughters. We can therefore write
\begin{align}
& P({\cal T}^0_0| W_0) = P({\cal T}^1_1| W_0) P({\cal T}^1_2| W_0) = \nonumber \\
&= \left \{ \left [ P({\cal T}^1_1| W^1_1) P(W|W) + P^{\rm GW}_{q}({\cal T}^1_1) P(M|W) \right ] (1-\delta_{{\cal T}^1_1;I}) + \delta_{{\cal T}^1_1;I} P(I)\right \} \nonumber\\
& \times \left \{ \left [ P({\cal T}^1_2| W^1_2) P(W|W) + P^{\rm GW}_{q}({\cal T}^1_2) P(M|W) \right ] (1-\delta_{{\cal T}^1_2;I}) + \delta_{{\cal T}^1_2;I} P(I) \right \}  \ ,
\end{align}
where, once again, we exploited the fact that if the daughter is mutated, the progeny obeys a GW process with probability $q$. In this expression we have indicated with $P(X_j^{k+1}|X_i^k)$ the transition probability that a mother with a given state at generation $k$ produces a daughter with another state at generation $k+1$. These probabilities do not depend on generation, and are easily expressed in terms of $p$ and $q$,
\beqa
P(W|W)&&=(1-p) \nonumber  \\
P(M|W)&&=p(1-q)\nonumber \\
P(I|W)&&=qp \nonumber\\
 P(W|M)&&=0 \nonumber \\ 
 P(M|M)&&=(1-q)\nonumber \\
  P(I|M)&&=q \nonumber \ .
\eeqa
This relation can be easily iterated, giving 
\begin{equation}
P({\cal T}^k_i| W^k_i) 
= \prod_{j\in D(i)}\left \{ \left [ P({\cal T}^{k+1}_j| W^{k+1}_j) P(W|W) + P^{\rm GW}_{q}({\cal T}^{k+1}_j) P(M|W) \right ] (1-\delta_{{\cal T}^{k+1}_j;I}) + \delta_{{\cal T}^{k+1}_j;I} P(I)\right \} \nonumber\\
\end{equation}
where $D(i)$ indicates the set of the (two) daughters of $i$.
What remains to be done at this point, is to write the boundary conditions at the bottom of the tree. If the last observed cell generation is $\kmax$, this means that we can only assess the activity/inactivity phenotype for cells of $k=\kmax-1$, this is therefore the effective depth of the phenotypic lineage under scrutiny. The sub-tree originating in any one of these cells is made of one cell only. If the cell is active, the probability to observe such tree is the identity, therefore we have $P({\cal T}^{k_{max}-1}_i| W^{k_{max}-1}_i)=1$. Once this last condition is fixed, these equations can be solved starting from the bottom up to the root, building step by step the conditional probabilities, and finally computing the likelihood of the complete tree. The maximum likelihood estimators for $p$ and $q$ can be then obtained by numerically maximizing the log-likelihood.

Before applying MLE to fit the SKH model to the experimental data, we test MLE on a set of lineages generated through the SKH model itself. We generate $10^3$ SKH trees with $p=0.1$, $q=0.46$ and we average the log-likelihood over these trees; we then maximize the mean log-likelihood and find
$p_\mathrm{MLE}=0.10$, $q_\mathrm{MLE}=0.45$. We can therefore be confident that the likelihood calculation in SKH is correct and therefore proceed to use MLE to fit the SKH model to the experimental data.

\begin{figure}[h!]
\centering
\includegraphics[width=0.8 \textwidth]{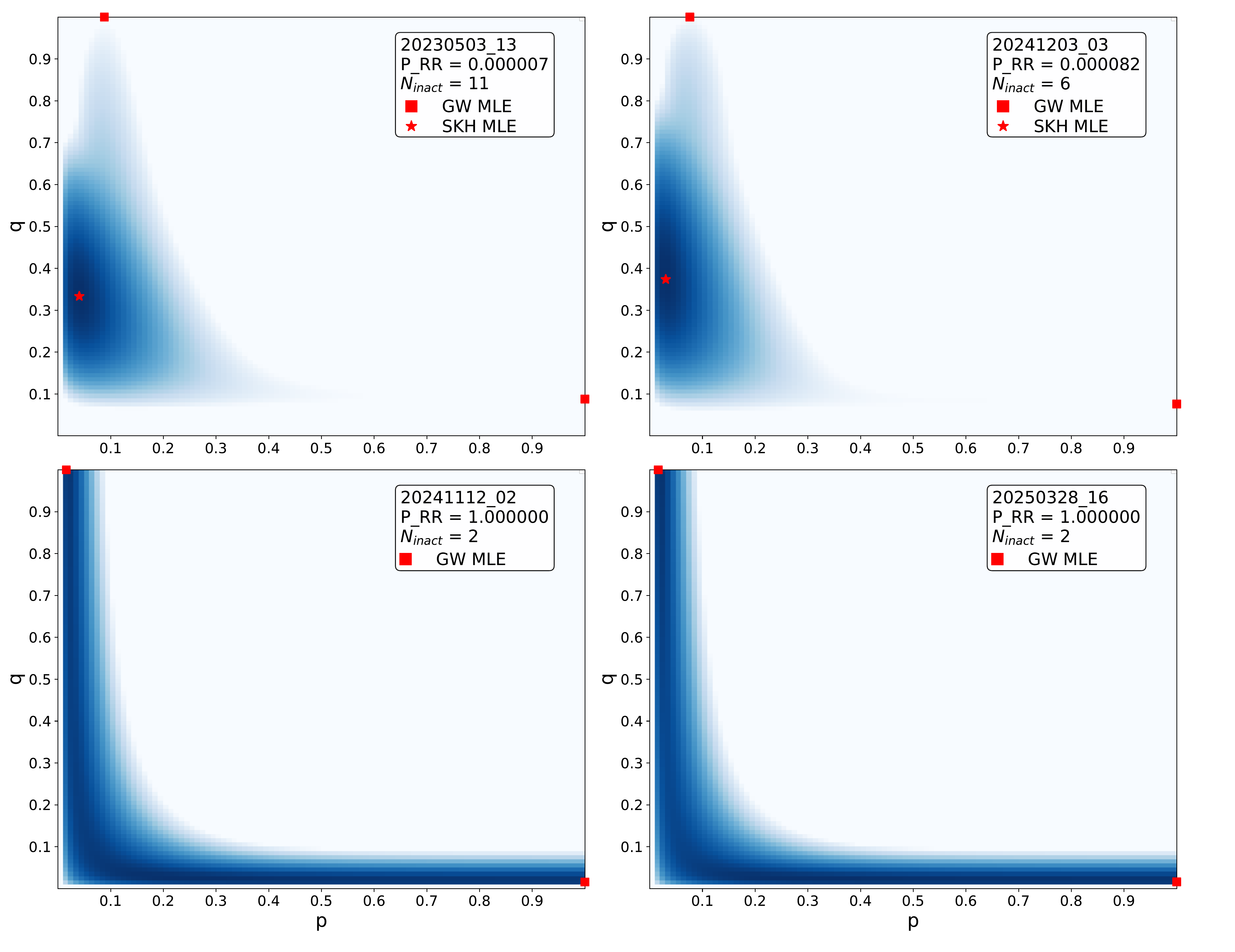}
\caption{Heat-maps representing the SKH likelihoods in four experimental BMSC colonies. Each panel reports, together with the map, the identity of the colony, the P-value from the entropy inheritance test, and the number of inactive cells. The red star represents the location of the maximum (i.e. the MLE for $q$ and $p$), which is well-defined only in lineages with a large enough number of inactive cells (top). When inactive cells are few (bottom) the likelihood is ver flat, making it hard to accurately identify the maximum. The red squares report the values of the MLE performed with the GW model.}
\label{like_maps}
\end{figure}

We report in Fig.\ref{like_maps} the likelihood landscape in the $(p,q)$ plane of SKH, for four experimental BMSC colonies. We see that in the two top colonies, which have large numbers of inactive cells and therefore strongly significant inheritance tests, the likelihood has a well-defined maximum. On the other hand, the two bottom colonies, which have few inactive cells and therefore a non-significant inheritance test, produce a very flat profile of the SKH likelihood, such that it becomes impossible to accurately locate the position of the maximum. This is the reason why we perform the MLE fit to the SKH model only in BMSC colonies for which the inheritance test gives a positive result.

%%%%%%%%%%%%%%%%%%%%%%%%%%%%%%%%%%%%%%%%%%%%%%%%%
%%%%%%%%%%%%%%%%%%%%%%%%%%%%%%%%%%%%%%%%%%%%%%%

We stress that there is in general no direct correlation between the outcome of the inheritance test and the flatness of the likelihood. The true correlation is between the total number of inactive cells in the colony, $\Nina$, and the outcome of both methods, i.e. the entropy-based inheritance test and the MLE fit to the SKH model: when $\Nina$ is too small both methods are blunt essentially because they do not have enough inactive cells to work with. The flatness of the likelihood for multi-parameter inference due to finite size effects, and the role of the number of informative events (in our case inactive cells) on MLE has been recently observed in a different context in \cite{aragon2026learning}.
On the other hand, when $\Nina$ is large, the entropy-based test has many inactive cells to reshuffle, hence it becomes sensitive to the actual structure of the tree, and it can either give $P<0.05$ or $P>0.05$, depending on the hereditary correlations in the tree.
For what concerns the likelihood, its maximum becomes well-shaped when $\Nina$ increases, but it could fall in either region of SKH, hereditary or not. 
In our BMSC dataset -- which is strongly hereditary -- it turns out that all trees that  pass the entropy inheritance test have a large number of inactive cells and therefore have a well-defined maximum of the likelihood (and thus of MLE parameters). But there might be non-hereditary cases in which the number of inactive cells is large, so that the maximum of the likelihood is well-defined, but the inheritance test is negative.

\subsubsection{MLE restricted to the mutated branch}
Within the SKH model, each mutation generates a sub-tree which is statistically identical to a GW tree with inactivity probability $q$; hence, within that subtree, one can use MLE applied to GW to fit this parameter.  As shown in Section 1, this amounts to compute $q$ as the ratio between the number of inactive cells and the total number of cells in the mutated sub-tree, namely,
\begin{equation}
  q= \left. \frac{\Nina}{\Nina+\Nacta}
  \right\rvert_\mathrm{mutated \ subtree}    .
  \label{zuppi}
\end{equation}
In the experimental lineage trees that pass the inheritance test, we are able to identify the mutation, and the relative mutated sub-tree, through the inactivity imbalance (see Sec.~V,B--C of main text). Therefore, we can use \eqref{zuppi} to fit the value of $q_\tau$ for each hereditary biological tree $\cal T$. Finally, to fit $p$ at fixed $q_\tau$, we can match the total number of inactive and missing cells of the biological tree $\cal T$ to the analytic expectation values of these quantities for SKH, which we computed in Sections 5.3 and 5.4, by minimizing the cost function,
\begin{equation}
  H(p) = 
  \Delta N_\mathrm{inact}(p,q_\tau)^2 + \Delta N_\mathrm{miss}(p,q_\tau)^2 \ ,
  \label{goulda}
\end{equation}
where $\Delta N_\mathrm{inact}(p,q_\tau)=N^\mathrm{SKH}_\mathrm{inact}(p,q_\tau) - N^\tau_\mathrm{inact}$ and similarly for the missing cells. The minimum of $H(p)$ is well-defined for each biological colony $\tau$, hence giving the fitted value $p_\tau$.

\subsubsection{Fitting the mutation lag}

A third way of doing the fit is to note that the mutation lag of the SKH model depends only on $q$ (Sec.5.5).  Hence one can determine this parameter for each biological tree $\cal T$ by simply matching the tree's mutation lag to that of the model.  Inverting relation~\eqref{eq:mutation-lag-scale},
we can determine the value of $q_\tau$ for a biological tree $\cal T$ characterized by a mutation lag ${l}_\tau$,
\begin{equation}
    q_\tau = 1 - e^{-1/{l}_\tau} \ .
\end{equation}
In this way we obtain the parameter $q_\tau$ for each biological tree $\cal T$ for which we have a significant inheritance test. For the determination of $p_\tau$, at fixed value of $q_\tau$, we can proceed as in the previous case by minimizing the cost function $H$ in~\eqref{goulda}.

\section{Likelihood-based inheritance test} \label{sec:like_ratio}

\begin{figure}[h!]
\centering
\includegraphics[width=0.6 \textwidth]{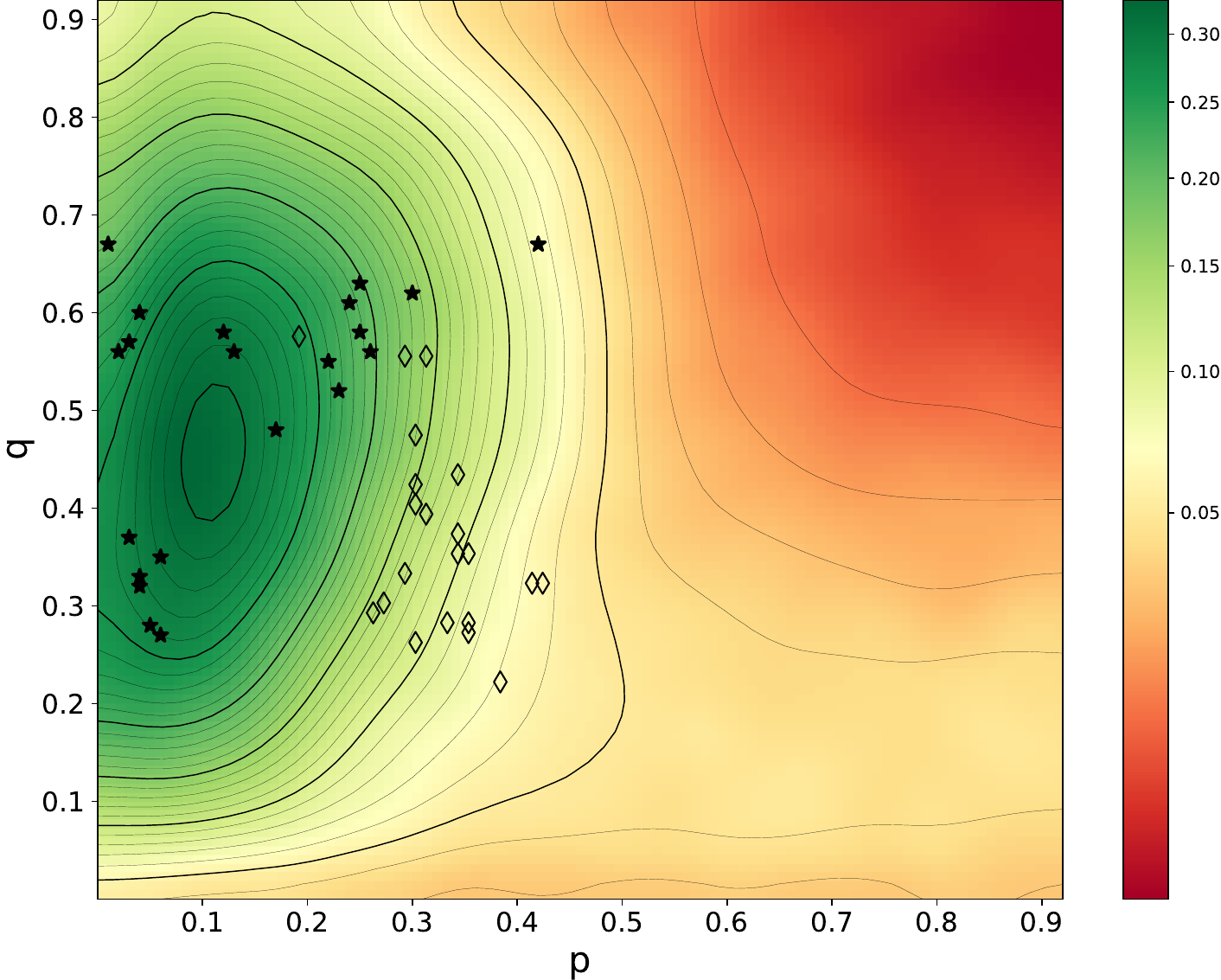}
\caption{MLE parameters of the SKH model fitted to $21$ biological trees (stars) and to $21$ synthetic non-hereditary trees (diamonds), both reported on the   hereditary SKH map. Biological trees are from colonies that pass the hereditary test. Non-hereditary trees are generated from the NHGW ensemble with $q_k = 0.001 \exp(k)$ (see text).}
\label{fig:contro}
\end{figure}

In the main text we reported that, for the $21$ BMSC colonies that pass the inheritance test, the MLE parameters obtained from the SKH likelihood all lie in the hereditary sector of the model.  This is illustrated explicitly in Fig.~\ref{fig:contro}, where the MLE values for the individual biological lineages are represented by black stars. This finding might suggest that MLE provide an independent way to assess the hereditary nature of a lineage: it would suffice to compute the SKH likelihood for the trial tree, and if the maximum is in the hereditary region (far from the GW edges) this means that the tree is hereditary (i.e. SKH is a better model than GW). This is, from the point of view of model comparison, a very naive procedure that can lead to wrong conclusions. 

\begin{figure}[h!]
\centering
\includegraphics[width=0.8 \textwidth]{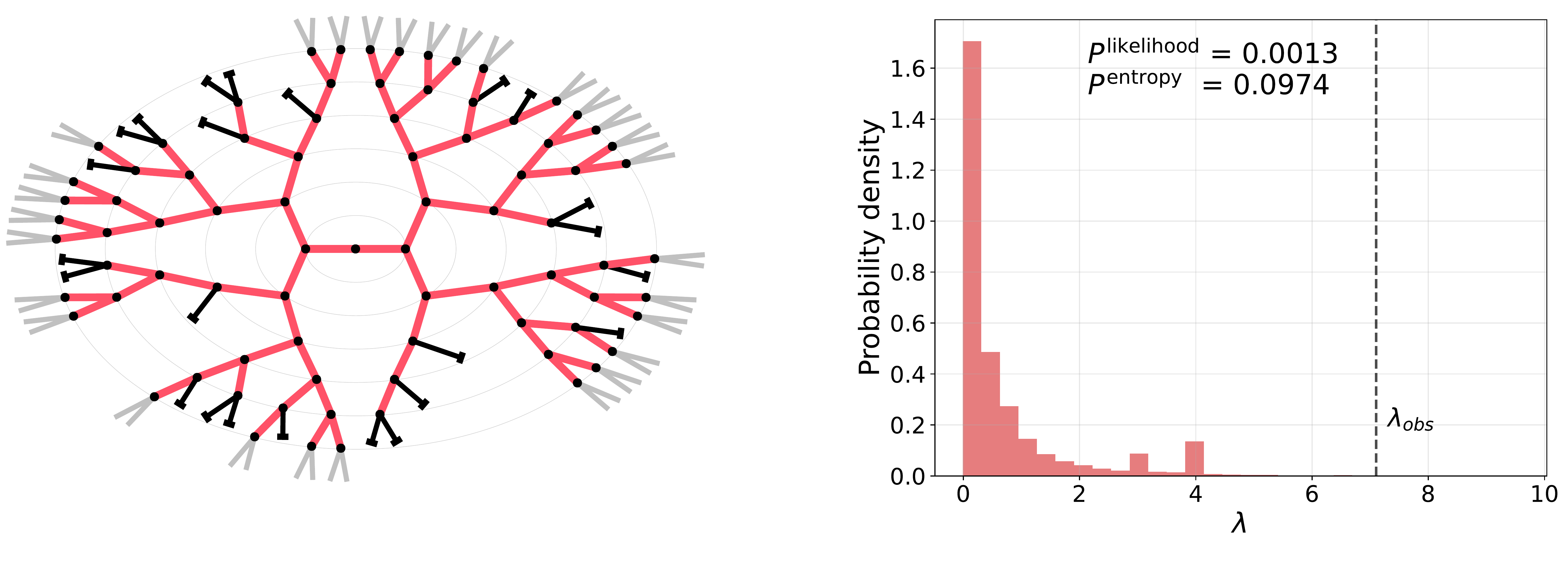}
\caption{Likelihood-based inheritance test for a NHGW non-hereditary tree. Left:  Typical lineage drawn with the NHGW model, with mutation probability as given in the text: because of the non-homogeneous nature of the tree, inactive cells only emerge in the last three generations. Right:  Distribution of the log-likelihood ratio, $\lambda$, computed over ${\cal R}=10^4$ GW samples. The whole distribution lies well below the observed log-likelihood ratio, wrongly indicating that the tree is hereditary.}
\label{fig:contro1}
\end{figure}

To illustrate this point, we consider a simple counter-example. We use the NHGW model parametrized by the inactivity probability vector $q_k = q_0\exp(k/k_0)$, with $q_0=0.001$ and  and $k_0=1$, to generate a set of $21$ non-hereditary synthetic trees with the same size $k_{\rm max}=7$ as the biological ones. 
The entropy-based inheritance test confirms that these trees are non-hereditary: the fraction of these NHGW lineages passing the RR test is $0.03$ (calculated over 100 trees). 
These trees have many inactive cells (see Fig.~\ref{fig:contro1}, left), hence the SKH likelihoods exhibit well-shaped (non-flat) maxima, leading to well-defined MLE for $q$ and $p$. These values are reported as open diamonds in Fig.~\ref{fig:contro}. We see that they all lie in proximity of the hereditary region of the model. Had we used the positions of these fits within the SKH hereditary map to assess their inheritance properties, we would have erroneously concluded that these trees are hereditary, while in fact they are not.

To understand the reason behind this finding, we remind that MLE is based on a fit of the SKH model to the data. If the trees are strongly hereditary, as it happens for the biological ones, the model is nicely suited to the structure of the data, and the fit will tune the parameters to reproduce such structure in the best possible way. In this respect, the fact that biological MLE lie in the hereditary region is not accidental, and it represents a consistency check of our analysis. On the contrary, the NHGW trees are more difficult to account for within the SKH ensemble. Due to the trend in the $q_k$, they tend to have many inactive cells at the last generations (see Fig.~\ref{fig:contro1}, left). To reproduce such structure, the best MLE-SKH can do is to choose the parameters in the hereditary region, since in the homogeneous non-hereditary sector the model can only generate many inactive cells by placing them at early generations, producing small trees very different from the real probed one.

%%%%%%%%%%%%%%%%%%%%%%%%%
%%%%%%%%%%%%%%%%%%%%%%%%%%%%%%%%%%%%%%%%%%%%%%%

However, merely looking at MLE is not the proper way to try to assess inheritance using the likelihood. Rephrasing the problem in correct terms, what one wants to do is to compare two models, the GW (non-hereditary) model, and the SKH hereditary one, and their ability to reproduce the data. GW - the null model - is nested  within the SKH ensemble, as a special case with one parameter instead of two. As a consequence, the MLE obtained with SKH will trivially correspond to a higher likelihood than the one obtained with GW. The crucial point is to assess whether this difference in likelihoods is significative (in which case the `alternative' SKH model truly is a better model for the data), or it is consistent with statistical fluctuations of the null ensemble. Likelihood ratio tests \cite{wilks1938large} precisely carry out this task, providing a P-value for the hereditary hypothesis, which is completely independent from imbalance, entropy and all that.

In our case, since we deal with finite samples and the null model is on the boundary of the alternative SKH one, it is convenient to use a parametric bootstrap likelihood-ratio test \cite{tibshirani1993introduction,self1987asymptotic,davison1997bootstrap}. Let us call $\cal T$ the tree whose hereditary structure needs to be tested. The first thing to do is to calculate the likelihood of $\cal T$ using GW and SKH, ${\cal L}_\mathrm{GW}^{(\tau)}(q)$ and ${\cal L}_\mathrm{SKH}^{(\tau)}(p,q)$; then, we need to maximize them and find the positions of the maxima, $q^\mathrm{GW}_\tau$ and $(p^\mathrm{SKH}_\tau, q^\mathrm{SKH}_\tau)$. From this, we define the log-likelihood ratio,
\beq
\lambda_\tau = 2\left[ \log {\cal L}_\mathrm{SKH}^{(\tau)}(p^\mathrm{SKH}_\tau,q^\mathrm{SKH}_\tau) -  \log {\cal L}_\mathrm{GW}^{(\tau)}(q^\mathrm{GW}_\tau) \right] \ .
\eeq
Next, we generate a large number $\cal R$ of trees with the GW model at $q=q^\mathrm{GW}_\tau$. For each one of these trees, $\sigma$, we calculate and maximize both likelihoods, and compute the log-likelihood ratio,
\beq
\lambda_\sigma = 2\left[ \log {\cal L}_\mathrm{SKH}^{(\sigma)}(p^\mathrm{SKH}_\sigma,q^\mathrm{SKH}_\sigma) -  \log {\cal L}_\mathrm{GW}^{(\sigma)}(q^\mathrm{GW}_\sigma) \right] \ .
\eeq
Then, the P-value is simply the fraction of these GW trees that have log-likelihood ratio $\lambda_\sigma $ larger than or equal to the observed one, $\lambda_{\rm obs}=\lambda_\tau$,
\beq
P^\mathrm{likelihood}_\mathrm{GW}= \frac{1}{\cal R} \sum_\sigma \Theta(\lambda_\sigma - \lambda_\tau)  \ ,
\eeq
where the subscript GW indicates that we are using it as a null model. 

To test the performance and the predictive power of this likelihood-based inheritance test, we apply it to the tree in Fig.\ref{fig:contro1}: this lineage has been generated by the NHGW model discussed at the beginning of this Section and problematic in terms of MLE location on the SKH inheritance map. 
The entropy-based inheritance tests for this tree give,
\beq 
P^\mathrm{entropy}_\mathrm{RR}= 0.097 \quad , \quad P^\mathrm{entropy}_\mathrm{GW}=0.97 \quad , \quad P^\mathrm{entropy}_\mathrm{NHGW}=0.08 \ ,
\eeq
Hence, {\it all} entropy-based tests correctly indicate that this tree is {\it not} hereditary.  Does the rigorous likelihood-based inheritance test we described in this Section  succeed in correctly identifying the non-hereditary nature of this tree? The answer is no; in fact, 
\beq 
P^\mathrm{likelihood}_\mathrm{GW} = 0.0013 \ ,
\eeq
wrongly suggesting that the tree is hereditary (see Fig.\ref{fig:contro1}, right). This is not a rare accident: on this kind of NHGW non-hereditary ensemble the fraction of trees that pass the likelihood-based inheritance test is $0.55$, while we recall that the fraction of trees passing the entropy-based test is $0.03$. As an aside, we note that the run time to calculate $P^\mathrm{likelihood}_\mathrm{GW}$ is approximately a factor $10^4$ larger than the run time to calculate $P^\mathrm{entropy}_\mathrm{RR}$.

%%%%%%%%%%%%%%%%%%%%%%%%%
%%%%%%%%%%%%%%%%%%%%%%%%%%%%%%%%%%%%%%%%%%%%%%%

Why does the likelihood-based inheritance test fail on this type of non-hereditary trees? The reason is connected with what we discussed when commenting MLE results at the beginning of this Section. The likelihood-based test needs  two models to run: a null one (in our case the non-hereditary GW) and an alternative model (in our case the hereditary SKH), whose choice is rather arbitrary. What may happen is that either one of these two models, or both, may be quite inadequate to capture the actual hereditary structure of the data. This is what happens here: the tree we are probing has a structure which is very unusual for the GW ensemble, due to its non-homogeneous nature. Hence, an SKH fit is far more likely than a GW fit, not because of any considerations of inheritance, but simply because that particular structure-through-generations of the inactive cells is much more likely within the SKH ensemble than the GW ensemble, irrespective of their hereditary correlations. Of course, one could object that it might be possible to conjure better models to run the likelihood-based inheritance test, but that is hardly the point: when we are confronted with a generic lineage of unknown hereditary origin, we really need a general model-agnostic method, rather than model-guessing.

%%%%%%%%%%%%%%%%%%%%%%%%%%%%%%%%%%%%%%%%%%%%%%%%%%%%%%%%%%%%%%%%%
\section{Topological distance vs physical distance}
% physical space vs tree space
Is it possible that the clustering of G$_0$ cells on the same branches of the lineage tree, which we interpret as an hereditary character, is in fact an epiphenomenon of spatial correlation in the physical space of the dish? Namely, is it possible that chemicals/nutrient heterogeneities, local crowding, or other spatial factors might influence the propensity to enter the G$_0$ phase in a way that is then reflected in the lineage topology? This is an old hypothesis, dating back to the sixties \cite{froese1964distribution}, shortly after the first evidence of correlations between division times of sister cells in bacterial colonies emerged \cite{powell1955some}, correlation that suggested to some authors that generation times could be affected by inherited factors \cite{kubitschek1962normal}. 

Our data, however, show that our results are {\it not} due to spatial proximity, as we are going to explain below. In what follows we denote as ``topological distance'' the distance on the tree, namely the number of links separating two different cells in the lineage; on the other hand, the ``physical distance'' is the actual distance between the two cells on the dish.

%%%%%%%%%%%%%%%%%%%%%%%%%%%%%%%%%%%%%%%%%%%%%%%%%
\begin{figure}[h!]
\centering
\includegraphics[width=0.35 \textwidth]{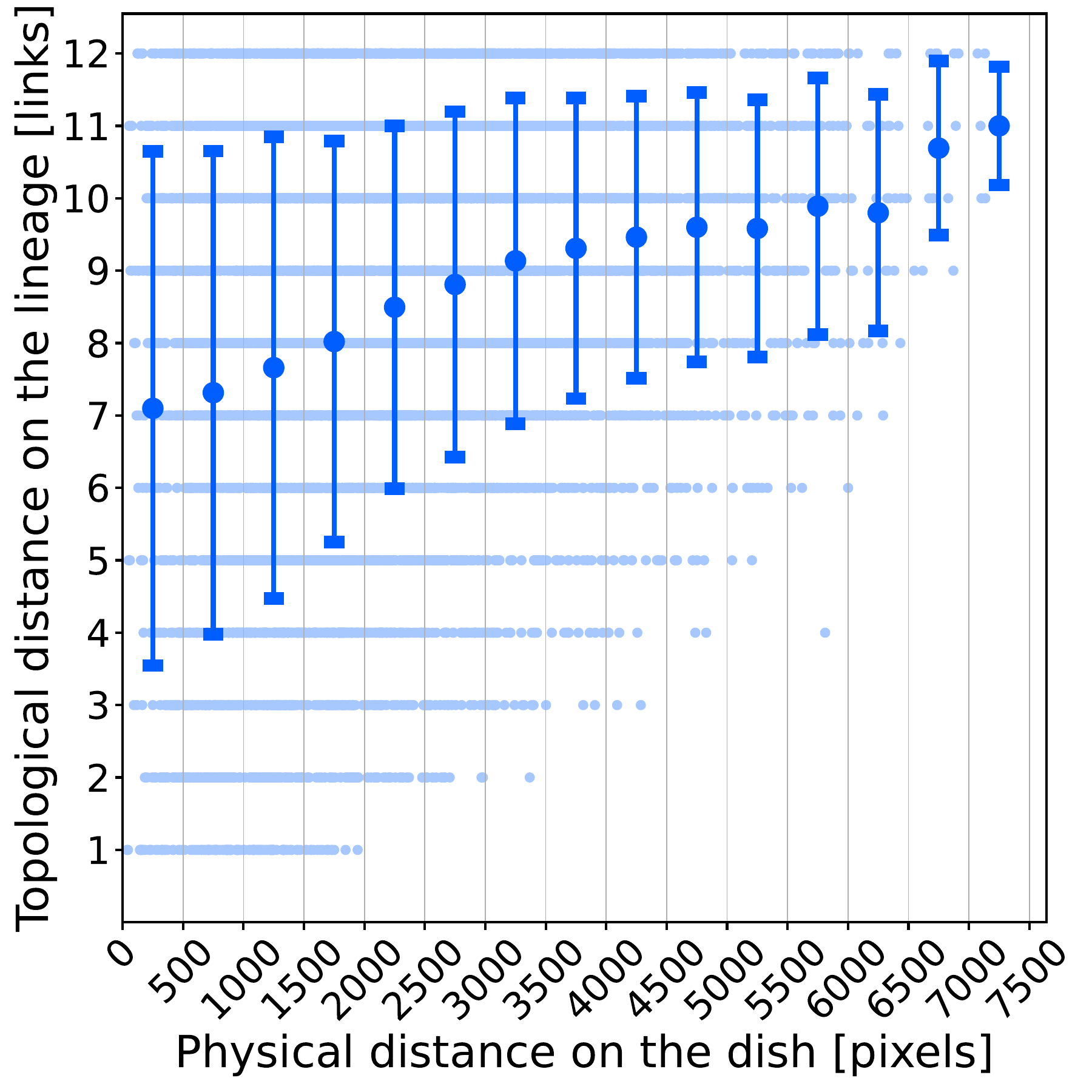}
\caption{Mutual distances between all mitosis in a sample colony, \texttt{20241017\_02} (light blue points); topological distance on the tree is measured in number of links separating two mitosis/nodes, while physical distance on the dish is measured in pixels separating the two mitosis. Correlation between the two types of distance is weak, with Spearman correlation coefficient $\rho=0.24$ (P-value $< 10^{-6}$). Even small physical distances correspond to a very wide range of topological distances; this is clear if we group together all physical distances in bins of $500$ pixels and average in each bin the relative topological distances (dark large points --- error bars are standard deviations). This plot demonstrates that short topological distances correspond to short physical distances, but that short physical distances do not correspond to short topological distances. In other words, a small spot on the dish is mapped onto a very large ``region'' on the tree: a spot of 250 pixels maps to an average distance on the tree of 7 links, which is very large in genealogical terms.}
\end{figure}
%%%%%%%%%%%%%%%%%%%%%%%%%%%%%%%%%%%%%%%%%%%%%%%

Proximity in the physical space is not the same as proximity on the tree. More precisely, points that are close on the tree tend also to be close in physical space, since cells derived from a common close progenitor start their life closer in space than cells derived by far progenitors; but the crucial point is that the {\it vice-versa} is not true, namely points close in physical space are not necessarily close in the lineage (Fig.6). 
For the sake of argument, let us assume that there exist on the Petri dish a localised spot of linear size $R$ where spatial factors promote the G$_0$ state; if a mitosis occurs in that  G$_0$-promoting spot and if the colony is already quite crowded --- and hence mobility is limited --- it is likely that also the two daughters' mitosis will occur in that same spot. If these were the only mitosis taking place in the spot, this could explain the clustering that we observe in the lineage trees without the need to invoke an hereditary mechanism. However, this is not what happens, as clearly illustrated in Fig.6:  the range of topological distances between the points belonging to a physical region of size $R$ is {\it very} wide; even a physical spot of a few hundred pixels, which is about the size of a cell, has a typical size on the tree of over $7$ links, and it is very likely to find within this spot points with topological distance of up to $12$ links, which is the maximum amplitude of any lineage. This result indicates that any hypothetical G$_0$-promoting spatial spot --- even a {\it very} small one --- would contain many mitosis with {\it large} topological distances from each other, which would imply that the G$_0$ cells emerging within the spot would be scattered all over the tree and not clustered within some branches, which is instead what we observe.  We conclude that the G$_0$ structure we discovered is very unlikely to be of spatial origin and that the hereditary explanation is the most plausible one.

%%%%%%%%%%%%%%%%%%%%%%%%%%%%%%%%

\section{Table 1 -- BMSC experimental data}
%%%%%%%%%%%%%%%%%%%%%%%%%%%%%%%%%%%%%%%%%%%%%%%%%
\begin{figure}[h!]
\centering
\includegraphics[width=1 \textwidth]{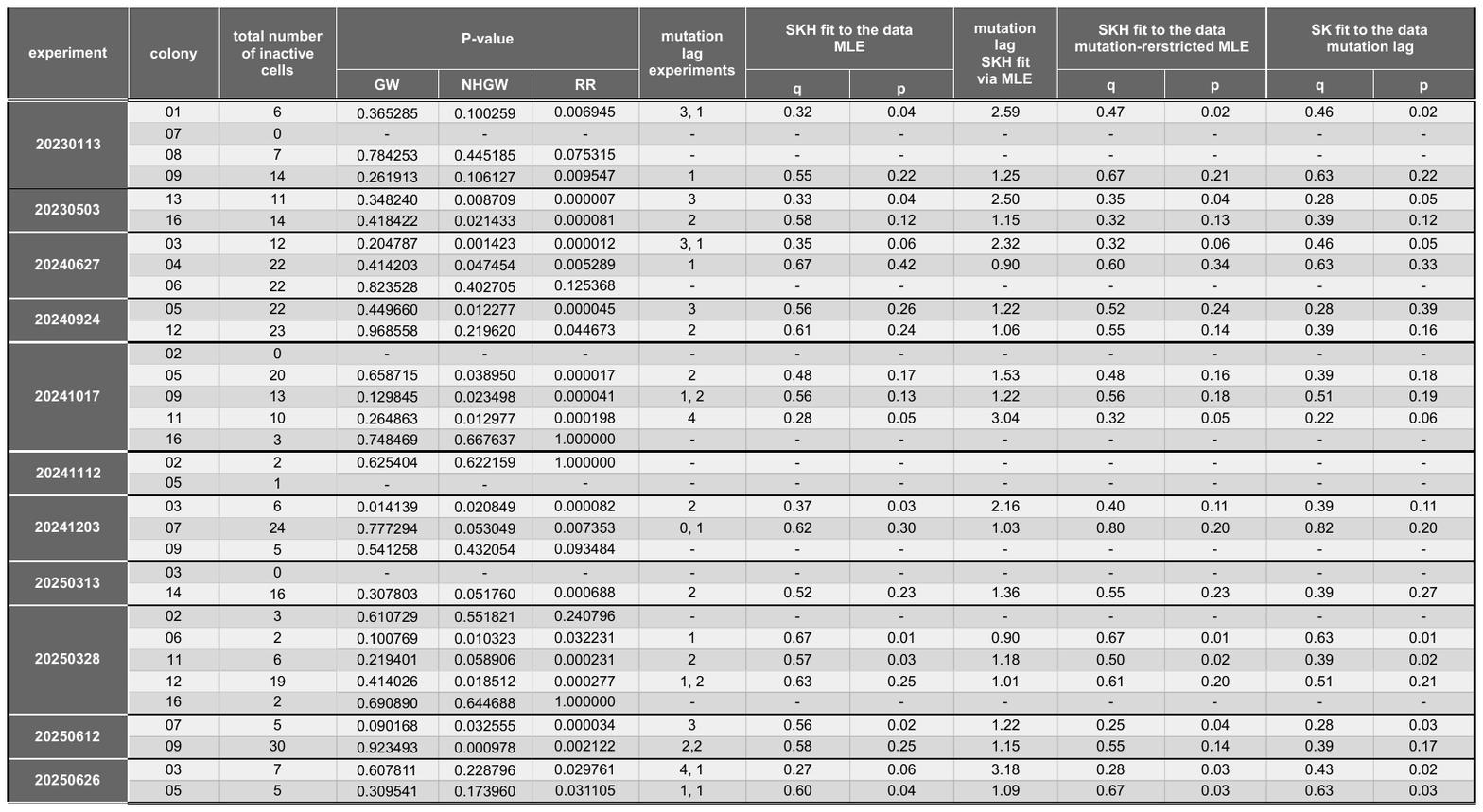}
\begin{justify}
{\bf Supplemental Table 1.} For each BMSC colony we report the outcome of the inheritance tests with the three methods we scrutinized, $P_\mathrm{GW}$, $P_\mathrm{NHGW}$ and $P_\mathrm{RR}$, where only this last one has been used to actually assess the inheritance level of the biological dataset.  These P-values have been calculated by generating ${\cal R}=10^6$ trees of the relative null ensembles. In the Table we report also the number of inactive cells of each colony: when $N_\mathrm{inact}\leq 1$, we do not report the (trivial) inheritance test for that lineage. We also report the mutation lag for each experimental colony that passed the RR inheritance test; when two mutations are present, we report both mutation lags. Then we report the SKH values of $p$ and $q$ fitted with the three different methods described in the main text. Finally, we report the mutation lag calculated with the exact relations of the SKH model, using as an input the parameters $(p,q)$ fitted to that clone by MLE. 
\end{justify}
\end{figure}
%%%%%%%%%%%%%%%%%%%%%%%%%%%%%%%%%%%%%%%%%%%%%%%

%%%%%%%%%%%%%%%%%%%%%%%%%%%%%%%%
\section{Lineage zoology}

In this Section we report a very diverse selection of lineage trees (all with $\kmax=7$) together their P-values according to the different inheritance tests, so that the reader can get a feeling of why and how a certain method performs well or not on a given tree.  Every P-value that does not agree with the nature of the generating model ($P<0.05$ on a non-hereditary tree, or $P\geq 0.05$ on a hereditary tree) is marked in red.  The criteria of choice of each set are explained in the relative captions.  We stress that we are {\it not} using this selection of trees to quantitatively benchmark the inheritance test, but merely to illustrate where and how things go wrong for certain methods with respect to others.  The quantitative benchmarking of the inheritance tests rests completely on the parameters $\Phi$ (the fraction of trees that pass the test) reported in Fig.~4 of the main text.  At the end of the Section we finally report the lineage trees of all 32 biological BMSC colonies.

%%%%%%%%%%%%%%%%%%%%%%%%%%%%%%%%%%%%%%%%%%%%%%%%%
\begin{figure}
\centering
\includegraphics[width=0.8 \textwidth]{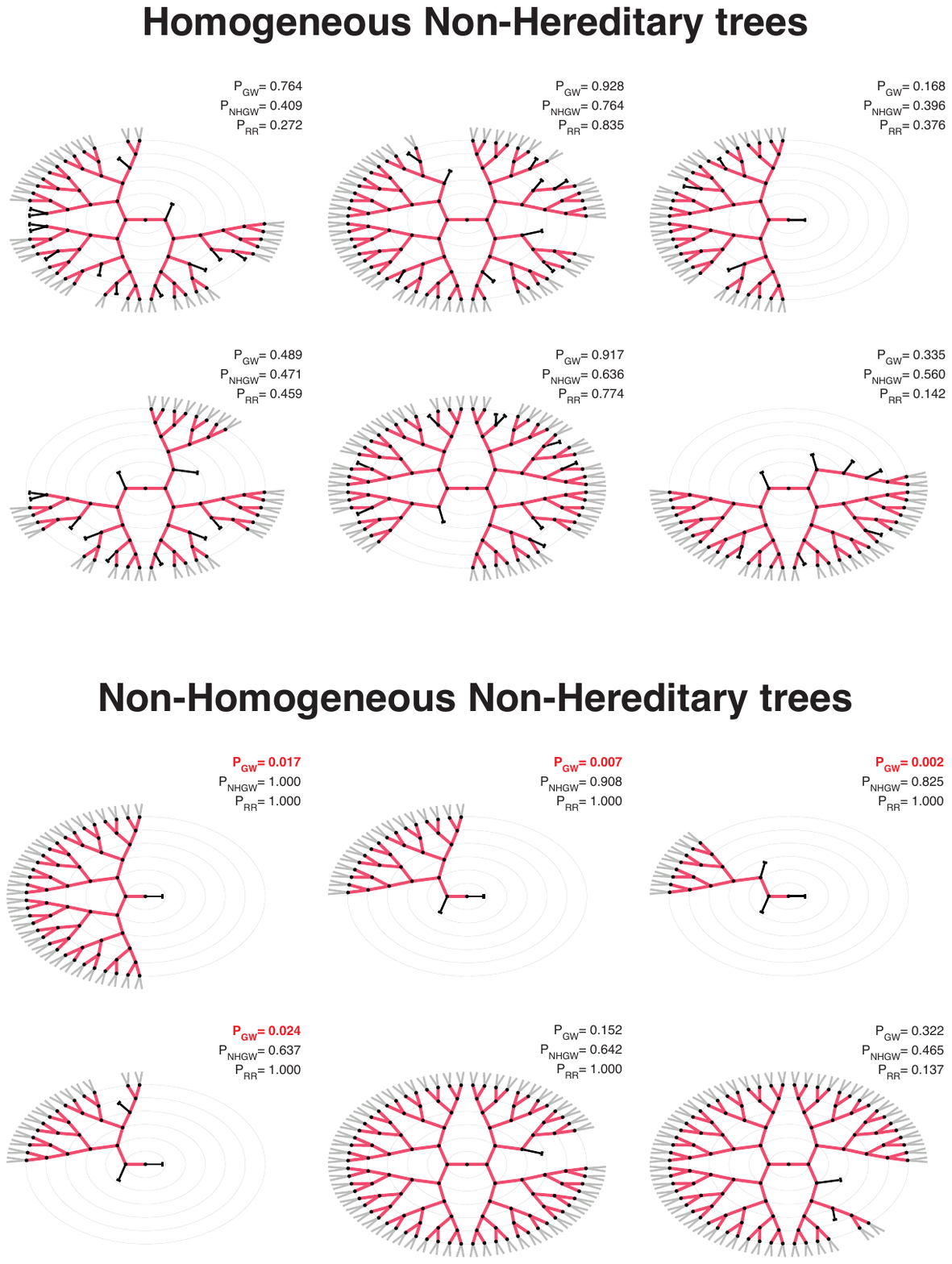}
\caption{
{\bf Simulated -- Non-Hereditary trees.} 
{\it Top: Homogeneous} -- These six trees have been produced by the GW model with a probability of inactivity $q=0.1$. This value produces a number of inactive cells quite similar (on average) to our biological dataset.  Apart from this, we have used no other criterion to select the six trees, as we know that all methods perform well on the homogeneous non-hereditary trees, namely they all give a very small fraction of significant trees.
{\it Bottom: Non-Homogeneous} -- These six trees have been produced by the NHGW model with a generation-dependent probability of inactivity given by the profile, $q_k=q_0 \exp(-k/k_0)$, with $q_o=0.8$ and $k_0=1.1$. In the case of the non-homogeneous non-hereditary trees, we have a very large arbitrariness  in the choice of $q_k$. Because we have no a priori expectation on the profile $q_k$ for our biological systems, we preferred selecting a profile that produced some instructive {\it false positives}  for the GW method. As we explained, GW is unable to recognise the $k$-dependent nature of the probability of inactivity, hence it deems a tree very unlikely, even though the tree is perfectly non-hereditary. On the other hand, both NHGW and RR perform well on the non-homogeneous non-hereditary trees. 
}
\end{figure}
%%%%%%%%%%%%%%%%%%%%%%%%%%%%%%%%%%%%%%%%%%%%%%%

%%%%%%%%%%%%%%%%%%%%%%%%%%%%%%%%%%%%%%%%%%%%%%%%%
\begin{figure}
\centering
\includegraphics[width=0.8 \textwidth]{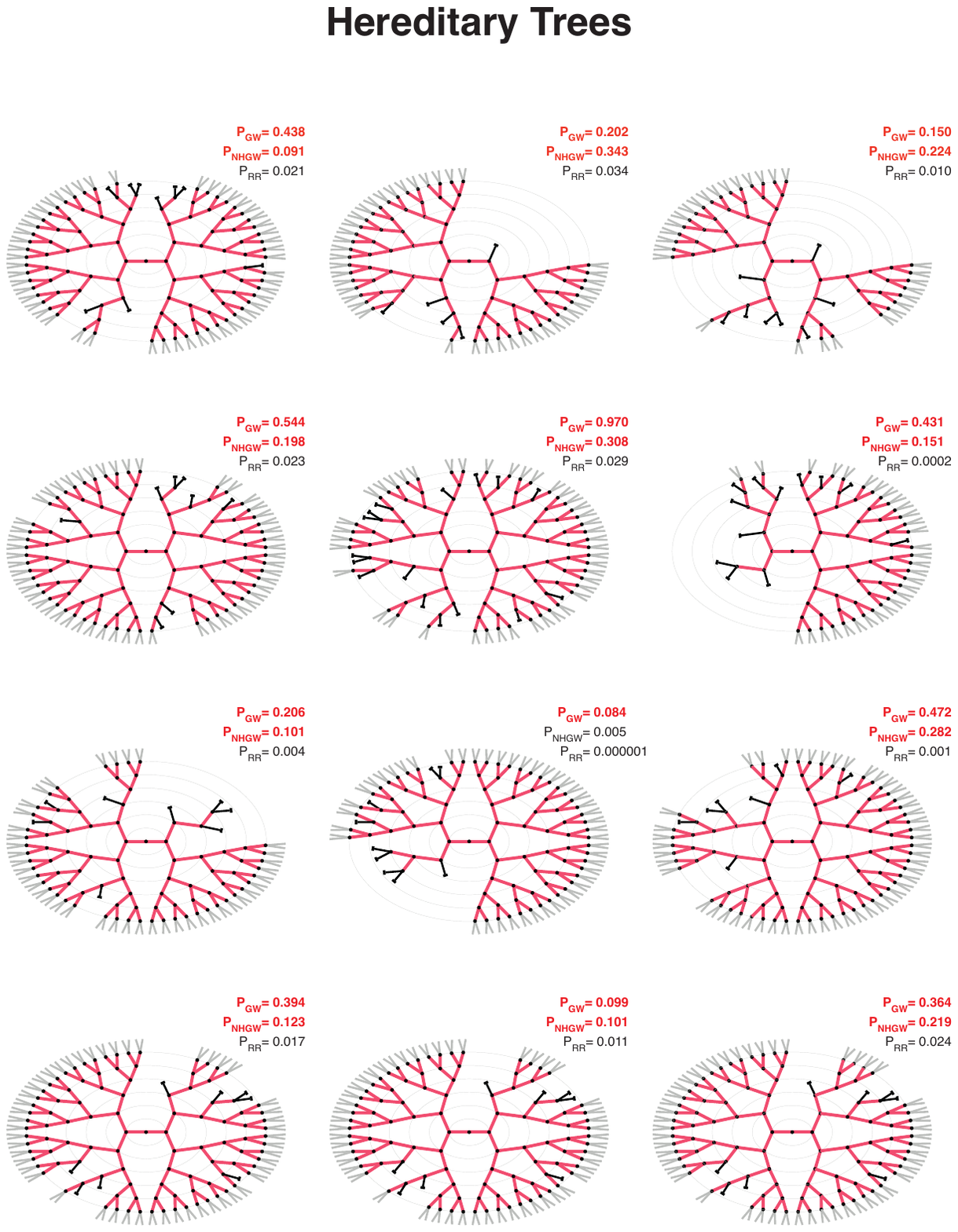}
\caption{
{\bf Simulated -- Hereditary trees.}
This is the most interesting and illuminating case, hence we present 12 hereditary trees produced with the SKH model. The parameters are $p=0.1$ and $q=0.46$, the mean MLE values fitted on the biological dataset.  We have seen in the benchmarking Section of the main text that --- when applied to the SKH hereditary set of trees --- NHGW gives a fraction of hereditary trees much smaller than RR, and much smaller than the theoretical value in the asymptotic limit, $\kmax\to\infty$. Therefore, we have chosen these SKH trees to illustrate this point, namely to show practical cases in which NHGW flags the tree as non-hereditary, while RR says that it is.  For $\kmax=7$ we {\it never} found a tree for which $P_\mathrm{NHGW} < 0.05$ and $P_\mathrm{RR}>0.05$ (and we have extracted millions of lineages), so we present here trees that illustrate the other disagreement, that is $P_\mathrm{NHGW} > 0.05$ with $P_\mathrm{RR} < 0.05$.
}
\end{figure}
%%%%%%%%%%%%%%%%%%%%%%%%%%%%%%%%%%%%%%%%%%%%%%%
%%%%%%%%%%%%%%%%%%%%%%%%%%%%%%%%%%%%%%%%%%%%%%%%%
\begin{figure}
\centering
\includegraphics[width=0.8 \textwidth]{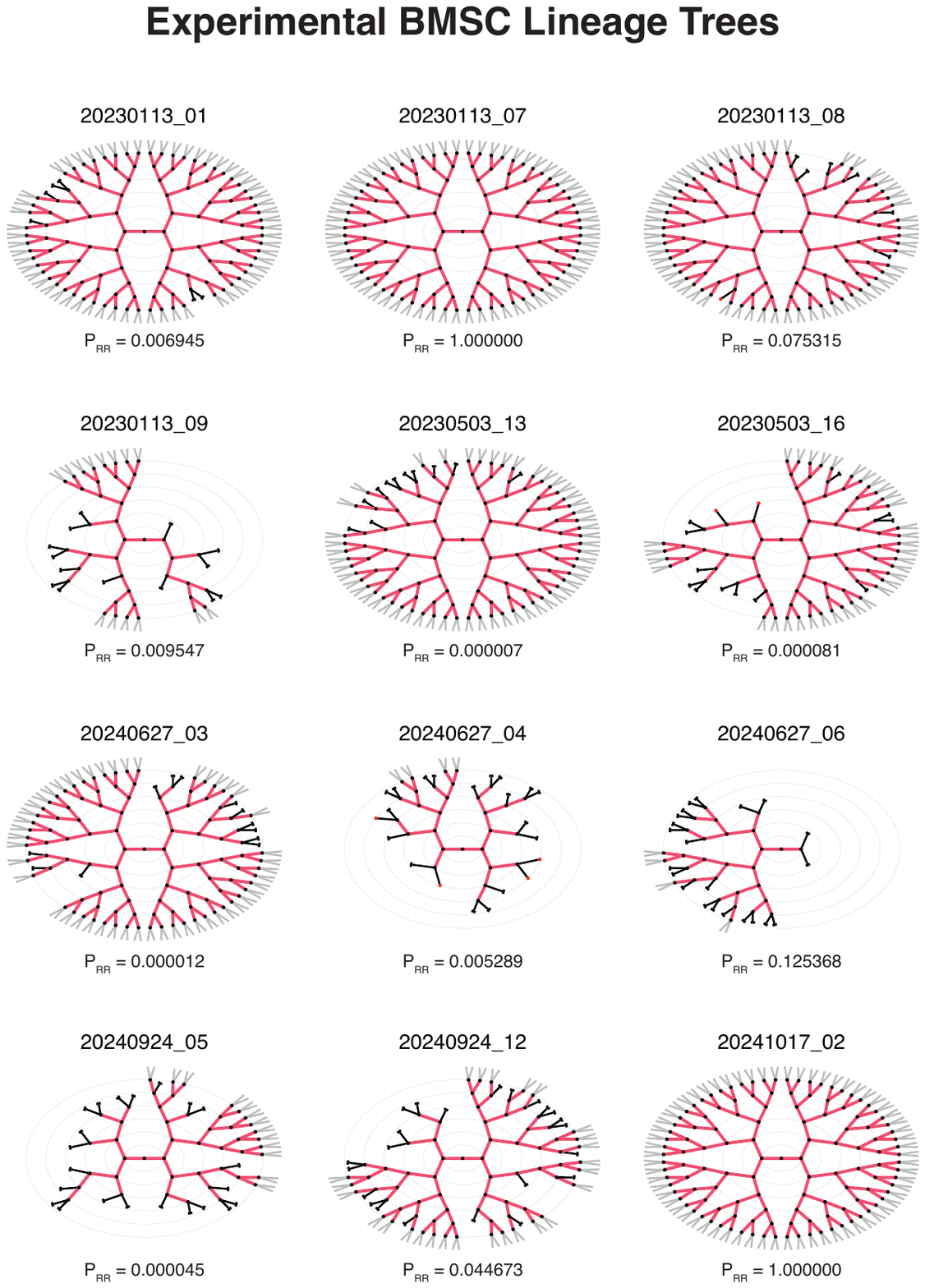}
\caption{
{\bf Experimental  BMSC trees -- Panel 1/3.}
In this and in the following two figures we report all 32 biological trees in our dataset. Beneath each one of the biological BMSC trees with more than one inactive cell, we report only $P_\mathrm{RR}$, which is the test selected and used in the main text to make the final inheritance assessment. The other inheritance test P-values can be found in Table 1.
}
\end{figure}
%%%%%%%%%%%%%%%%%%%%%%%%%%%%%%%%%%%%%%%%%%%%%%%
%%%%%%%%%%%%%%%%%%%%%%%%%%%%%%%%%%%%%%%%%%%%%%%%%
\begin{figure}
\centering
\includegraphics[width=0.8 \textwidth]{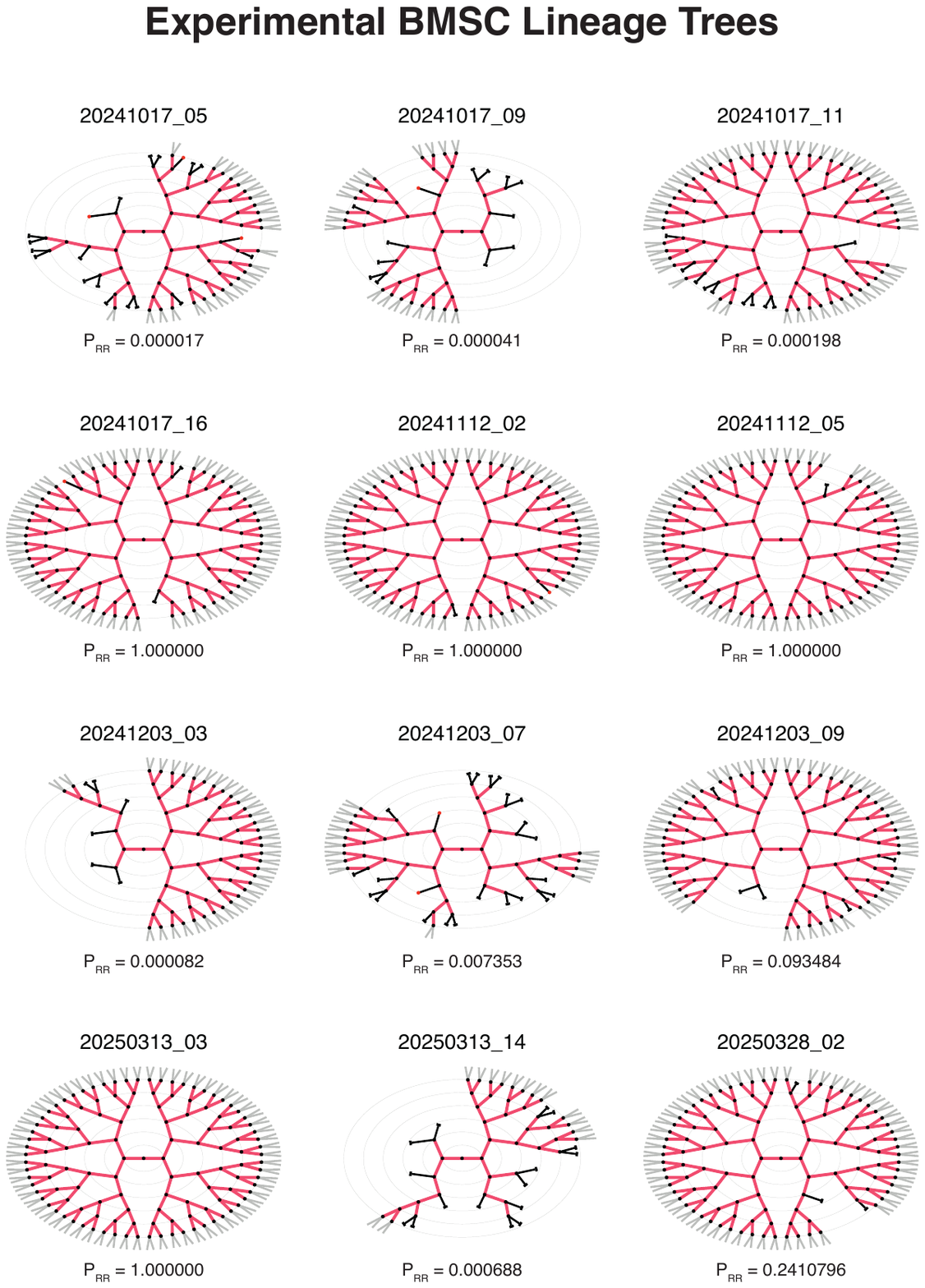}
\caption{
{\bf Experimental BMSC trees -- Panel 2/3.} 
}
\end{figure}
%%%%%%%%%%%%%%%%%%%%%%%%%%%%%%%%%%%%%%%%%%%%%%%
%%%%%%%%%%%%%%%%%%%%%%%%%%%%%%%%%%%%%%%%%%%%%%%%%
\begin{figure}
\centering
\includegraphics[width=0.8 \textwidth]{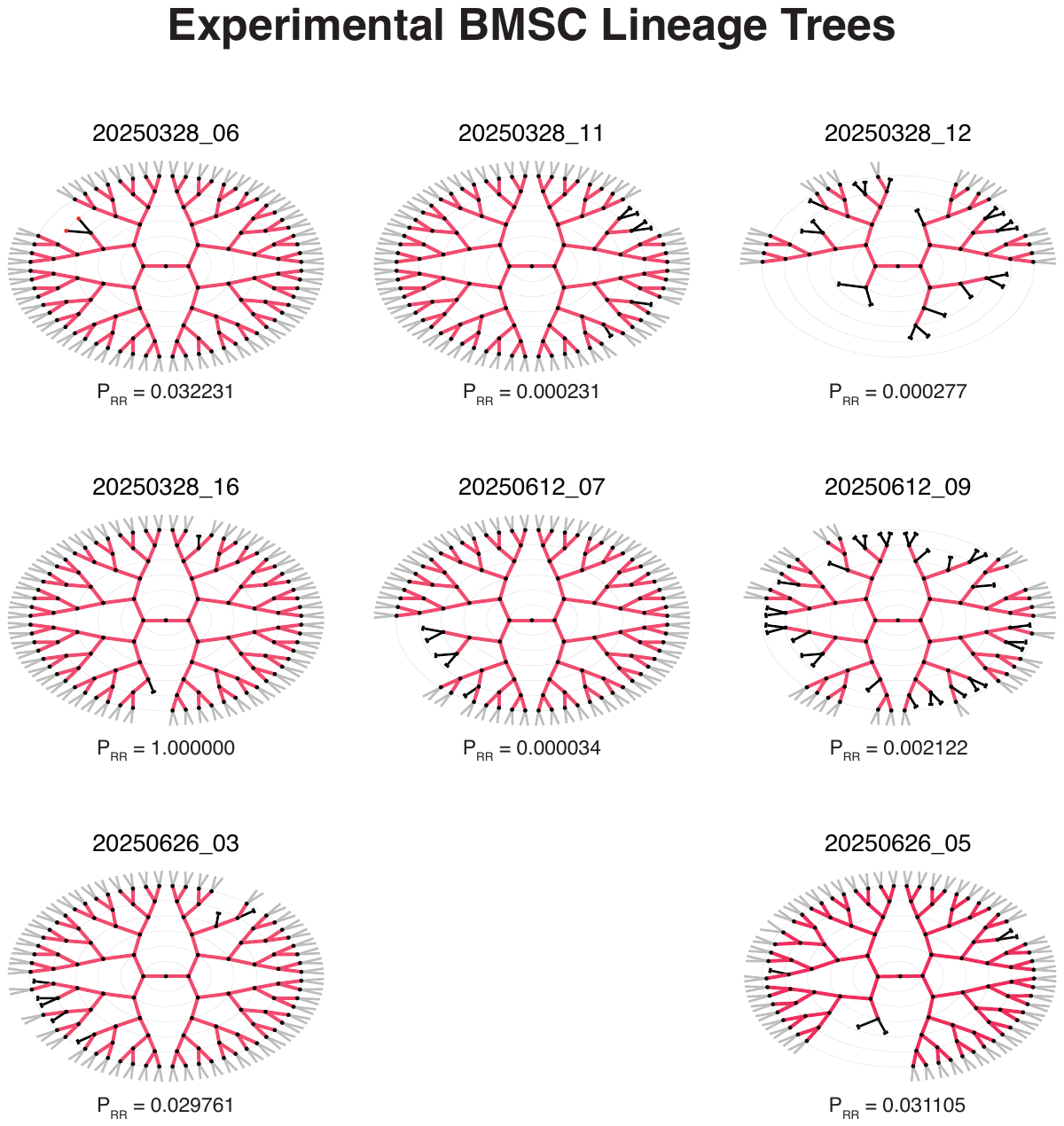}
\caption{
{\bf Experimental BMSC trees -- Panel 3/3.} 
}
\end{figure}
%%%%%%%%%%%%%%%%%%%%%%%%%%%%%%%%%%%%%%%%%%%%%%%

\newpage

\clearpage

%%%%%%%%%%%%%%%%%%%%%%%%%%%%%%%%%%%%%%%%%%%%%%%
%%%%%%%%%%%%%%%%%%%%%%%%%%%%%%%%%%%%%%%%%%%%%%%
\bibliographystyle{apsrev4-1}
\bibliography{bibliography-stem-cells-cobbs}
%%%%%%%%%%%%%%%%%%%%%%%%%%%%%%%%%%%%%%%%%%%%%%%
%%%%%%%%%%%%%%%%%%%%%%%%%%%%%%%%%%%%%%%%%%%%%%%